\documentclass[pdflatex,sn-nature]{sn-jnl}

\usepackage[utf8]{inputenc}
\usepackage[T1]{fontenc}
\usepackage{graphicx}%
\usepackage{multirow}%
\usepackage{amsmath,amssymb,amsfonts}%
\usepackage{amsthm}%
\usepackage{mathrsfs}%
\usepackage[title]{appendix}%
\usepackage{xcolor}%
\usepackage{textcomp}%
\usepackage{manyfoot}%
\usepackage{lscape}
\usepackage{booktabs}%
\usepackage{gensymb}
\usepackage{algorithm}%
\usepackage{algorithmicx}%
\usepackage{algpseudocode}%
\usepackage{listings}%

\theoremstyle{thmstyleone}%
\theoremstyle{thmstyletwo}%

\theoremstyle{thmstylethree}%

\usepackage{anyfontsize} 

\begin{document}

\title{Failed jet breakout in the metal-poor broad-lined type Ic supernova 2026gzf}


\author*[1]{\fnm{Antonio} \sur{Martin-Carrillo}}\email{antonio.martin-carrillo@ucd.ie}
\affil[1]{\orgdiv{School of Physics and Centre for Space Research, University College Dublin}, \orgname{UCD}, \orgaddress{\street{Belfield}, \city{Dublin}, \postcode{D04 V1W8}, \country{Ireland}}}\label{aff:UCD}

\author[2]{\fnm{Christina C.} \sur{Th\"one}}
\affil[2]{\orgname{E. Kharadze Georgian National Astrophysical Observatory}, \orgaddress{\street{Mt. Kanobili}, \city{Abastumani, Adigeni}, \postcode{0301}, \country{Georgia}}}\label{aff:AbAO}

\author[3,4,5]{\fnm{James K.} \sur{Leung}}
\affil[3]{\orgdiv{David A. Dunlap Department of Astronomy and Astrophysics}, \orgname{University of Toronto}, \orgaddress{\street{50 St. George Street}, \city{Toronto}, \postcode{M5S 3H4}, \state{Ontario}, \country{Canada}}}
\affil[4]{\orgdiv{Dunlap Institute for Astronomy and Astrophysics}, \orgname{University of Toronto}, \orgaddress{\street{50 St. George Street}, \city{Toronto}, \postcode{M5S 3H4}, \state{Ontario}, \country{Canada}}}
\affil[5]{\orgdiv{Racah Institute of Physics}, \orgname{The Hebrew University of Jerusalem}, \orgaddress{\city{Jerusalem}, \postcode{91904}, \country{Israel}}}

\author[1]{\fnm{Gregory} \sur{Corcoran}}

\author[6]{\fnm{Antonio} \sur{de Ugarte Postigo}}
\affil[6]{\orgdiv{Aix Marseille Univ.}, \orgname{CNRS, CNES, LAM}, \orgaddress{\street{38, rue Frédéric Joliot-Curie}, \city{Marseille}, \postcode{13388}, \country{France}}}

\author[7]{\fnm{Peter G.} \sur{Jonker}}
\affil[7]{\orgdiv{Department of Astrophysics/IMAPP}, \orgname{Radboud University}, \orgaddress{\street{Houtlaan 4}, \city{Nijmegen}, \postcode{6525 XZ}, \country{The Netherlands}}}\label{aff:Radboud}

\author[8,9]{\fnm{Luca} \sur{Izzo}}
\affil[8]{\orgname{INAF}, \orgdiv{Osservatorio Astronomico di Capodimonte}, \orgaddress{\street{Salita Moiariello 16}, \postcode{I-80121} \city{Naples}, \country{Italy}}}\label{aff:INAF-OAC}
\affil[9]{\orgdiv{DARK, Niels Bohr Institute}, \orgname{University of Copenhagen}, \street{Jagtvej 128}, \postcode{2200} \city{Copenhagen}, \country{Denmark}}\label{aff:DARK}

\author[7, 10]{\fnm{Andrew J.} \sur{Levan}}
\affil[10]{\orgdiv{Department of Physics}, \orgname{University of Warwick}, \orgaddress{\street{Coventry, CV4 7AL}, \country{UK}}}

\author[11, 12]{\fnm{Benjamin P.} \sur{Gompertz}}
\affil[11]{\orgdiv{School of Physics and Astronomy}, \orgname{University of Birmingham}, \orgaddress{\street{Edgbaston}, \city{Birmingham}, \postcode{B15 2TT}, \state{West Midlands}, \country{UK}}}\label{aff:Birmingham1}
\affil[12]{\orgdiv{Institute for Gravitational Wave Astronomy}, \orgname{University of Birmingham}, \orgaddress{\street{Edgbaston}, \city{Birmingham}, \postcode{B15 2TT}, \state{West Midlands}, \country{UK}}}\label{aff:Birmingham2}

\author[6]{\fnm{St\'ephane} \sur{Basa}}

\author[13, 14]{\fnm{Nikhil} \sur{Sarin}}
\affil[13]{\orgdiv{Kavli Institute for Cosmology}, \orgname{University of Cambridge}, \orgaddress{\street{Madingley Road}, \city{Cambridge}, \postcode{CB3 0HA}, \country{UK}}}
\affil[14]{\orgdiv{Institute of Astronomy}, \orgname{University of Cambridge}, \orgaddress{\street{Madingley Road}, \city{Cambridge}, \postcode{CB3 0HA}, \country{UK}}}

\author[7]{\fnm{Jonathan} \sur{Quirola-V\'asquez}}

\author[15]{\fnm{Rob A. J.} \sur{Eyles-Ferris}}
\affil[15]{\orgdiv{School of Physics and Astronomy}, \orgname{University of Leicester}, \orgaddress{\street{University Road}, \city{Leicester}, \postcode{LE1 7RH}, \country{UK}}}\label{aff:Leicester}

\author[16]{\fnm{Riccardo} \sur{Brivio}}
\affil[16]{\orgdiv{INAF–Osservatorio Astronomico di Brera}, \orgaddress{\street{Via E. Bianchi 46}, \city{Merate (LC)}, \postcode{23807}, \country{Italy}}}\label{aff:INAF-OAB}

\author[17]{\fnm{Alan M.} \sur{Watson}}
\affil[17]{\orgdiv{Instituto de Astronom\'ia}, \orgname{Universidad Nacional Aut\'onoma de M\'exico}, \orgaddress{\street{A.P. 70-264}, \postcode{04510}, \state{Ciudad de M\'exico}, \country{Mexico}}}

\author[1]{\fnm{Laura} \sur{Cotter}}

\author[18]{\fnm{Jennifer Alexandra} \sur{Chac\'on}}
\affil[18]{\orgdiv{Instituto de Astrof{\'{\i}}sica}, \orgname{Facultad de F{\'{\i}}sica, Pontificia Universidad Cat{\'o}lica de Chile}, \orgaddress{\street{Campus San Joaqu{\'{\i}}n, Av. Vicu\~na Mackenna 4860}, \city{Santiago}, \postcode{7820436}, \country{Chile}}}\label{aff:PUC}

\author[19]{\fnm{Andrea} \sur{Rossi}}
\affil[19]{\orgdiv{INAF - Osservatorio di Astrofisica e Scienza dello Spazio},\orgaddress{\street{via Piero Gobetti 93/3}, \city{Bologna}, \postcode{I-40129},\, \country{Italy}}} 

\author[20]{\fnm{Andrea} \sur{Melandri}}
\affil[20]{\orgdiv{INAF - Osservatorio Astronomico di Roma},\orgaddress{\street{via Frascati 33}, \city{Monte Porzio Catone}, \postcode{I-00078},\, \country{Italy}}} 

\author[21, 22]{\fnm{Piramon} \sur{Kumnurdmanee}}
\affil[21]{\orgdiv{Deparment of Physics and Astronomy}, \orgname{Northwestern University}, \orgaddress{\street{2145 Sheridan Road}, \city{Evanston}, \postcode{60208}, \state{Illinois}, \country{USA}}}
\affil[22]{\orgdiv{Center for Interdisciplinary Exploration and Research in Astrophysics (CIERA)}, \orgname{Northwestern University}, \orgaddress{\street{1800 Sherman Ave., 8th Floor}, \city{Evanston}, \postcode{60208}, \state{Illinois}, \country{USA}}}

\author[15]{\fnm{Nial R.} \sur{Tanvir}}

\author[23, 24]{\fnm{Anshika} \sur{Gupta}}
 \affil[23]{\orgname{Aryabhatta Research Institute of Observational Sciences (ARIES)}, \orgaddress{\city{Manora Peak, Nainital}, \postcode{263001}, \country{India}}}\label{aff:ARIES}
\affil[24]{\orgname{Department of Physics, Indian Institute of Technology Roorkee }, \orgaddress{\city{Roorkee}, \postcode{247667}, \country{India}}}\label{aff:IITR}

\author[25]{\fnm{Franz E.} \sur{Bauer}}
\affil[25]{\orgdiv{Instituto de Alta Investigaci{\'{o}}n}, \orgname{Universidad de Tarapac{\'{a}}}, \orgaddress{\street{Casilla 7D}, \city{Arica|, \postcode{1010069}, \country{Chile}}}}\label{aff:Tarapaca}

\author[26]{\fnm{Jean-Gr\'egoire} \sur{Ducoin}}
\affil[26]{\orgdiv{CPPM}, \orgname{Aix Marseille Univ, CNRS/IN2P3}, \orgaddress{\street{163, Av. de Luminy}, \city{Marseille}, \postcode{13288}, \state{Marseille}, \country{France}}}\label{aff:CPPM}

\author[27, 16]{\fnm{Andrea} \sur{Reguitti}}
\affil[27]{\orgname{INAF – Osservatorio Astronomico di Padova}, \orgaddress{Vicolo dell'Osservatorio 5}, \city{Padova}, \postcode{35122}, \country{Italy}}

\author[23]{\fnm{Kuntal} \sur{Misra}}

\author[28]{\fnm{Dong} \sur{Xu}}
\affil[28]{\orgname{National Astronomical Observatories, Chinese Academy of Sciences}, \orgaddress{\city{Beijing}, \postcode{100101}, \country{China}}}

\author[29]{\fnm{Susanna D.} \sur{Vergani}}
\affil[29]{\orgdiv{LUX}, \orgname{Observatoire de Paris, Université PSL, CNRS, Sorbonne Uni-
versité}, \orgaddress{\postcode{92190}, \city{Meudon}, \country{France}}}

\author[21, 22]{\fnm{Wen-fai} \sur{Fong}}

\author[10]{\fnm{Kendall} \sur{Ackley}}
\author[30]{\fnm{Edilberto} \sur{Aguilar-Ruiz}}
\author[31, 32]{\fnm{Dalya} \sur{Akl}}
\author[33, 34]{\fnm{Miguel \'Angel} \sur{Aloy}}
\author[28]{\fnm{Jie} \sur{An}}
\author[17]{\fnm{Camila} \sur{Angulo-Valdez}}
\author[35]{\fnm{Sarah} \sur{Antier}}
\author[36]{\fnm{Jean-Luc} \sur{Atteia}}
\author[17]{\fnm{Rosa L.} \sur{Becerra}}
\author[37]{\fnm{Rene P.} \sur{Breton}}
\author[38]{\fnm{Nathaniel R.} \sur{Butler}}
\author[16]{\fnm{Sergio} \sur{Campana}}
\author[20]{\fnm{Francesco} \sur{Carotenuto}}
\author[39]{\fnm{Jorge} \sur{Casares Vel\'azquez}}
\author[7,40]{\fnm{Ashley A.} \sur{Chrimes}}
\author[41]{\fnm{Valerio} \sur{D'Elia}}
\author[7]{\fnm{Joyce N. D. van} \sur{Dalen}}
\author[42]{\fnm{Fabio De} \sur{Colle}}
\author[43]{\fnm{Massimiliano De} \sur{Pasquale}}
\author[44, 39]{\fnm{Vik S.} \sur{Dhillon}}
\author[26]{\fnm{Damien} \sur{Dornic}}
\author[44, 45]{\fnm{Martin J.} \sur{Dyer}}
\author[16]{\fnm{Matteo} \sur{Ferro}}
\author[1]{\fnm{Morgan} \sur{Fraser}}
\author[46]{\fnm{Andrew S.} \sur{Fruchter}}
\author[36]{\fnm{Francis} \sur{Fortin}}
\author[47,48]{\fnm{Duncan K.} \sur{Galloway}}
\author[49]{\fnm{Leonardo} \sur{Garc\'ia-Garc\'ia}}
\author[39, 50]{\fnm{Stefan} \sur{Geier}}
\author[30]{\fnm{Ramandeep} \sur{Gill}}
\author[49]{\fnm{No\'emie} \sur{Globus}}
\author[19]{\fnm{Roberto} \sur{Gualandi}}
\author[26]{\fnm{Marion} \sur{Guelfand}}
\author[16, 51]{\fnm{Francesco} \sur{Guidolin}}
\author[52]{\fnm{Dieter H.} \sur{Hartmann}}
\author[7]{\fnm{Agnes P. C. van} \sur{Hoof}}
\author[53]{\fnm{Pall} \sur{Jakobsson}}
\author[23, 54]{\fnm{Divyanshu} \sur{Janghel}}
\author[10]{\fnm{Tom L.} \sur{Killestein}}
\author[55]{\fnm{Sylvio} \sur{Klose}}
\author[56]{\fnm{Shiho} \sur{Kobayashi}}
\author[57]{\fnm{Rubina} \sur{Kotak}}
\author[58]{\fnm{Amit} \sur{Kumar}}
\author[30]{\fnm{Asuka} \sur{Kuwata}}
\author[59]{\fnm{Tanmoy} \sur{Laskar}}
\author[17]{\fnm{William H.} \sur{Lee}}
\author[26]{\fnm{Massimiliano} \sur{Lincetto}}
\author[39,50]{\fnm{Gianluca} \sur{Lombardi}}
\author[42]{\fnm{Diego} \sur{L\'opez-C\'amara}}
\author[10]{\fnm{Joseph D.} \sur{Lyman}}
\author[19]{\fnm{Elisabetta} \sur{Maiorano}}
\author[60]{\fnm{Keiichi} \sur{Maeda}}
\author[6]{\fnm{Nikos} \sur{Mandarakas}}
\author[27]{\fnm{Francesco} \sur{Magnani}}
\author[61]{\fnm{Jirong} \sur{Mao}}
\author[62]{\fnm{Enrique} \sur{Moreno M\'endez}}
\author[55]{\fnm{Ana Mar\'ia} \sur{Nicuesa Guelbenzu}}
\author[63]{\fnm{Kanthanakorn} \sur{Noysena}}
\author[64]{\fnm{Laura K.} \sur{Nuttall}}
\author[15]{\fnm{Paul T.} \sur{O'Brien}}
\author[11]{\fnm{David} \sur{O'Neill}}
\author[27, 65]{\fnm{Paolo} \sur{Ochner}}
\author[49, 66]{\fnm{Margarita} \sur{Pereyra}}
\author[67]{\fnm{Giovanna} \sur{Pugliese}}
\author[68]{\fnm{Gavin} \sur{Ramsay}}
\author[69,70]{\fnm{Lauren} \sur{Rhodes}}
\author[71, 72]{\fnm{Andrea} \sur{Saccardi}}
\author[73]{\fnm{Ruben} \sur{Salvaterra}}
\author[17]{\fnm{Fredd} \sur{S\'anchez \'Alvarez}}
\author[6]{\fnm{Benjamin} \sur{Schneider}}
\author[74]{\fnm{Steve} \sur{Schulze}}
\author[15]{\fnm{Rhaana L. C.} \sur{Starling}}
\author[10]{\fnm{Danny} \sur{Steeghs}}
\author[10]{\fnm{Kzrysztof} \sur{Ulaczyk}}
\author[20]{\fnm{Chiara} \sur{Ventura}}
\author[75,76]{\fnm{Tayyaba} \sur{Zafar}}
\author[28]{\fnm{Zi-Pei} \sur{Zhu}}


\affil[30]{\orgdiv{Instituto de Radioastronom\'ia y Astrof\'isica}, \orgname{Universidad Nacional Aut\'onoma de M\'exico}, \orgaddress{\street{Antigua Carretera a P\'atzcuaro \# 8701,\\ Ex-Hda. San Jos\'e de la Huerta}, \city{Morelia}, \postcode{58089}, \state{Michoac\'an}, \country{Mexico}}}
\affil[31]{\orgname{New York University Abu Dhabi}, \orgaddress{\street{Saadiyat Island}, \city{Abu Dhabi}, \postcode{PO Box 129188}, \country{UAE}}}
\affil[32]{\orgdiv{Center for Astrophysics and Space Science (CASS)}, \orgname{New York University Abu Dhabi}, \orgaddress{\street{Saadiyat Island},\city{Abu Dhabi}, \postcode{PO Box 129188}, \country{UAE}}}
\affil[33]{\orgname{Departament d'Astronomia i Astrof\'isica, Universitat de Val\`encia}, \orgaddress{\street{Av. Vincent Andr\'es Estell\'es, 19}, \city{Burjassot}, \postcode{46100}, \country{Spain}}}
\affil[34]{\orgname{Observatori Astron\`omic, Universitat de Val\`encia}, \orgaddress{\city{Paterna}, \postcode{46980}, \country{Spain}}}
\affil[35]{\orgname{IJCLab}, \orgaddress{\street{15 Rue Georges Clemenceau}, \city{Orsay}, \postcode{91400}, \country{France}}}
\affil[36]{\orgdiv{Univ Toulouse, CNES, CNRS, IRAP}, \orgaddress{\street{9, Av. du Colonel Roche}, \city{Toulouse}, \postcode{31028}, \country{France}}}
\affil[37]{\orgdiv{Jodrell Bank Centre for Astrophysics, Department of Physics and Astronomy}, \orgname{The University of Manchester}, \orgaddress{\city{Manchester}, \postcode{M13 9PL}, \state{State}, \country{UK}}}
\affil[38]{\orgdiv{School of Earth \& Space Exploration}, \orgname{Arizona State University}, \orgaddress{\postcode{85287}, \state{AZ}, \country{USA}}}
\affil[39]{\orgname{Instituto de Astrof\'isica de Canarias}, \orgaddress{\city{La Laguna}, \postcode{E-38205}, \state{Tenerife}, \country{Spain}}}
\affil[40]{\orgdiv{European Space Research and Technology Centre (ESTEC)}, \orgname{European Space Agency (ESA)}, \orgaddress{\street{Keplerlaan 1}, \city{Noordwijk}, \postcode{2201 AZ}, \country{the Netherlands}}}
\affil[41]{\orgdiv{Space Science Data Centre}, \orgname{Italian Space Agency}, \orgaddress{\street{Via del Politecnico snc}, \city{Rome}, \postcode{00133}, \country{Italy}}}
\affil[42]{\orgdiv{Instituto de Ciencias Nucleares}, \orgname{Universidad Nacional Aut{\'o}noma de M{\'e}xico}, \orgaddress{\street{Circuito Exterior S/N}, \city{Coyoacan}, \postcode{70-543 04510}, \country{Mexico}}}
\affil[43]{\orgdiv{Department of Mathematics and Computer Sciences, Physical Sciences and Earth Sciences}, \orgname{University of Messina}, \orgaddress{Via F. S. D'Alcontres 31}, \city{Messina}, \postcode{98166}, \country{Italy}}
\affil[44]{\orgdiv{Astrophysics Research Cluster, School of Mathematical and Physical Sciences}, \orgname{University of Sheffield}, \orgaddress{\city{Sheffield}, \postcode{S3 7RH}, \state{State}, \country{UK}}}
\affil[45]{\orgdiv{Research Software Engineering}, \orgname{University of Sheffield}, \orgaddress{\city{Sheffield}, \postcode{S1 4DP}, \country{UK}}}
\affil[46]{\orgdiv{Department}, \orgname{Space Telescope Science Institute}, \orgaddress{\street{3700 San Martin Drive}, \city{Baltimore}, \postcode{21218}, \state{MD}, \country{United Sates of America}}}
\affil[47]{\orgdiv{School of Physics \& Astronomy}, \orgname{Monash University}, \orgaddress{\city{Clayton}, \postcode{VIC 3800}, \country{Australia}}}
\affil[48]{\orgname{Institute for Globally Distributed Open Research and Education (IGDORE)}}
\affil[49]{\orgdiv{Instituto de Astronom{\'\i}a}, \orgname{Universidad Nacional Aut\'onoma de M\'exico}, \orgaddress{\street{km 107 Carretera Tijuana-Ensenada}, \city{Ensenada}, \postcode{22860}, \state{Baja California}, \country{Mexico}}}
\affil[50]{\orgdiv{Gran Telescopio CANARIAS}, \orgaddress{\street{C/ Cuesta de San Jos\'e s/n}, \city{Bre\~{n}a Baja}, \postcode{38712}, \country{Spain}}}
\affil[51]{\orgdiv{Dipartimento di Scienza e Alta Tecnologia}, \orgname{Universit\'a degli studi dell'Insubria}, \orgaddress{\street{Via Valleggio 11}, \city{Como}, \postcode{22100}, \country{Italy}}}
\affil[52]{\orgdiv{Department of Physics and Astronomy}, \orgname{Clemson University}, \orgaddress{\street{Kinard Lab of Physics}, \city{Clemson}, \postcode{SC 29634-0978}, \country{USA}}}
\affil[53]{\orgdiv{Centre for Astrophysics and Cosmology, Science Institute}, \orgname{University of Iceland}, \orgaddress{\street{Dunhagi 5}, \city{Reykjavik, \postcode{107}, \country{Iceland}}}}
\affil[54]{\orgname{Mahatma Jyotiba Phule Rohilkhand University}, \orgaddress{\city{Bareilly}, \postcode{243006}, \country{India}}}
\affil[55]{\orgname{Th\"uringer Landessternwarte Tautenburg}, \orgaddress{\street{Sternwarte 5}, \city{Tautenburg}, \postcode{07778}, \country{Germany}}}
\affil[56]{\orgdiv{Astrophysics Research Institute}, \orgname{Liverpool John Moores University}, \orgaddress{\street{146 Brownlow Hill}, \city{Liverpool}, \postcode{L3 5RF}, \country{UK}}}
\affil[57]{\orgdiv{Department of Physics \& Astronomy}, \orgname{University of Turku}, \orgaddress{\street{Vesilinnantie 5}, \city{Turku}, \postcode{FI-20014}, \country{Finland}}}
\affil[58]{\orgdiv{Centre for Electronic Imaging, School of Physical Sciences}, \orgname{The Open University}, \orgaddress{\street{Walton Hall}, \city{Milton Keynes}, \postcode{MK7 6AA}, \country{UK}}}
\affil[59]{\orgdiv{{Department of Physics \& Astronomy}, \orgname{University of Utah}, \orgaddress{\city{Salt Lake City}, \postcode{84112}, \state{Utah}, \country{USA}}}}
\affil[60]{\orgdiv{Department of Astronomy}, \orgname{Kyoto University}, \orgaddress{\street{Kitashirakawa-Oiwake-cho, Sakyo-ku}, \city{Kyoto}, \postcode{606-8502}, \, \country{Japan}}}
\affil[61]{\orgdiv{Yunnan Observatories, Chinese Academy of Sciences}, \orgaddress{\street{396 YangFangWang}, \city{Kunming}, \postcode{650216},\, \country{China}}}
\affil[62]{\orgdiv{Facultad de Ciencias}, \orgname{Universidad Nacional Aut\'onoma de M\'exico}, \orgaddress{\street{Circuito Exterior S/N}, \city{Coyoac\'an}, \postcode{Apartado Postal 70-264, 04510}, \city{CDMX}, \country{Mexico}}}
\affil[63]{\orgname{National Astronomical Research Institute of Thailand}, \orgaddress{\city{Chiang Mai}, \postcode{50180}, \country{Thailand}}}
\affil[64]{\orgdiv{Institute of Cosmology and Gravitation}, \orgname{University of Portsmouth}, \orgaddress{\city{Portsmouth}, \postcode{PO1 3FX}, \country{UK}}}
\affil[65]{\orgdiv{Department of Physics and Astronomy}, \orgname{University of Padova}, \orgaddress{\street{Vicolo dell'Osservatorio 3}, \city{Padova}, \postcode{35122}, \country{Italy}}}
\affil[66]{\orgname{Secretar\'ia de Ciencia, Humanidades, Tecnolog\'ia, e Innovaci\'on}}
\affil[67]{\orgdiv{Anton Pannekoek Institute of Astronomy, University of Amsterdam}, \orgaddress{\street{Science Park 904}, \city{ Amsterdam},  \postcode{1098 XH}, \country{The Netherlands}}}
\affil[68]{\orgname{Armagh Observatory \& Planetarium}, \orgaddress{\street{College Hill}, \city{Armagh}, \postcode{BT61 9DG}, \country{UK}}}
\affil[69]{\orgdiv{Trottier Space Institute at McGill, 3550 Rue University, Montreal, Quebec H3A 2A7, Canada}}
\affil[70]{\orgdiv{Department of Physics, McGill University, 3600 Rue University, Montreal, Quebec H3A 2T8, Canada}}
\affil[71]{\orgdiv{Université Paris-Saclay, Université Paris Cité}, \orgname{CEA, CNRS, AIM}, \orgaddress{\street{91191}, \city{Gif-sur-Yvette}, \country{France}}}
\affil[72]{\orgdiv{Centre national d’études spatiales (CNES)},\orgaddress{\city{Paris}, \country{France}}}
\affil[73]{INAF-Istituto di Astrofisica Spaziale e Fisica Cosmica di Milano, Via A. Corti 12, 20133 Milano, Italy}
\affil[74]{\orgdiv{Department of Particle Physics and Astrophysics}, \orgname{Weizmann Institute of Science}, \orgaddress{\street{234 Herzl St}, \city{Rehovot}, \postcode{76100}, \country{Israel}}}
\affil[75]{\orgdiv{School of Mathematical and Physical Sciences}, \orgname{Macquarie University}, \orgaddress{ \city{Sydney}, \postcode{2109}, \state{NSW}, \country{Australia}}}
\affil[76]{\orgdiv{Astrophysics and Space Technologies Research Centre}, \orgname{Macquarie University}, \orgaddress{\city{Sydney}, \postcode{2109}, \state{NSW}, \country{Australia}}}


\abstract{A long-standing question in the death of massive stars is the role of relativistic jets. While many $\gamma$-ray bursts and some fast X-ray transients seem to be associated with broad-lined type Ic supernovae \cite[Ic-BL SNe,][]{cano17, finneran25a, vanDalen25}, the opposite is not true. The lack of observable jet emission in those Ic-BL SNe can be explained by invoking off-axis jets\cite{corsi23, schroeder25}, choked jets that inject all their energy into the stellar envelope \cite{piran19, Hamidani25},  baryon-loaded jets for which the prompt high-energy emission is strongly suppressed \cite{huang02}, or non-jetted SNe \cite{Hamidani25}. The lack of exact explosion time in the majority of SNe presents an obstacle to distinguish between these scenarios. Here we report the properties of SN\,2026gzf associated with the X-ray thermal \emph{Einstein Probe} shock-breakout EP260321a at $z$\,$=$\,0.0343. The absence of compelling shocked cocoon and radio emission up to 54 days, combined with initial expansion velocities of $\sim$30,000\,km\,s$^{-1}$ and a circumstellar shell of $\sim$\,0.07\,M$_\odot$, favour a scenario for SN\,2026gzf in which a jet was choked in the circumstellar shell. Our high-spatial resolution images of the SN environment show that the progenitor was located between two highly star-forming regions with a metallicity lower than any previously known Ic-BL SN. As the first case of a Ic-BL SN associated with high-energy prompt emission without the signature of a jet, SN\,2026gzf provides a unique perspective to understand the successful launch of relativistic jets during the deaths of massive stars.}




\maketitle

\section{Introduction}\label{sec1}

Supernovae (SNe) discovered through the detection of high-energy transients offer a unique opportunity to study the death of massive stars from the exact moment when the explosion occurs. Typically, this has been done through $\gamma$-ray instruments, the so-called $\gamma$-ray burst SN (hereafter GRB-SN) population, with SN\,1998bw \cite{Galama1998} as the archetype. In most of these cases, the early emission is dominated by the GRB afterglow, with the SN component emerging after a few days and dominating, in the optical/NIR bands, during its peak time at $\sim$\,15\,--20\, days since the explosion in the rest-frame of the source. In others, such as GRB\,060218/SN\,2006aj \cite{Campana2006}, GRB\,100316D/SN\,2010bh \cite{Starling2011} and GRB\,171205A/SN\,2017iuk \cite{Izzo2019}, while the X-ray and radio wavelengths show evidence of relativistic jets, the optical is dominated by an initial shocked cocoon, lasting $\sim 1$\,--\,5\,days, and followed by a Nickel-powered SN bump. Notably, all GRB-SN cases consist of H- and He-free stripped envelope SN with broad-lines (hereafter Ic-BL SNe) \cite{cano17, corsi23}. 

The launch of \emph{Einstein Probe} (EP) in January 2024, opened a window to study the origin of fast X-ray transients (FXTs). Since then, a handful of FXTs have been associated with Ic-BL SNe. Mirroring the GRB-SN population, some optical counterparts of FXTs seem to show a combination of emission from relativitic jets and SN \cite[e.g.][]{vanDalen25, Srivastav25, Sun25}, while others, such as SN\,2025kg \cite{eylesferris25, rastinejad25, srinivasaragavan25a}, SN\,2025fhm \cite{cotter26}, SN\,2025wkm \cite{srinivasaragavan25b}, show optical emission dominated by the shocked cocoon and the Nickel-powered SN bump. 

Regardless of the properties of the relativistic jets and thus its direct presence in the optical/NIR light curves, in all cases, the $\gamma$/X-ray prompt emission consists of a non-thermal component attributed to emission from the relativistic jet. In those cases where jet emission seems to be missing in optical/NIR, a thermal component is also seen, often linked to the SN shock-breakout \cite[e.g][]{Campana2006, Starling2011, Izzo2019, cotter26}. 

SN\,2008D, associated with XRF\,080109 \cite[e.g.][]{Soderberg2006, Mazzali08, modjaz09}, is an outlier in the population of SNe with prompt emission. This event is generally considered the first SN shock-breakout ever observed. However, the prompt emission of SN\,2008D, discovered by \emph{Swift}, seems to show two spectral components, one thermal and one non-thermal, similarly to the typical GRB/FXT-SN cases, suggesting the presence of possible jetted emission \cite{Mazzali08}. In contrast, the spectral monitoring campaign revealed that SN\,2008D fits better as a type Ib SN, instead of the usual Ic-BL. Despite being significantly fainter than any previously known GRB/FXT-SN and without detection of the relativistic jet after the prompt emission ended, SN\,2008D also exhibited an initial early bump in its light curve that dominated during the $\sim$\,4\,d after the explosion, which is likely associated with a shocked cocoon phase, and thus would point to the existence of a jet \cite{Soderberg08, Mazzali08}.

Although the picture of Ic-BL SNe associated with high-energy prompt emission seems to be quite consistent, the question of whether the larger population of Ic-BLs with no explosion time emission is capable of forming and launching relativistic jets is still not clear. Studies of the temporal and spectroscopic properties of this ``normal" Ic-BL population do not reveal major differences compared to those with high-energy emission \cite[e.g.][]{modjaz16, srinivasaragavan24, finneran2025b}. Even though some ``normal" Ic-BLs are expected to launch jets seen off-axis, e.g. SN\,2020bvc \cite{Izzo2020}, dedicated radio campaigns, which provide the most optimal wavelength for these types of search, only seem to find them in a handful of cases \cite{schroeder25}. This is likely to hint at a limit on the number of SNe capable of, at least successfully, launching a relativistic jet. In the optical band, the shocked cocoon, created from the energy deposited by the jet when breaking through the stellar envelope, is expected to be seen even when the jet is off-axis \cite{zheng25, zheng26}, providing an additional tool to reveal the existence of the jet, even if not detected directly. Even if the jet is fully choked and is not capable of escaping the stellar envelope, the energy deposited in this layer is believed to provide the necessary momentum to generate ejecta velocities, $v$\,$>$\,0.1$c$ \cite{piran19}.

\section{\texorpdfstring{Properties of SN\,2026gzf}{Properties of SN 2026gzf}}\label{sec2}
On March 21, 2026 at 12:30:18 (UTC, referred to as T$_{0}$), the EP mission reported a new X-ray transient EP260321a \citep{gcn44068, gcn44075}. The prompt emission lasted $\sim$\,432\,s and consisted only of thermal emission with a temperature of $\sim$\,0.16\,keV and no additional non-thermal component, making EP260321a a strong SN shock-breakout candidate, as studied in detail in \cite{Yuan26}. An optical counterpart candidate was found at a location consistent with a bright blue knot in a nearby galaxy at $z$\,=\,0.0343 \cite{gcn44070}. Notably, the location was also consistent with the position of an optical transient, AT2018mtl, reported eight years prior \cite{astronote18}. Pre-explosion variability was suggested through recent Rubin observations \cite{gcn44084}, with long time variability presented in \cite{janet26}. We used the Gravitational-wave Optical Transient Observer (GOTO, see Methods) telescope to search for pre-explosion variability up to $\sim$\,2.4\,years before the EP trigger. During this time, no significant detection is found using the wide-band optical filter $L$ (wavelength range of 400\,nm\,--\,700\,nm) up to $L\,>\,20.16$, corresponding to an absolute magnitude of $M_L\,>\,-15.76$.

Our monitoring campaign started at $\sim T_0 + 6.48$\,hr, revealing the candidate source towards the North-West of its host galaxy (Fig.~\ref{fig:snimage}). This initial observation was quickly followed by VLT FORS2 (imaging) and MUSE (spectroscopy) observations at $T_0+12.23$\,hr and $T_0+16.98$\,hr, respectively, which confirmed the variability of the optical counterpart and established its distance at a common redshift to the underlying galaxy, $z$\,=\,0.0343 (see Methods). 

As shown in the inset of Fig.~\ref{fig:snimage}, our high spatial resolution VLT/MUSE narrow field mode (NFM) observation, taken at $T_0+21.52$\,days, resolved the SN location, revealing a complex environment in which the SN falls in between two bright star formation regions with an offset from the centre of the known blue knot (dominated by the much brighter Eastern star forming region) of $\Delta \text{RA} = 0.192\pm0.024^{\prime\prime}$ and $\Delta\text{Dec} = 0.031\pm0.026^{\prime\prime}$. This is equivalent to a projected distance of 134\,$\pm$\,24\,pc. Our ability to disentangle the SN from the H\,\textsc{ii} region allows us to precisely locate the SN at $\text{RA} = 09$:59:42.870 and $\text{Dec} = +00$:25:06.39 with an error of 0.12$^{\prime\prime}$ (see Methods).

Our rapid response to the EP trigger allowed us to catch the end of a short-lived phase in which the optical SN seems to initially fade, only to quickly start rising at $\sim T_0+12$\,hr after the EP trigger (see Fig.~\ref{fig:snlightcurve}). The optical/NIR light curve is largely consistent with the emission from $^{56}$Ni radioactive decay. However, there is a clear excess in the emission during the full rising phase (see Fig.~\ref{fig:snlightcurve}, panel c) that requires an additional component, which we interpret as circumstellar material (CSM) caused by mass-loss from the progenitor during its final stages (see Methods). We find that a total CSM mass of $\sim$0.07\,M$_{\odot}$ in a shell with inner radius $\sim$\,3\,$\times$\,10$^{13}$\,cm, can explain the pre-peak behaviour of SN\,2026gzf. These values are quite consistent with the CSM shell required in the case of the Ic-BL SN\,2018gep \cite{ho19}. With our model, we derive a SN kinetic energy of $\sim$\,2\,$\times$\,10$^{52}$\,erg, indicative of a potential central engine, an ejecta mass of $\sim$\,7\,M$_{\odot}$ and a Nickel mass of $\sim$\,0.28\,M$_{\odot}$. We also find that the initial fading seen before the rising phase kicks off could be explained by the cooling of the SN shock-breakout (see Fig.~\ref{fig:snlightcurve}, panel b and Methods). 

Our first spectroscopic observation at $T_0+16.98$\,hr reveals a featureless thermal continuum of temperature $\sim$\,15,000\,K (Fig.~\ref{fig:spectra}), consistent with the spectra typically seen during the shock cooling. It should be noted that at the time of this first spectrum, the SN was already starting to rise, and thus the shock cooling and the CSM + $^{56}$ Ni SN components were both in play at this time. After this initial epoch, the SN features start to emerge with the clear presence of the typical SN Ic-BL absorption features such as the Ca\,\textsc{ii} doublet, Mg\,\textsc{ii}, Fe\,\textsc{ii} and Si\,\textsc{ii}. Additionally, from $T_0+2.5$\,d (2.42\,d in rest frame time) the spectra seem to exhibit an absorption feature consistent with either He\,\textsc{i} and/or Na\,\textsc{i}a. In reality, the high velocity of the SN ejecta blends these two lines, and thus, it is likely that what we see is a mixture of both components, similar to what it was seen in the Ic-BL SNe SN\,2009bb \cite{Pignata11} and SN\,2016coi \cite{yamanaka17, Terreran19}. Using the methodology described in \cite{finneran2025b}, we measure an initial expansion velocity of the ejecta of $\sim$\,33,000\,km\,s$^{-1}$ that declines over the first 60 days since the explosion time, as a power-law with a decay index of $-0.30$\,$\pm$\,0.03. As shown in Fig.~\ref{fig:spectra}, up to $\sim$\,24\,d, the SN is estimated to expand outwards linearly at a speed of $\sim$\,0.06$c$. During this time, the SN cools very rapidly, reaching a recombination temperature of carbon and oxygen of $\sim$\,5000\,K after which the linear radius expansion evolution seems to stop as shown in Fig~\ref{fig:spectra} panel d.

Our two VLA radio observations, spanning until $T_0+54$\,d, set tight limits on any radio emission component associated with SN\,2026gzf. Extrapolating the results of our CSM+SN model to the radio bands, we see that our upper limits are well consistent with the model predictions, confirming the validity of the model obtained through the optical/NIR data. We then explore what limits these radio observations impose to any unseen relativistic jet (see Methods). The two radio upper limits are capable of putting important constraints to the existence of a relativistic jet, ruling out an on-axis jet for a large number of beamed jet kinetic energies and the full range of densities explored (Fig.~\ref{fig:afterglow}, left panel for an ISM model and Extended Data Fig.~\ref{fig:afterglow_wind} in the case of a wind model). The analysis shows multiple possible scenarios in which a jet could be ruled out even at off-axis angles of $\sim 50^{\circ}$. 

In particular, our radio observations show that a relativistic jet can exist only for off-axis viewing angles of $>14^{\circ}$ in the case of a weak EP250304a-like afterglow \cite{cotter26}, or $>34^{\circ}$ more typical values of jet kinetic energy and medium density \cite[e.g.][]{laskar15}. The limits derived for an ISM environment are overall similar to those found using a wind environment (see Extended Data Fig.~\ref{fig:afterglow_wind}). Invoking a structured jet geometry would add more sideways emission, resulting in limits constraining larger off-axis angles.

\section{Host galaxy and environment}
As part of our observing campaign we obtained integral field spectroscopy (IFS) using MUSE at the VLT in both wide-field (WFM) and narrow-field mode (NFM) on March 23 and April 13, 2026, respectively. WFM and NFM have spatial samplings of $0.2^{\prime\prime}$ and $0.025^{\prime\prime}$, respectively. 

\begin{figure}[!ht]
\centering
\includegraphics[angle=0,width=1\textwidth]{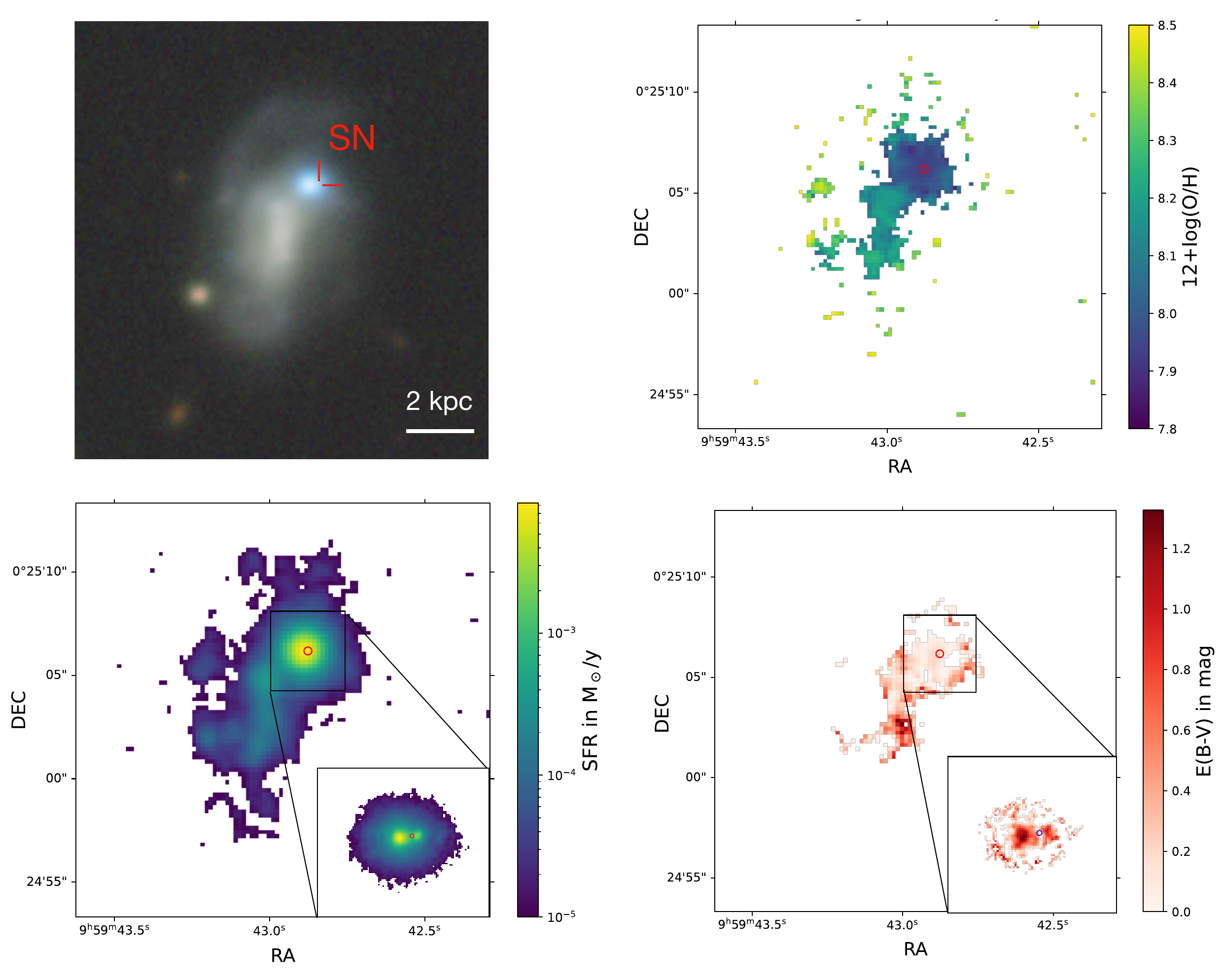}
\caption{\textbf{Top left}: Color image of the host galaxy made from gri images taken from the HSC-SSP survey \cite{HSC_SSP}. \textbf{Top right:} Metallicity map of the host galaxy obtained with the MUSE-WFM observation. \textbf{Bottom left:} Star formation map of the host galaxy from MUSE-WFM. The inset image shows the star formation in the local region of the SN as observed with MUSE-NFM. \textbf{Bottom right}: Extinction map of the host galaxy from MUSE-WFM. The inset image shows the extinction in the local region of the SN as observed with MUSE-NFM. In all images, the location of SN\,2026gzf is shown as a red circle (blue in the bottom right).}
\label{fig:sn_maps}
\end{figure}

The host galaxy is a star-forming dwarf galaxy ($\log (M_{*}/M_\odot) = 8.45^{+0.18}_{-0.13}$) as obtained from an SED fit to photometric data of the host, see Methods), possibly a dwarf spiral with an extended bar, with three to four spiral arms, or an irregular galaxy with a bar. The SN is located in a very bright and blue star-forming (SF) region in the Northern spiral arm with a diameter of $\sim1$\,kpc, reminiscent of 30\,Dor in the Large Magellanic Cloud. Our NFM data show that this SF region actually consists of two knots, one very bright, dominating the total flux to the East, and a smaller, dimmer one to the West. SN\,2026gzf is located in between these two SF regions, right at the edge of the fainter, western, SF region (see Figs.~\ref{fig:snimage} and \ref{fig:sn_maps}). Both regions show a large number of bright emission lines; the Eastern region also includes emission of [Ar\,\textsc{iv}] and a tentative detection of He\,\textsc{ii} indicative of a hard radiation field, likely from a very young stellar population (see Fig.~\ref{fig:sn_maps}). An X-shooter spectrum of the site of the SN furthermore shows Balmer lines in emission down to H20 (see Methods).

We create 2D maps of star formation rate (SFR) from H$\alpha$, extinction from the Balmer decrement (H$\alpha$/H$\beta$) and metallicity using the O3N2 parameter in the calibration of \cite{Marino13}. The bright SF region close to the SN has the lowest metallicity in the host as well as the highest SFR (the SFR in Fig.~\ref{fig:sn_maps} is plotted in logarithmic scale to distinguish the variate of SF regions within the host) and also a very high specific SFR (SSFR, see Fig.\ref{fig:sn_metallicity}). The Eastern SF region has a factor of $\sim$10 higher SFR than the one to the West. An integrated spectrum of the SN site (see Methods) results in a metallicity of $12+\log(\text{O/H}) = 7.97\pm0.05$. The extinction at the actual SN site is consistent with zero, the bright SF regions East and West do show $E(B-V) \sim 1$~mag in the centre of each region.


\section{Discussion}
\begin{figure}[!ht]
\includegraphics[angle=0,width=1\columnwidth]{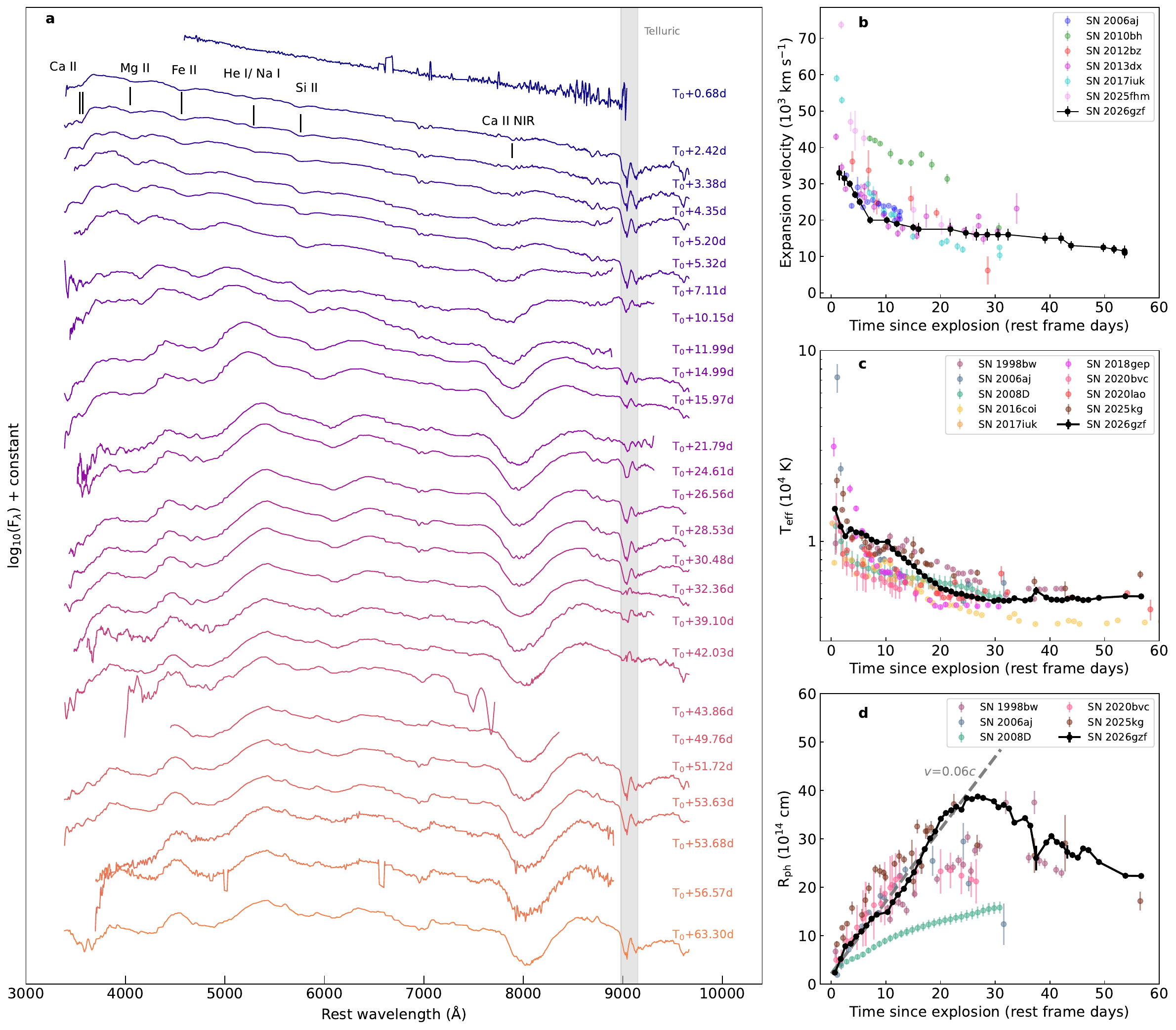}
\caption{\textbf{(a)} Spectroscopic sequence of SN\,2026gzf with rest-frame time since the EP trigger shown on the right side. The spectra have been smoothed with host emission lines and A and B telluric features been masked for clarity. Additional strong telluric features at $\sim$\,9000\,\AA\, have been highlighted in grey. The location of main spectroscopic features are indicated with vertical lines.\textbf{(b)} Expansion velocity evolution of the ejecta features seen in the SN spectra of SN\,2026gzf in comparison with a sample of GRB-SNe from \cite{finneran2025b, cotter26}. It's evolution is consistent with a single power-law decay. \textbf{(c)} Blackbody temperature evolution as a function of time in the rest frame from SED blackbody fitting in comparison to Ic-BL SNe with and without GRB associations from \cite{modjaz09, ho19, finneran25a, stritzinger26}. \textbf{(d)} Photosphere radius evolution as a function of time in the rest frame from SED blackbody fitting in comparison to Ic-BL SNe with and without GRB association. The dashed line shows the linear expansion velocity at $\sim$\,0.06\,\textit{c}.}
\label{fig:spectra}
\end{figure}

To date, all known GRB/FXT\,--\,SN associations in which the observed relativistic jet was too faint in the UV/optical bands are characterized by light curves with distinct two-stage behaviour. The early hotter phase, lasting $\sim$\,1\,--\,5\,days, is typically driven by emission from the shocked cocoon produced by the interplay between the relativistic jet successfully breaking through the circumstellar envelope surrounding the progenitor. The second, and longer phase, is instead powered by the radioactive decay of $^{56}${Ni}. As shown in the Extended Data Fig.~\ref{fig:sn_luminosity}, the early phase manifests itself in the blackbody bolometric luminosity evolution as an initial luminous fast evolving component which often rivals the peak luminosity of the $^{56}${Ni} powered bump. Furthermore, the high-energy prompt emission of these particular cases of GRB/FXT\,--\,SN displayed a non-thermal component associated with a relativistic jet, which is also revealed in their X-ray and radio light curves \citep[e.g.][]{Campana2006, Izzo2019, eylesferris25, rastinejad25, cotter26, Mazzali08}. These events also show extremely high velocities, reaching speeds of $\sim$\,50,000\,--\,70,000 km\,s$^{-1}$ in the first few days and high temperatures of up to $\sim$\,30,000\,K as shown in Fig.~\ref{fig:spectra}, panels b and c, respectively.

From the group of Ic-BL SNe with no association with high-energy prompt emission, SN\,2020bvc clearly shows an early phase compatible with a shocked cocoon, consistent with the progenitor successfully launching an off-axis relativistic jet as suggested by \cite{Izzo2020}. Interestingly, SN\,2009bb \cite{Pignata11} has also been proposed to have launched an off-axis jet on the basis of the extraordinary radio emission observed (although it could also be cocoon emission from a failed jet \cite{decolle18}). As the SN\,2009bb photometric monitoring seems to have missed the initial $\sim$\,3\,d since the true explosion time, the initial shocked cocoon phase was completely missed, being impossible to truly assess its existence, highlighting the limitations of not having a high-energy trigger.

In this context, the Extended Data Fig.~\ref{fig:sn_luminosity} shows that SN\,2026gzf clearly lacks this first initial phase with a luminosity at $\sim T_0+1$\,d $\sim0.003\times$  that of SN\,2006aj, despite being $\sim1.5\times$ brighter at the SN peak time, suggesting a lack of shocked cocoon emission. In comparison to SN\,2017iuk, the Ic-BL case with the best confirmed shocked cocoon \citep{Izzo2019}, SN\,2026gzf seems more luminous throughout its light curve, with the exception again of the very early phase, with a luminosity $\sim$\,0.3\,$\times$ that of SN\,2017iuk.

We note that the lack of a shocked cocoon phase was also suggested in the cases of SN\,2016coi \citep{yamanaka17, Terreran19} and SN\,2020lao \citep{stritzinger26}, two Ic-BL SNe without high-energy prompt emission and for which the uncertainty in the true explosion time created a significant hurdle to set tight constraints in the properties of the early light curve. When compared to SN\,2026gzf, these SNe look remarkably similar, specially SN\,2020lao. SN\,2016coi seems much less luminous throughout its entire light curve, but the shape of its luminosity evolution is a good match to SN\,2026gzf. Overall, it also seems to be much cooler with initial temperatures not reaching 10,000\,K. This may be an indication that the derived explosion time may indeed be underestimated by a few days or some slighly different physical process in play. In the case of SN\,2020lao the match to SN\,2026gzf in terms of its luminosity and temperature is excellent, although we note a similar issue regarding its initial temperature.

The moderate initial temperature of $\sim$\,15,000 \,K seems to further highlight the particular early behaviour of SN\,2026gzf at early times. We notice that, while the temperature evolution of SN\,2026gzf does not seem fully unique, there is a much more moderate cooling evolution during the first 10\,days than typically seen in other Ic-BL SNe, with a temperature $\sim$\,10,000\,K at $T_0+10$\,d, when the majority of Ic-BL SNe seem to cluster at $\sim$\,6000\,K \citep[see Fig.~\ref{fig:spectra}, panel c, and e.g.][]{Taddia19, srinivasaragavan24}. After $T_0+10$\,d, SN\,2026gzf seems to experience a phase of faster cooling, reaching a floor temperature of $\sim$\,5,000\,K consistent with most Ic-BL SNe. Interestingly, all Ic-BLs seem quite consistent in the observed photospheric radius evolution with an average speed of $\sim$0.06\,$c$ (see Fig.~\ref{fig:spectra}, panel b), suggesting a common engine among Ic-BL SNe. In this parameter space, SN\,2008D clearly stands out as the slowest expanding SN despite it having been proposed that a mildly relativistic jet may also have been successfully launched \cite{Mazzali08}.

\subsection{ZAMS mass estimation}
Our light curve model suggests that the Nickel powered SN phase consists of $\sim$\,7\,M$_{\odot}$ of ejected mass and $\sim$\,0.28\,M$_{\odot}$ of Nickel mass. These values are well within the typical range seen in many other Ic-BL SNe with and without high-energy prompt emission \citep[e.g.][]{cano17, finneran25a, srinivasaragavan24}. In the case of the ejected mass, the value seems to fall more in line with the population of GRB/FXT\,--\,SNe.

With these two values, we explore the possible range of ZAMS masses of the progenitor following a methodology similar to \cite{rastinejad25} using several approximations that should return a final value accurate within a factor of a few. For this estimation, we do not consider the effects of binary mass transfer, ejection, or irregular mass-loss rate episodes. Our calculations are heavily based on the single star model described in \cite{woosley2002} for a black hole accretion disk model. As shown in Fig.~\ref{fig:spectra}, SN\,2026gzf is consistent with a fully stripped envelope SN where all its H and He have been lost at the time of explosion. Thus, we start by setting the mass of the star at explosion to the C/O core mass. In this case, the C/O mass is the combination of the total ejecta mass obtained from the measured ejecta mass (6.7\,--\,7.4\,M$_\odot$) and Nickel mass (0.2\,--\,0.4\,\,M$_\odot$), the disk mass and the compact object remnant mass for which, in this case, we consider a black hole of 3\,--\,5\,\,M$_\odot$. As described in \cite[][and references therein]{rastinejad25}, for a jet-driven explosion, the total mass of Nickel mass ejected corresponds to $\sim$\,5\,$\times$ the one measured through the SN explosion. Thus, to eject 0.2\,--\,0.4\,\,M$_\odot$ of Nickel, about 1\,--\,2\,\,M$_\odot$ would have to be processed, leading to a total C/O core of 10\,--\,15\,M$_\odot$, when combining it with the ejected mass and the black hole mass. Using the C/O star masses for low-metallicity stars from \cite{woosley2002}, we estimate a ZAMS progenitor mass range of 33\,--\,42\,M$_\odot$. This mass range is consistent with the expectations for the progenitors of black holes \cite{fryer99}. However, we note that the range could be close to the limit in which a direct collapse into a black hole, without a SN, could exists (although rotation could move this limit towards greater masses, \cite{fryer99}). Ultimately, a detailed study of the nebular phase $>$\,200\,d since explosion, should provide a much more precise measurement of the ZAMS mass of the progenitor.

\subsection{Evidence of a failed jet}
The high kinetic energy, $\sim$\,2\,$\times$\,10$^{52}$\,erg, derived from our model is consistent with that of typical GRB/FXT-SNe events and points towards the existence of a central engine. Additionally, the high initial velocities observed of $\sim$ \,30,000\,km\,s$^{-1}$, also support the need for a source of additional energy. Reference \cite{piran19} suggests that a relativistic jet that fails to break through the stellar envelope and deposits most of its energy in it (also known as choked jets) can explain the observed properties of several Ic-BL SNe for which a relativistic jet was not seen, like in SN\,2026gzf. The deep radio limits obtained, the lack of non-thermal emission in the prompt EP emission and the lack of evidence for a shocked cocoon in the early phases of SN\,2026gzf (as it fails to work together with the CSM+SN model) strongly suggest that a choked jet could be responsible for the measured kinetic energy, expansion velocities, and initial hot temperatures. Future radio observations spanning 100s\,--\,1000s\,days after the explosion will allow us to extend our radio limits to rule out larger off-axis angles and further confirm our results.  

Reference \cite{Hamidani25} proposes a detailed formalism to predict whether a jet can be successfully launched, heavily choked, fully choked or even not produced at all based on the mass and radius of the surrounding CSM. Our SN modelling required CSM interaction to explain the SN pre-peak behaviour, giving us the necessary tools to test these predictions and build a stronger evidence for a choked jet. From our model, we measure $\sim$0.07\,M$_{\odot}$ in a shell with inner radius $\sim$\,3\,$\times$\,10$^{13}$\,cm, where the radius is not obtained from the best fit model, but from the X-ray light curve presented in \cite{Yuan26}. With these values, SN\,2026gzf falls into the category described in \cite{Hamidani25} as \textit{LLGRB}, which represents a failed jet. Within this case, \cite{Hamidani25} predicts a shock-breakout which produces soft to hard X-ray photons, like the one seen by EP \cite{Yuan26}, with SN features similar to those of Ic-BL like SN\,2006aj, as we have seen for SN\,2026gzf, and faint afterglows, consistent with the lack of radio detection. As mentioned in \cite{Hamidani25}, to date, there had not been confirmed cases of this category, making SN\,2026gzf potentially the first ever Ic-BL SN with a confimed failed jet.

\subsection{The mass-loss rate}
Our model suggests that the progenitor of SN\,2026gzf deposited $\sim$\,0.07\,M$_\odot$ into a shell with inner radius $\sim$\,3\,$\times$\,10$^{13}$\,cm. We can then use equation 8 in \cite{ho19}, to estimate the average mass-loss rate prior to the explosion, resulting in $\Dot{M}$\,$\sim$\,7.5\,$v_{\rm wind, 1000}$ M$_{\odot}$\,yr$^{-1}$, where $v_{\rm wind, 1000}$ is the wind velocity in units of 1,000\,km\,s$^{-1}$. If we assume a typical wind velocity of 1,000\,km\,s$^{-1}$, the derived mass-loss rate would be $\Dot{M}$\,$\sim$\,7.5\,M$_{\odot}$\,yr$^{-1}$, 10\,$\times$ that of SN\,2018gep \cite{ho19} and 5\,--\,7 orders of magnitude higher than what is typically expected for Ic-BL SNe \cite{smith14}. We note that to obtain a mass-loss rate similar to SN\,2018gep, we would need wind velocities of $\sim$\,100\,km\,s$^{-1}$, considerably lower than those expected in H- and He- free Wolf-Rayet progenitors \cite[e.g.][]{grafener17}. This extreme high value seems to suggest an intense late mass-loss episode not expected of ordinary steady Wolf-Rayet winds progenitors.

\subsection{The SN environment}
\begin{figure}[!ht]
\centering
\includegraphics[angle=0,width=0.49\textwidth]{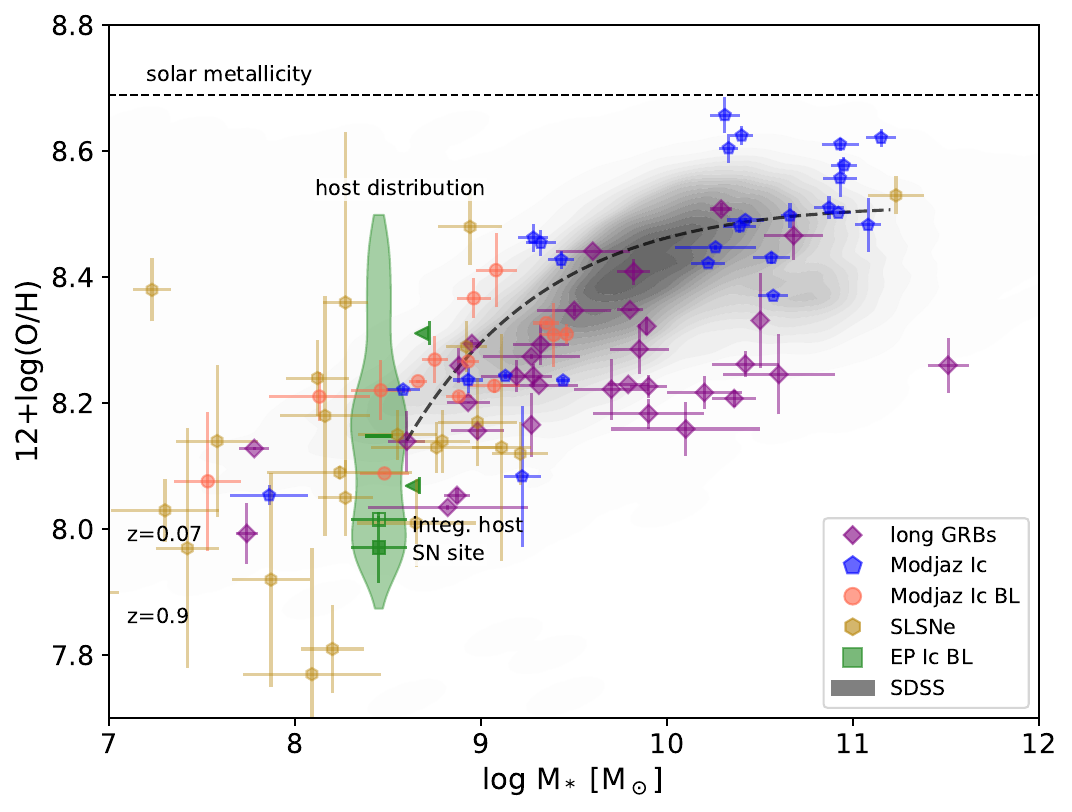}
\includegraphics[angle=0,width=0.49\textwidth]{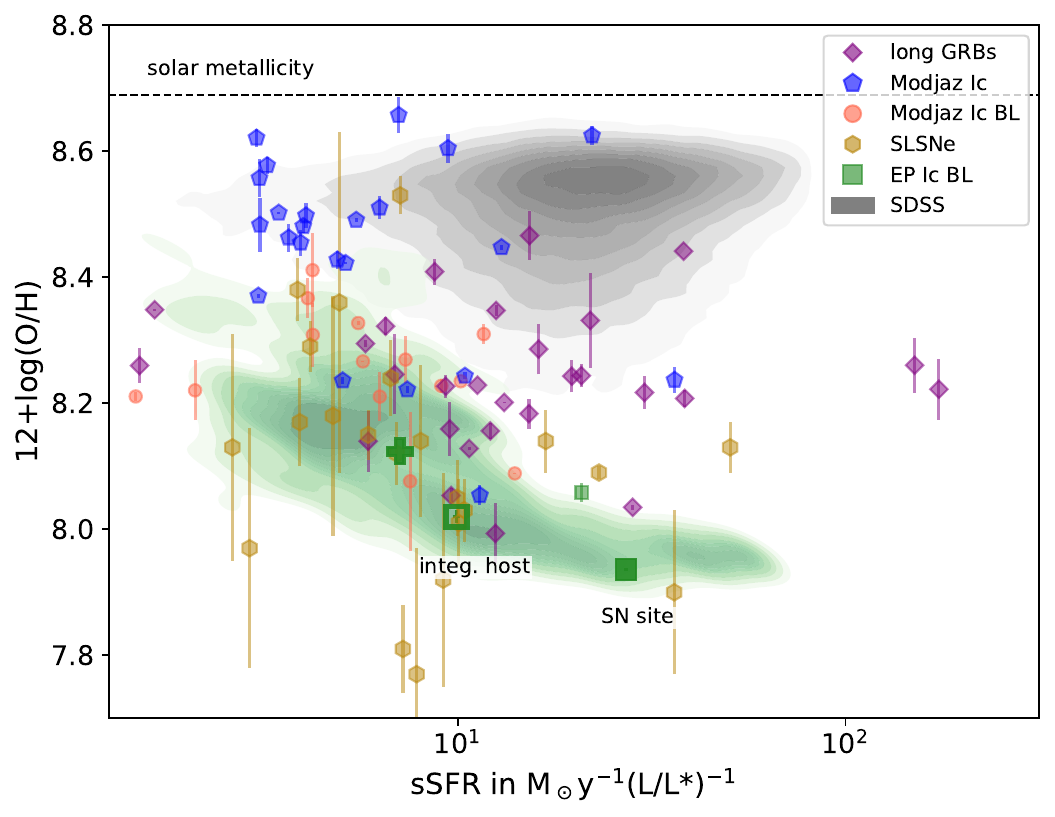}
\caption{\textbf{Left:} Mass-metallicity ($M$--$Z$) relation showing the distribution of metallicities in spaxels across the MUSE-WFM datacube (green violin plot, the thick horizontal line indicates the median). The black dashed line is the fit to the $M$--$Z$ relation in the \cite{Marino13} calibration used for all samples from \cite{Sanchez19}. Comparison samples are local metallicities (and global masses) from fully stripped enveloped SNe (Ic, Ic-BL, \cite{modjaz20}), global metallicities of long GRBs from \cite{Kruhler15, Han10, DellaValle06, Christensen08, Kelly13, Schulze14, Thoene14, Izzo17, Heintz18, Cano17b, deUgarte18, Melandri19, deUgarte20, Thoene2024} and stellar masses from the GHostS database (http://www.grbhosts.org) and \cite{Palmerio19}, superluminous SN Type I host \cite{Leloudas15, Perley16, Schulze18} and two other \emph{Einstein Probe} events with Ic-BL associations (SN\,2025kg \cite{rastinejad25}, and SN\,2025wkm \cite{Corcoran26}, those galaxy masses are upper limits) and star-forming galaxies from the SDSS, DR16 \cite{SDSS_DR16} with a $\text{S/N} > 5$ in the [N\,\textsc{ii}] line. \textbf{Right:} Luminosity weighted specific SFR (SSFR) for the same sample. The green density plot is the distribution from individual spaxels in the MUSE-WFM datacube, the value of the SN site is the average of 4 spaxels around the SN position. Empty green symbols are values from an integrated spectrum of the host, the large filled symbol the value of the SN site and the dashed line/plus the median value of the green shaded distributions of the host galaxy.}
\label{fig:sn_metallicity}
\end{figure}

The metallicity of the site of SN\,2026gzf is one of the lowest measured for any Ic-BL SN or even long GRBs associated to Ic-BL SNe (see Fig.~\ref{fig:sn_metallicity}) and one of the lowest metallicity regions in the host, while the median is fully consistent with a low redshift M-Z relation. It should be noted, though, that the global host spectrum is dominated by the bright H\,\textsc{ii} region and shows a metallicity only slightly higher than the SN site. In a comparison sample from \citep{modjaz20}, Ic SN sites tend to show higher metallicities than Ic-BL sites. The two other Ic-BL events related to EP triggers included in our comparison are consistent with the \citep{modjaz20} sample. Long GRBs are largely consistent with the Ic-BL metallicity distribution but their hosts tend to have higher masses; note, however, that the long GRB host sample spans a much larger redshift range (up to z$\sim$4), while all the hosts from \citep{modjaz20} are $z$\,$<$\,0.2. The peculiarity of the SN site and its H\,\textsc{ii} region becomes even more evident in a specific SFR (weighted by the luminosity) vs. metallicity comparison. While the bulk of the host is consistent with typical values for long GRB and Ic/Ic BL hosts and SN sites, the H\,\textsc{ii} region at the SN site shows a tail up to higher values of sSFR ($\sim$100 M$_\odot$y$^{-1}$(L/L*)$^{-1}$) at almost constant metallicity, the SN site is located within this tail with a SSFR of 27$\pm$5 M$_\odot$y$^{-1}$(L/L*)$^{-1}$. Hence this largely uniform region of the galaxy seems to have experienced some recent burst of SF hosting a number of (very) massive stars. Given the high progenitor mass estimate of SN\,2026gzf, this SN could have been one of the first stars after this recent starburst to explode as a SN, as the region has not yet been enriched by those SN explosions. 

\subsection{\texorpdfstring{Rates of SN\,2026gzf-like events}{Rates of SN 2026gzf-like events}}
It is an important question if many, or even all SN Ic-BL may create soft X-ray events similar to EP260321a upon core collapse. Small number statistics makes rate calculations difficult. However, based on this single detection we infer a rate of $0.8^{+1.3}_{-0.6}\,\times\,10^{-6}$\,Mpc$^{-3}$\,yr$^{-1}$ (see Methods). There are few estimates of the rates of SN Ic-BL, with the ASAS-SN local sample estimating $3.3^{3.3}_{-2.2}\,\times\,10^{-6}$\,Mpc$^{-3}$\,yr$^{-1}$, which is consistent with all Ic-BL creating explosions akin to EP260321a.

\begin{figure}[!ht]
\centering
\includegraphics[angle=0,width=1\columnwidth]{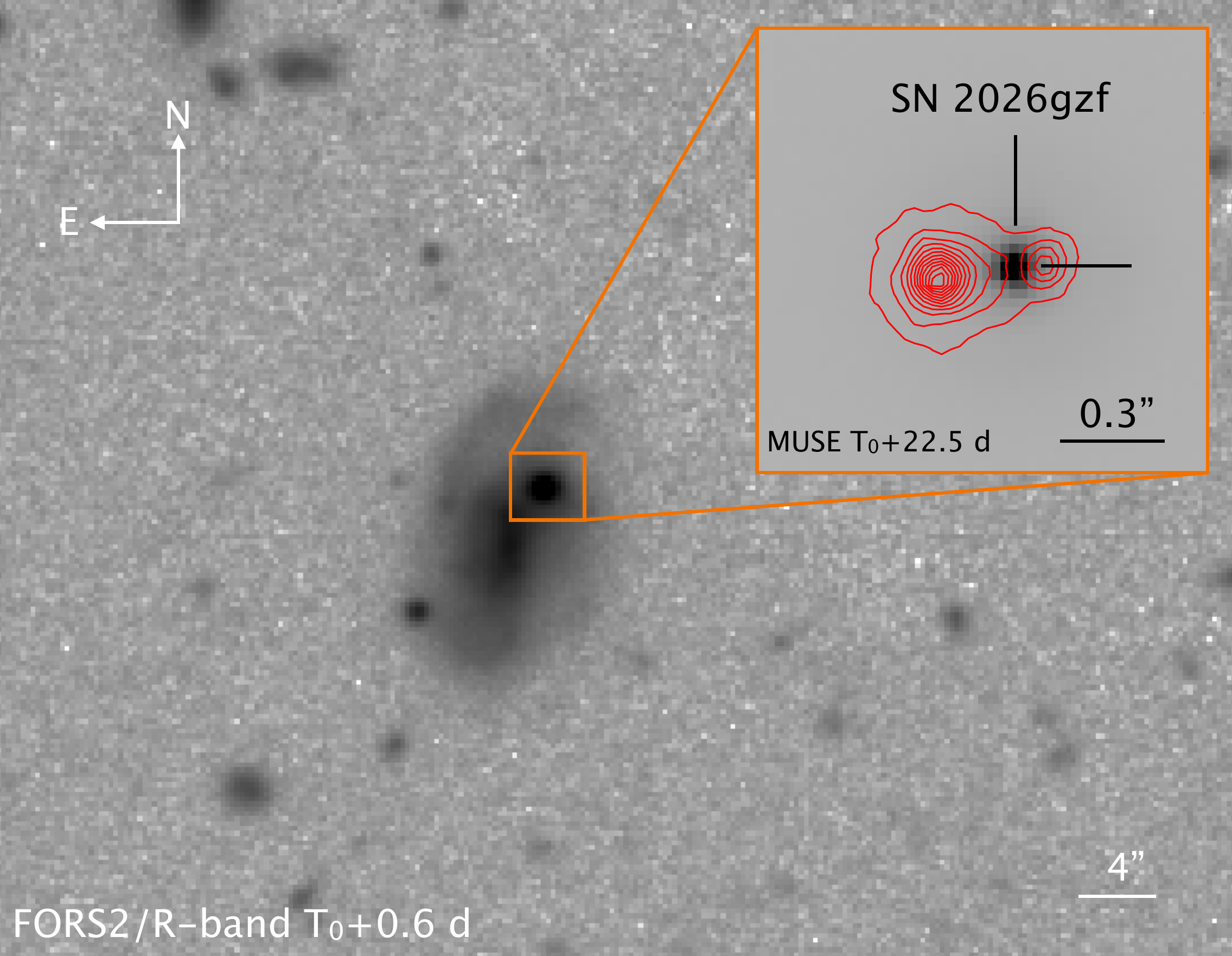}
\caption{VLT/FORS2 image in the $R$ band obtained 0.6\,d after the explosion with the SN SN\,2026gzf marked by the orange square. The inset image shows a higher spatial resolution close-up look at the region of the SN through our MUSE NFM image obtained 22.5\,d after the explosion. The red contours show two regions of high H${\alpha}$ emission surrounding the location of SN\,2026gzf.}
\label{fig:snimage}
\end{figure}

\begin{figure}[!ht]
\centering
\includegraphics[angle=0,width=1\columnwidth]{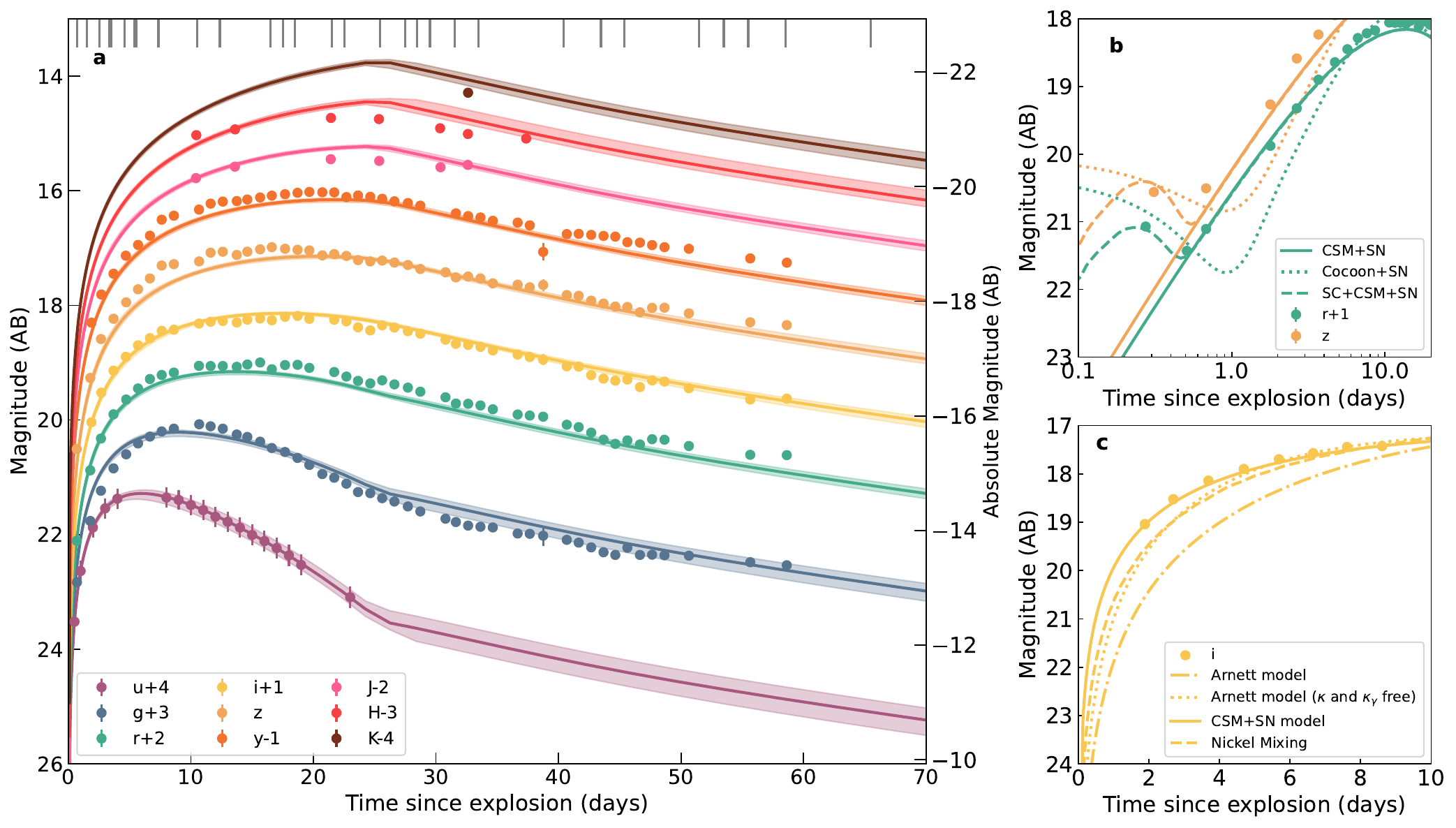}
\caption{\textbf{(a)} Optical/NIR multi-band light curve of SN\,2026gzf in AB magnitudes (apparent on the left and absolute on the right) with offsets applied to each band for clarity. Solid lines show the best fit of the CSM interaction\,+\,SN powered by $^{56}$Ni radioactive decay model. Shaded areas show the posterior region drawn from a sample of 100 random values. Grey vertical lines on  the top axis mark the epochs of our spectroscopic observations. \textbf{(b)} Detailed view of the beginning of the observing campaign showing that the radioactive decay powered phase becomes detectable at $\sim T_0 +0.4$\,d. The solid line shows the best-fit CSM+SN model excluding the first $r$ and $z$ data point. The dotted line corresponds to a model including a shocked cocoon (cocoon+SN) while the dashed line includes a shock cooling (SC) from the initial shock-breakout (see Methods). \textbf{(c)} Detailed view of the SN rising phase showing that neither a pure Arnett model (dashed-dot line) nor a Nickel mixing model (dashed line) is sufficient to explain the observed behaviour of SN\,2026gzf even if the opacities ($\kappa$ and $\kappa_{\gamma}$) are left free (dotted line). The addition of CSM interaction to the Arnett model (solid line) seems to explain better the earlier behaviour seen (see Methods).}
\label{fig:snlightcurve}
\end{figure}

\begin{figure}[!ht]
\centering
\includegraphics[angle=0,width=0.49\textwidth]{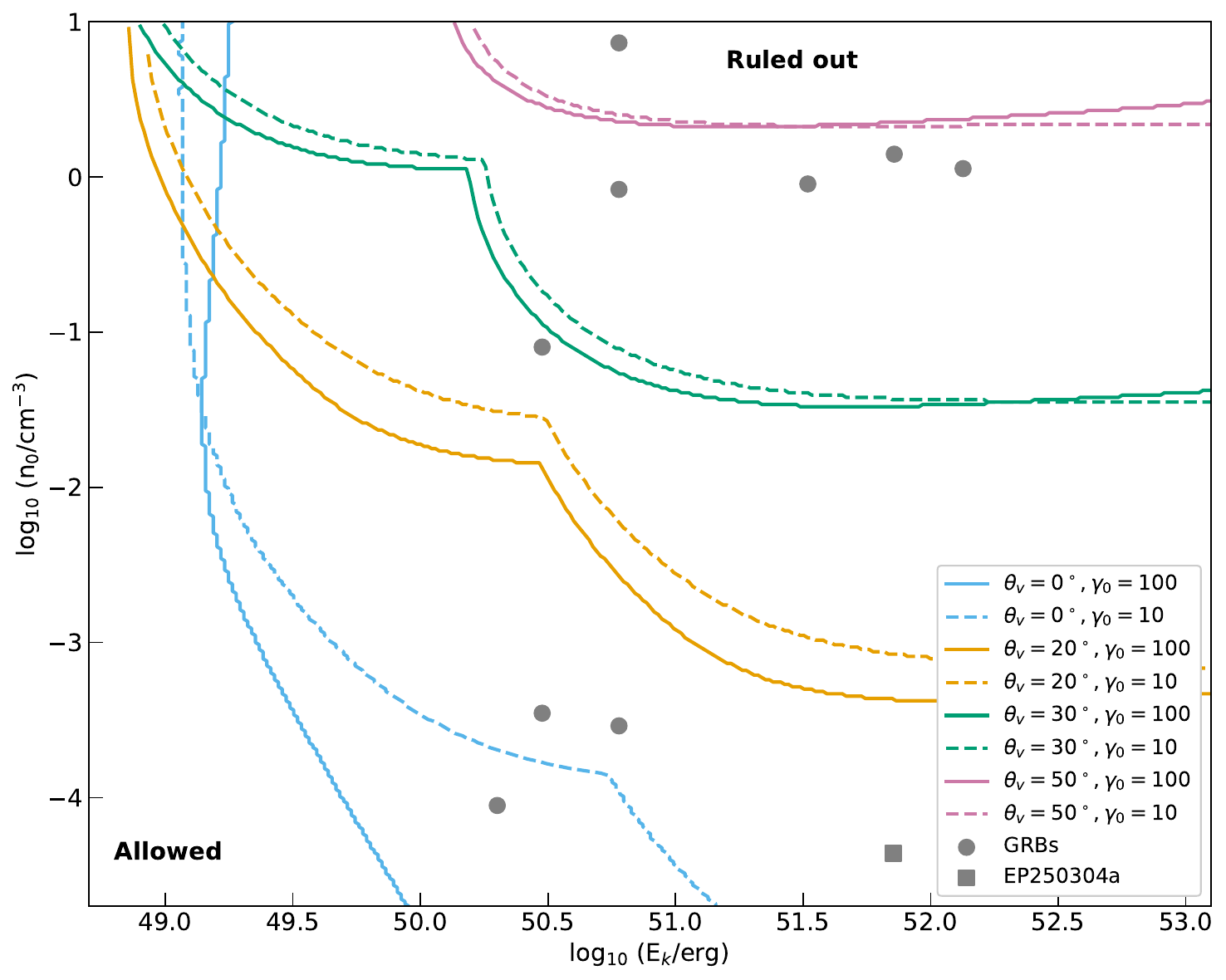}
\includegraphics[angle=0,width=0.49\textwidth]{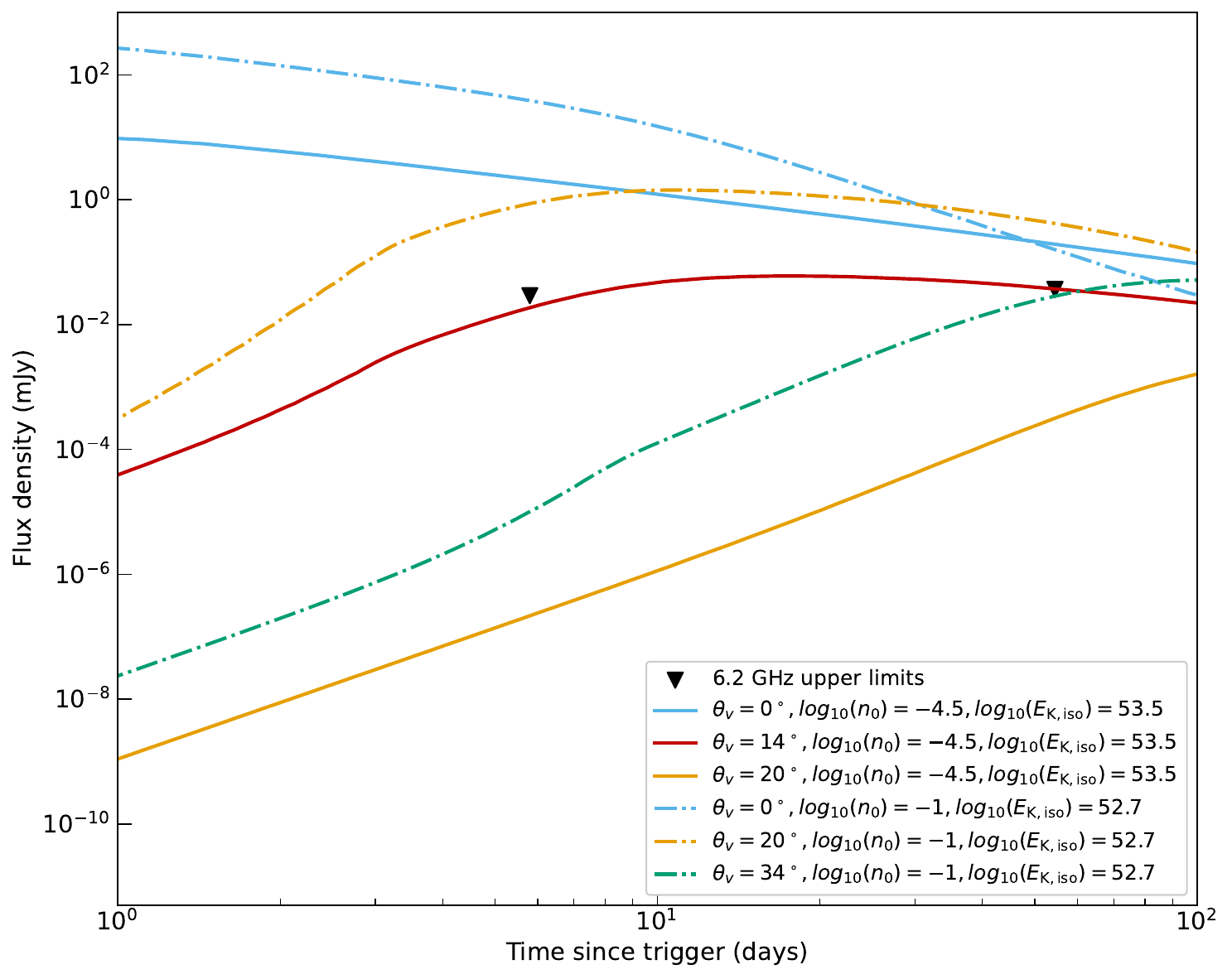}
\caption{\textbf{Left}: The valid parameter space for a relativistic jet in the ISM medium density ($n_{0}$) versus beamed corrected kinetic energy of a jet ($E_{\rm K}$) plane obtained from models generated with the \texttt{Redback} top-hat model for different viewing angles ($\theta_{\rm v}$) and two bulk Lorentz factors ($\gamma_0$). The lines represent the $n_{0}$\,--\,$E_{\rm K}$ boundary limits of \textit{allowed} (below) and \textit{ruled out} (above) from our radio VLA upper limits. The grey circles show values measured for a selection of on-axis ISM GRBs \cite{kangas2021}. The grey square corresponds to EP250304a as an example of faint jets associated with a sub-group of GRB\,--\,SN for which the optical emission is dominated by the SN \cite{cotter26}. Figure adapted from Fig. 3 \cite{schroeder25}. \textbf{Right}: Predicted radio afterglow lightcurves at 6.2 GHz from a top-hat model relativistic jet for different viewing angles, medium densities and kinetic energies and fixed $\epsilon_e = 0.1$ and $\epsilon_B = 0.001$. Downward triangles mark our VLA observational upper limits. Within the range of kinetic energies explored in the left panel, a possible unseen relativistic jet associated with EP260321a is always excluded in the on-axis scenario, and can only exist for off-axis viewing angles $>14^{\circ}$ in the case of a weak EP250304a-like afterglow, or $>34^{\circ}$, in the case of a typical GRB afterglow.}
\label{fig:afterglow}
\end{figure}

\clearpage

\section{Methods}\label{sec11}
Throughout this work, we assume a $\Lambda$CDM cosmology based on the Planck parameters
\citep{planck18}. Errors are given at the 1\,$\sigma$ level unless otherwise stated.

\subsection{Pre-explosion archival data}
We investigated the position of EP260321a for evidence of pre-trigger activity in the archival data of the Gravitational-wave Optical Transient Observer \citep[GOTO;][]{Steeghs22,Dyer24}. GOTO is a modular multi-node robotic telescope network with nodes at the Observatorio del Roque de los Muchachos in La Palma, Spain, and at Siding Spring Observatory, Australia. It consists of 8 Unit Telescopes (UTs) per mount, with 2 mounts per node and a total field of view of 44 square degrees per pointing. In addition to responding to GW \citep{Gompertz20} and GRB \citep{Belkin24,Kumar25} alerts, GOTO performs an all-sky survey with $4\times45$\,s exposures on a cadence of 2\,--\,3 days.

We queried the GOTO forced photometry server (Jarvis et al., in prep) with the EP260321a coordinates. Photometry was performed using the GOTO pipeline on template subtracted images \citep{Lyman26}. No compelling pre-trigger variable emission is detected in the available data beginning on 2023-10-27 at 06:18:42.5 UT ($T_0-876$\,days), with typical 3$\sigma$ limiting magnitudes of $L=20.16\pm0.61$ mag (mean $\pm\,1\sigma$). EP260321a was first detected by GOTO in images taken at 11:25:46.2 on 2026-03-22 ($T_0+0.955$\,days) with a magnitude of $L=19.51\pm0.10$. The previous epoch, taken at 11:32:09.3 UT on 2026-03-20, yielded a $3\sigma$ upper limit of $L>20.91$.

\subsection{Photometric observations}
After the initial trigger of EP260321a and subsequent identification of the optical counterpart, we started an optical/NIR follow-up campaign to monitor the evolution of the source. Table~\ref{tab:optical_observations} provides an overall log of all telescopes involved in this work, indicating the filters used and the number of epochs in each filter used for this paper. For eas of use and practicality, the complete final optical/NIR photometry will be made available via \texttt{CDS} portal\footnote{\url{https://portal.cds.unistra.fr}} in machine readable format.

Our observing campaign includes data in the period of March 21st to June 2nd, 2026. During this time, we observed SN\,2026gzf in the optical using the Las Cumbres Observatory (LCO) global telescope network \cite{Brown13} with several of their 1\,m telescopes around the world (CLN2026A-001; PI J. Chac\'on), contributing to our $ugriz$ coverage. Additional coverage was also obtained via the Planetary Defense (PD) 1\,m telescope at Cerro Tololo Inter-American Observatory, Chile (CN2026A-67; PI J. Chac\'on), and the ESO VLT Survey Telescope (VST) at Cerro Paranal, Chile (ESO programme 116.29H5; PI J. Chac\'on), providing $ugri$ coverage that complemented our LCO monitoring.

We also obtained multi-band $grizy$ images with the DDRAGO wide-field imager on the COLIBR\'I\footnote{\url{https://www.colibri-obs.org/}} telescope (PI S. Basa), a Franco-Mexican robotic 1.3\,m telescope operated by the Observatorio Astron\'omico Nacional (OAN) in the Sierra de San Pedro Mártir, Baja California \citep{Basa2022, Langarica2024}.

Additional optical epochs were obtained with the 2\,m Liverpool Telescope (LT) \cite{steele2004} located at the Roque de Los Muchachos observatory, La Palma, Spain (XPL25A25/PL26A03; PI R. Eyles-Ferris); the 0.8\,m Joan Or\'o Telescope (TJO) located at the Catalan Pyrenees (PI A. Martin-Carrillo); the Rapid Eye Mount (REM) 0.6\,m telescope \cite{zerbi01} located at La Silla, Chile (PI M. Ferro); the Asiago Schmidt 67/92\,cm telescope at Mount Ekar on the Asiago Plateau of the Veneto region, Italy (PI A. Reguitti); and the ESO 8\,m Very Large Telescope (VLT) at the Cerro Paranal observatory (114.27PZ; PIs N. R. Tanvir, S. D. Vergani and D. B. Malesani), Chile.

In all cases, the images were reduced and stacked by custom made pipelines for each telescope using the most up-to-date calibration files. Due to the significant brightness of the host galaxy, and in particular the blue knot region located almost under the location of the SN, we performed PSF template subtracted photometry in all of our images. For consistency, we used, in all cases, archival images obtained with the Hyper Suprime-Cam \citep[HSC,][]{HSC_SSP} instrument on the Subaru telescope from the HSC\,--\,SSP programme. These images provide great depth, good filter matching, and good spatial resolution, allowing us to create good and reliable template subtracted images. In the case of the $u$-band images, we used archival SDSS templates which offer a good match to our filter as well as good depth.

We obtained near-infrared (NIR) observations of SN\,2026gzf with the 3.58\,m Telescopio Nazionale Galileo (TNG) at the Roque de los Muchachos observatory, La Palma, Spain (A53TAC\_6; PI A. Melandri) and the 8.4\,m Large Binocular Telescope (LBT) on Mt. Graham, Arizona, USA (IT-2025B-023; PI E. Maiorano). We analysed the NIR images after modelling and subtracting the host galaxy, since we could not rely on a good enough template image to be used as a reference. The resulting fit of the galaxy profile is good resulting in a reliable subtraction. We note that, at times, the resulting images may be slightly over subtracted which would lead to slightly fainter magnitude values. However, this is expected to be well accounted for in our errors.

\subsection{Spectroscopic observations}
We performed spectroscopic observations that ranged from March 21st to May 25, 2026. Table~\ref{tab:speclog} shows a list of all epochs obtained in this period and the telescopes that have contributed to the spectral monitoring campaign presented in this work.

We obtained spectra with the ESO VLT telescope at Paranal, Chile, with multiple instruments: two MUSE, two X-shooter, one UVES (all under the programme 114.27PZ; PIs N. R. Tanvir, S. Vergani, D. B. Malesani) and 15 FORS2 (116.297W; PI A. Martin-Carrillo). All VLT spectra were reduced using the standard \texttt{EsoReflex} and \texttt{EDPS} pipelines \cite{EDPS, fors_recipe}. The MUSE spectra of the target were extracted from the data cube using the MUSE Python Data Analysis Framework (MPDAF) python package \citep{mpdaf}. 

We used the Nordic Optical Telescope (P71-506; PIs D. B. Malesani, J. Fynbo, D. Xu) at Roque de los Muchachos observatory, La Palma, Spain, to obtain 5 additional spectra. These observations were reduced using a set of Python scripts for the reduction of long-slit spectra available on Github (\texttt{Pylongslit}, \cite{Valeckas2025}). 

We also obtained two observations with the Gran Telescopcio Canarias at the Roque de los Muchachos observatory, La Palma, Spain (GTCMULTIPLE3G-26A; PI S. Geier), and another two with the MMT (PI. W-f. Fong) at the Fred Lawrence Whipple Observatory, Arizona, US. The data were reduced using custom specific pipelines for each telescope.

Additionally, we obtained one spectroscopic observation with the Liverpool Telescope at the Roque de los Muchachos observatory, La Palma, Spain (XPL25A25/PL26A03; PI R. Eyles-Ferris), the 3.6\,m Devasthal Optical Telescope located at the Devasthal campus of Aryabhatta Research Institute of Observational Sciences, India (PI Anshika Gupta), and the 1.82\,m Asiago Copernico telescope at Mount Ekar on the Asiago Plateau of the Veneto region, Italy (PI A. Reguitti).


\subsection{Astrometric analysis}
\label{methods:astrometry}
The MUSE-NFM observation provides a unique opportunity to disentangle the location of the SN from the blue knot, and thus produce its precise coordinates. Unfortunately, the MUSE-NFM image did not include any nearby field star to calibrate its astrometry. To solve this issue, we proceeded to calculate the off-set between the centre of the blue knot (H\,\textsc{ii} region) and the SN itself. This off-set was then applied it to the refined coordinates of the blue knot to calculate the true coordinates of the SN. Below we described the steps and considerations followed in this process:

\begin{enumerate}
\item Using a $g$-band image from the Legacy Sky Survey \cite{LegacySurvey}, in which the blue knot (H\,\textsc{ii} region) is well detected, we measured its centre coordinates ($\text{RA} = 09$:59:42.883, $\text{Dec} = +00$:25:06.36)
\item We then created a H$\alpha$ image from the MUSE-NFM observation to measure the position of the H\,\textsc{ii} region in it. As we show in Fig.~\ref{fig:snimage}, the location of the H\,\textsc{ii} region is not contaminated by the bright SN, making this measurement reliable.
\item We created also a white light image from the MUSE-NFM observation in which the SN dominates over the H\,\textsc{ii} region, allowing us to accurately measure its position.
\item We then calculated the offsets between the positions of the H\,\textsc{ii} region and the SN. This step assumes that, while the astrometry is not calibrated in absolute terms, it remains constant across image slices, and is correct at pixel scale. From this step we measured an offset of $\delta \mbox{RA}$\,=\,-0.20$^{\prime\prime}$ and $\delta \mbox{Dec}$\,=\,0.03$^{\prime\prime}$.
\item We finally added the offsets to the position of the H\,\textsc{ii} region measured from the Legacy Sky Survey image to retrieve the final absolute coordinates of the SN, $\mbox{RA} = 09$:59:42.870, $\mbox{Dec} = +00$:25:06.39.
\item As an additional step, we repeated the whole process using an $i$-band image from COLIBR\'I at the SN peak, obtaining $\mbox{RA} = 09$:59:42.877, $\mbox{Dec} = +00$:25:06.45. These coordinates differ by 0.12$^{\prime\prime}$ with respect to those mentioned in the previous step. While MUSE-NFM provides higher spatial resolution, we take this as a conservative error in the final position of the SN.
\end{enumerate}


\subsection{SN comparison sample}
Throughout this paper, we compared the properties of SN\,2026gzf to that of several Ic-BL SNe with and without high-energy prompt emission, plus the particular case of SN\,2008D, a type Ib SN considered the first ever observation of a SN shock-breakout.

To date, there are $\sim$\,200 spectroscopically confirmed Ic-BL SNe\footnote{\url{https://www.wis-tns.org}} and $\sim$\,60 GRB/FXT-SN associations \cite{finneran25a}. For this paper, we attempted to build an SN sample that would be the most relevant to highlight the photometric and spectroscopic features seen in SN\,2026gzf, while still showing the diversity of Ic-BL SNe.

In the end our comparison sample includes: 
\begin{itemize}
    \item SN\,1998bw, as the GRB-SN archetype used as a comparison in all GRB/FXT-SN publications \cite{Galama1998},
    \item SN\,2008D, as the first SN shock-breakout \cite{Soderberg08, modjaz09, Mazzali08},
    \item the sub-population of Ic-BL SNe with GRB/FXT association for which the jet is faint and the optical emission is dominated from very early times by the shocked cocoon \cite{cotter26},
    \item SN\,2020bvc, as a SN without prompt emission but with strong evidence of an off-axis relativistic jet \cite{Izzo2020},
    \item a sub-group of Ic-BL SNe, namely SN\,2016coi, SN\,2018gep and SN\,2020lao, which share certain photometry and/or spectroscopic characteristics seen in SN\,2026gzf \citep{yamanaka17, Terreran19, stritzinger26, ho19}.
\end{itemize}

All these SNe share that they have rich datasets from very early times, which is extremely useful to fully exploit them as comparisons for SN\,2026gzf.

We ruled out including in this sample typical GRB-SN cases because any rigorous comparison would have required special care of the GRB afterglow emission and because any existing shocked cocoon would have been masked by the brighter afterglow component.

\subsection{SN modelling}
In all the modelling cases described in this section we used nested sampling through the Nessai sampler \cite{nessai, Williams21, Williams23}, via \texttt{Bilby}~\citep{Ashton19}. We considered a likelihood in which a systematic error is added in quadrature to our measured ones to take into account possible filter inconsistencies between the different telescopes and instruments as implemented in \texttt{Redback} v1.5.3 \citep{sarin24}. 

We started by considering the most common and simplest model to explain the SN temporal evolution which invokes a radioactive power source from the $^{56}$Ni radioactive decay and subsequent elements, the so-called `Arnett' model \cite{arnett82}. In this scenario, we assume a homologous expansion of ejecta with mass $M_{\rm ej}$ and velocity $v_{\rm ej}$, with some fraction of $^{56}$Ni, $f_{\rm Ni}$, powering it. The model, as implemented in \texttt{Redback} \cite{sarin24}, accounts for the grey opacity $\kappa$ and the $\gamma$-ray leakage, $\kappa_{\gamma}$.

For the modelling, we excluded the first initial $r$ and $z$ band epochs that seem to show an initial decay (see Fig.~\ref{fig:snlightcurve}, panel b). Our results (see Table~\ref{tab:arnett_modelling} show that the Arnett model fails at modelling the rising phase of the SN, suggesting an excess in emission with respect to the expected SN contribution. As shown in Fig.~\ref{fig:snlightcurve}, panel c, leaving $\kappa$ and $\kappa_{\gamma}$ free results in an unphysical opacity of $\kappa$\,$\sim$\,0.01\,cm$^{2}$\,g$^{-1}$, suggesting that an additional component may be at play. Forcing $\kappa$ and $\kappa_{\gamma}$ to their most commonly inferred values for Ic-BL SNe (0.07\,cm$^{2}$\,g$^{-1}$ and 0.03\,cm$^{2}$\,g$^{-1}$, respectively, \cite{Inserra2013,Wheeler2015,Nagy2018}), makes the issue of failing to match the observed SN rising even stronger. As a result, the ejecta masses derived from these pure Arnett models are significantly underestimated, and the fraction of Nickel is significantly higher than expected. In Fig.~\ref{fig:contours_arnett}, we show the corner plots with the obtained posteriors for the model parameters.

Reference \cite{piro13} presented a theoretical framework in which a shallower Nickel deposit (i.e. higher Nickel mixing) can explain a $^{56}$Ni powered light curve starting to rise sooner, overlapping with the shock-breakout cooling phase, in excess to a case with a deeper Nickel deposit. We thus explore this possibility by fitting our data with the Nickel mixing model available in \texttt{Redback} \cite{rastinejad25}. The results of our fit are shown in Table~\ref{tab:sn_modelling} with the corner plots of the posteriors in Fig.~\ref{fig:contours_models}. In this model we kept $\kappa$ and $\kappa_{\gamma}$ frozen. As shown in Fig.~\ref{fig:snlightcurve} panel c, the Nickel mixing model seems to provide a similar result during the rising phase of the SN as the Arnett model with all parameters free, solving the original issues of extremely low values for the opacity. The \texttt{log\_evidence} parameter obtained from the fit strongly suggests that this model is preferred over any of the pure Arnett models previously considered. We note that if $\kappa$ is left free in the Nickel mixing model, the model tends to attempt some balancing between the amount of Nickel mixing and the opacity, ultimately leading to lower than expected values of $\kappa$. The amount of Nickel mixing derived from the model is extremely large, $\sim$\,87\% and not well constrained, leading to potentially being 100\%. As a comparison, the same model suggested a Nickel mixing fraction of $\sim$\,62\% in the case of SN\,2025kg \cite{rastinejad25}. This level of mixing would imply the existence of Nickel throughout the full stellar interior, which could be explained if a relativistic jet is launched from the core, carrying Nickel towards the outer layers.

The inability of these models to fully interpret the rising phase of the SN made us consider additional models. From the report of possible pre-explosion variability, we decided to explore a CSM interaction model on top of the Nickel powered Arnett SN model, also available in \texttt{Redback} \cite{chatzopoulos13, villar17, jiang20}. For this model, we again explored the possibility of leaving $\kappa$ and $\kappa_{\gamma}$ completely free. While $\kappa_{\gamma}$ seems to behave normally, it seems that $\kappa$ is very sensitive to the behaviour during the early SN phase, taking over any other contributions and resulting in an unphysical value. Thus, similarly to our previous models, we kept $\kappa$ and $\kappa_{\gamma}$ frozen. Besides the parameters already discussed for the Arnett model, the CSM interaction introduces four new parameters to fit the data: the mass of the CSM shell, $M_{\rm CSM}$, the radius of the CSM shell, $R_{\rm CSM}$, the CSM density profile exponent, $\eta$ and the CSM density profile amplitude, $\rho$.

The results of the model are shown in Table~\ref{tab:sn_modelling} with the corner plots of the posteriors for the `CSM+SN-2' and Nickel mixing shown in Fig.~\ref{fig:contours_models}. The model denoted by us as `CSM+SN-1' corresponds to a study in which all the parameters are explored, with the exception of those previously mentioned ($\kappa$, $\kappa_{\gamma}$ and $R_{\rm CSM}$). We note that in these results, $\eta$ is quite unconstrained with a central value that seems to differ slightly from 2, the expected value for a steady mass-loss rate. To study the effect of this parameter on our model, we run a second model `CSM+SN-2' in which $\eta$ was frozen to 1.8. This value was chosen as it corresponds to the lower edge of the 1\,$\sigma$ contour obtained from the `CSM+SN-1' model in an attempt to fully break the degeneracy with the maximum value of 2 (expected from a steady stellar wind). The resulting fit is quite similar to the previous attempt with very similar \texttt{log\_evidence} values. While statistically speaking, we cannot differentiate between either model, an $\eta$\,$\sim$\,1.8 would suggest an increased mass-loss which could possibly lead to the variability  presented by \cite{janet26}. We note that the models used here do not capture the key difference in velocity profile that would let us discriminate between an eruptive mass-loss episode in the mass or a departure from a steady wind mass-loss rate~\citep{Sarin2026_csm}. 

This CSM+SN model offers the best representation of the rising phase of the SN as shown in Fig.~\ref{fig:snlightcurve}, panel c. The ejecta mass, Nickel fraction and kinetic energy obtained from this model are consistent with the Ic-BL SN population (with and without an association with high-energy emission) \citep[e.g.,][]{Lyman16, srinivasaragavan24}.

We note that this model is very sensitive to the range of CSM mass explored, with a preference to significantly increase the mass of the CSM shell and reduce both that of the ejected material and its Nickel fraction. We thus recommend caution when using this model, as the CSM can attempt to overtake the SN emission in an unphysical way.

\subsection{The pre-SN rising phase}

As shown in Fig.~\ref{fig:snlightcurve}, panel b, our first epoch in $r$ and $z$ bands is not consistent with the rise of the Nickel decay powered phase, suggesting that for the initial $<$\,12\,h, the SN emission was decaying. The lack of more data precludes us from doing a proper formal fitting of this very early emission. However, with the help of some toy models, we have attempted to put tight constraints on the possible nature of this initial component. Specifically, we have considered a shocked cocoon origin and a shock-breakout cooling phase. Both of these phases are common in many other SNe and thus, it feels natural to explore them here. This is true especially in the case of the shocked cocoon phase, which is always observed in GRB/FXT-SNe where the jet is not powerful enough to be observable in the optical/NIR range \citep[e.g. SN\,2006aj, SN\,2010bh, SN\,2017iuk, SN\,2025kg, SN\,2025fhm][respectively]{Campana2006, Starling2011, Izzo2019, eylesferris25, rastinejad25, cotter26}. In both scenarios, we add the extra early component to the CSM+SN model, which we already know satisfactorily fits the SN behavior from the rising phase.

We initially explored the shocked cocoon scenario, by invoking the \texttt{Redback} models \texttt{shocked\_cocoon\_csm\_and\_arnett} \cite{Hamidani25a, Hamidani25} and \texttt{shocked\_cocoon\_and\_arnett} \cite{Piro18}. We added the light curves of several GRB/FXT-SNe shown in Fig.~\ref{fig:sn_luminosity}, scaled to the redshift of SN\,2026gzf to visually explore the typical shocked cocoon parameters that satisfy the early light curve bumps, which in luminosity may appear as early excess over the expected pure bolometric black body light curve, observed in these other cases. As shown in Fig.~\ref{fig:sn_luminosity}, the timescales of these initial light curve bumps tend to be on the order of a few days. Another significant difference is the luminosity of this early phase compared to that seen in SN\,2026gzf. 

We start by fixing the parameters obtained from the previous model and let the model explore the shocked cocoon parameters. It seems that in both shocked cocoon models considered, the jet activity timescale is the key factor that allows us to create an extremely short-lived early bump, as required by our observations. However, the resulting jet activity timescale is $\ll0.01$\,s, which seems unrealistic. As shown in Fig.~\ref{fig:snlightcurve}, panel c, while a parameter space can be found to match the brightness of the initial values and end of the shocked cocoon very early, the interplay between the shocked cocoon component becoming less dominant and the rising SN affects the overall rising phase that the model predicts, failing to match the observed data. Even if the CSM and SN components are left free once again, we don't seem to find a solution that would work simultaneously for the first points and the emergence of the main SN bump. 

Before fully discarding the shocked cocoon emission, we considered the possible limitations that the theoretical framework used could impose in our analysis. We then took the observed early bumps of SN\,2020bvc (a proposed Ic-BL SN with a relativistic jet seen off-axis), SN\,2006aj (the archetype GRB-SN for cases in which the jet is not dominant in optical) and SN\,2008D (a type Ib SN with shock-breakout prompt emission) and applied them two scaling factors, brightness and stretching to control their width. 

The result is quite consistent with the previous step using \texttt{Redback}. While the initial points can be matched, the smooth transition seen between the shocked cocoon and the Nickel powered emission impacts on how the combined model predicts the rising of the SN, missing our observations during almost the first 10 days as shown in Fig~\ref{fig:snlightcurve} panel c.

These two methods seem to rule out a shocked cocoon emission phase typically seen in GRB/FXT-SN as a consequence of the jet successfully breaking through the stellar envelope.

Using \texttt{Redback}, we created a custom model function combining the \texttt{shock\_cooling} model \cite{Piro21} with the CSM+SN model used to fit the full light curve. This allows us to explore a parameter space in which all three components are interconnected and can affect each other. Similarly to before, we start by fixing the CSM+SN parameters to the values obtained from the fit. We then start exploring the different parameters in small steps to identify the effects that each of them has on the resulting light curve. As shown in Fig.~\ref{fig:snlightcurve} panel c, we seem to find a solution that satisfies sufficiently the brightness seen in the first $r$ and $z$ points and that would almost seamlessly integrate with the CSM+SN model without any effect on the rising phase of the SN.

Due to the degeneracy between the \texttt{shock\_cooling} parameters, it is very likely that the solution we find is not unique. Therefore, the values derived from this exploratory toy-model should not be used to draw significant conclusions beyond the point that a shock-breakout cooling could offer an explanation for our initial observations. The solution found seems to show parameters that are physically possible with a radius of $\sim$\,10$^{11}$\,cm and an energy of $\sim$\,10$^{51}$\,erg. However, we note that the density power-law slope tends to move towards higher values than what one would expect in the case of Wolf-Rayet progenitors. We caution again that the parameter degeneracy is not broken here, and thus other parameter space solutions may still exist. However, the high density power-law slope could also suggest that the shock-breakout cooling is not the final solution for this early phase. Unfortunately, the lack of early data in our photometry campaign prevents us from fully constraining the true nature of the initial component.

\subsection{Limits on the emission from a relativistic jet}
We obtained 2 epochs of radio observations with the Karl G. Jansky Very Large Array (VLA) in its A-configuration under the programme 26A-385 (PI J. Leung; see Table~\ref{tab:radio_observations}). The observations were centred on the position of the optical counterpart. These two epochs covered a wide range of radio frequencies from 6.2 to 22 GHz. No new radio source was detected in either epoch at the $5\sigma$-detection threshold. Since the SN is embedded in a region of high star formation, which can result in non-transient radio emission, we report the 5\,$\sigma$ upper limits in Table~\ref{tab:radio_observations} to avoid being affected by any faint non-varying source.

The lack of radio emission provides an opportunity to explore the GRB afterglow physical model parameter space to set limits on the properties of an unseen relativistic jet associated with SN\,2026gzf. We consider three possible scenarios: i) a jet seen off-axis with typical physical parameters seen in most GRBs; ii) an off-axis faint jet similar to those seen in other GRB/FXT SN associations in which the optical emission is dominated by the shocked cocoon and SN radioactive decay bump, while the jet is only seen in radio and X-rays \cite[SN\,2006aj, SN\,2010bh, SN\,2017iuk, SN\,2025fhm, e.g.][]{cotter26}; and iii) a mildly relativistic jet.

From the physics of GRB afterglows, the parameters with the largest impact on the final radio light curve seem to be the medium density and the jet kinetic energy \cite{granot2002}. Although the fraction energy that goes into the electrons ($\epsilon_e$) and magnetic field ($\epsilon_B$) can also have a non-negligible impact on the radio light curve, their contribution is less important for the typical range of values observed other GRBs \cite{zhang24}. For this analysis, we fixed their values to $\epsilon_e = 0.1$ and $\epsilon_B = 0.001$. We note that these values differ slightly from those used in \cite{schroeder25} in a sample of radio Ic-BL sNe. However, their resulting valid and ruled out parameter space seems consistent with the one we found for SN\,2026gzf.

We considered the top-hat jet implemented in \texttt{Redback} \cite{gavin18} expanding into an ISM or wind environments and explored the regimes of relativistic (Lorentz factor of 100) and mildly relativistic (Lorentz factor of 10) jets. In all cases, we assumed a jet opening angle of $\sim$\,10$^{\circ}$ \cite{zhang24}.

Fig.~\ref{fig:afterglow} and Extended Data Fig.~\ref{fig:afterglow_wind} show the medium density vs jet beamed kinetic energy parameter space for which a jet associated with SN\,2026gzf could have been successfully launched and not observed. The right panels of Fig.~\ref{fig:afterglow} and Extended Data Fig.~\ref{fig:afterglow_wind} show the resulting radio light curves for the parameter space explored.

In both cases, we find that a possible unseen relativistic jet associated with EP260321a/SN\,2026gzf is always excluded in the on-axis scenario and can only exist for off-axis viewing angles $>14^{\circ}$ in the case of a weak EP250304a-like afterglow \cite{cotter26}, or $>34^{\circ}$, in the case of a typical GRB afterglow.

The assumption of a top-hat geometry simplifies the number of assumptions we have to make in terms of the level of spreading that a structure jet would have. Since a structured jet would always have more signal detected outside the jet at any given time than a top hat, our off-axis limits can act as lower limits, with higher off-axis angles restricted in the case of structured jets.

\subsection{Environment and host galaxy properties}
\subsubsection{Line and property maps}
We derive emission line maps using both the WFM and NFM in a similar manner, however, the NFM data are only used to pinpoint the exact position of the SN (see also \ref{methods:astrometry}) and to derive the SFR, since the [N\,\textsc{ii}] line, crucial to derive metallicities, is not visible in individual spaxels. For all other maps we use the WFM data. Emission line maps are obtained by summing the flux in a 12 pixel window around the centre of each line and subtracting the continuum by fitting a straight line between a continuum region 30--40 pixels left and right of the emission line to account for the different slopes especially in the vicinity of the SN. For the S/N and the spectral resolution of MUSE, summing the flux is a more robust approach than fitting a Gaussian to the emission line since finding the line centre would be difficult in some spaxels, in particular for weaker lines. Property maps were then obtained simply using those obtained emission line maps, corrected for Galactic and intrinsic extinction. 

To account for the small offset in the WCS calibration as described in Sect.~\ref{methods:astrometry} we align the two datasets by matching the centroid of the bright H\,\textsc{ii} region at the SN location (for the WFM), which largely corresponds to the peak of the Eastern H\,\textsc{ii} region in the NFM since the unresolved H\,\textsc{ii} region is dominated by the flux of the brighter, Eastern, H\,\textsc{ii} region. The HSC-SSP g-band image (see below) is matched in the same way by determining the centroid of the H\,\textsc{ii} region after resampling to the resolution of the MUSE WFM cube.

\subsubsection{\texorpdfstring{Integrated H\,{\sc ii} region spectra}{Integrated H II region spectra}}

To compare directly different H\,\textsc{ii} regions across the host we extract several integrated spectra from different H\,\textsc{ii} regions (see Fig. \ref{fig:host_spectra}). From the WFM we can only extract the complete H\,\textsc{ii} region containing the SN, while in the NFM this splits into two distinct H\,\textsc{ii} regions, which we call ``H\,\textsc{ii}-region East''  (the brighter one) and ``H\,\textsc{ii}-region West''. The SN is located at the very Eastern edge of the Western SF region. We extract four further prominent H\,\textsc{ii} regions across the host from the WFM data, three of them south of the giant H\,\textsc{ii} region (H\,\textsc{ii} SE-1 through 3), H\,\textsc{ii} SE-2 might correspond to the galaxy bulge if this is indeed a spiral galaxy, H\,\textsc{ii} SE-4 is the brightest SF region in the South-Eastern spiral arm. 

All SF regions show typical nebular emission lines, the line fluxes for the integrated regions are listed in Tab.~\ref{tab:integratedspecs}. For non-detections, 3$\sigma$ flux limits are calculated at each wavelength, by multiplying the spectral error around the expected line centre and multiplying it by a window of three pixels (3.75 {\AA}, width of the expected line), times three to get a 3$\sigma$ limit. The emission lines all share the same redshift (see Fig. \ref{fig:host_spectra}), hence the giant H\,\textsc{ii} region and the SN site originate in that galaxy and that all H\,\textsc{ii} regions belong to the same galaxy. The bright H\,\textsc{ii} region (and its respective Eastern and Western parts) as well as the H\,\textsc{ii} region just south of it (H\,\textsc{ii} SE-1) show emission of He\,\textsc{i} $\lambda$4740, 4923, 5876 and 6680 and [O\,\textsc{i}]. H\,\textsc{ii} East and West furthermore have strong emissions of [Ar\,\textsc{iii}]+[\textsc{iv}], [S\,\textsc{iii}], [Fe\,\textsc{iii}] and there is a possible detection of He\,\textsc{ii} in the Eastern H\,\textsc{ii} region. The presence of He\,\textsc{ii} implies a very hard radiation field, here likely produced by a large number of young, massive stars, including Wolf-Rayet stars. Furthermore, our early X-shooter spectrum (see Tab.~\ref{tab:speclog}), with a wavelength coverage down to 3000 \AA, shows Balmer lines in emission down to $\sim$H20, and no indication of stellar absorption. All this indicates a very young stellar population at least in the H\,\textsc{ii}-East region, reminiscent of some extreme hosts of superluminous SNe (SLSNe) such as PTF12dam \cite{Thoene15}.

\subsubsection{SED fitting}
To determine masses we performed SED fitting of the galaxy and the hosts of EP250827b, EP250108a and EP250304a, connected to Ic-BL SNe. We retrieved images from public archives using Pan-STARRS DR2 \citep{Chambers2016} and the DESI Legacy Imaging Surveys DR10 \citep{LegacySurvey} at optical wavelengths and performed aperture photometry using \texttt{stdweb} \citep{stdweb}. The aperture size was chosen to encompass the full optical extent of each host galaxy. UV photometry was obtained from GALEX \citep{GALEX}, mid-infrared measurements from the AllWISE catalog \citep{Cutri2013}, using the aperture photometry products provided in the catalog by taking the same aperture as in optical, as closely as permitted by the available data products. The resulting host-galaxy photometry is presented in Table~\ref{tab:SEDfits}.

We then performed SED fitting using CIGALE\footnote{\url{http://cigale.lam.fr}} (Code Investigating GALaxy Emission; \cite{Boquien2019}), taking a parameter space similar to \cite{Corre2018}. We adopted a delayed star formation rate ($\text{SFR} \propto t/\tau_0^2 \cdot e^{-t/\tau_0}$), with a recent burst of star formation on top. The initial mass function of \cite{Chabrier03} is adopted with the stellar synthesis models of \cite{BC03}. The stellar emission absorbed by dust and re-emitted in the IR is modelled using IR templates from \cite{Dale14} and using $\alpha_{\rm IR} \in [1,3]$. To account for possible systematic uncertainties between the different photometric datasets and model assumptions, an additional uncertainty of 0.1 mag was added in quadrature to all photometric measurements, following the implementation available in CIGALE. The host galaxy SED results are provided in Table~\ref{tab:SEDfits}.

\subsubsection{Mass-metallicity and sSFR-metallicity relations}
For the host of SN\,2026gzf we used the metallicity distribution of all spaxels from the WFM with a S/N in [N\,\textsc{ii}] (the weakest line) of at least 3 and the O3N2 parameter as described above. For the SSFR we use the H$\alpha$ derived SFR and the $g$-band flux using a pre-explosion image from Hyper Suprime-Cam (HSC) on the Subaru telescope from the Hyper Suprime-Cam Subaru Strategic Survey (HSC-SSP) \cite{HSC_SSP}. The HSC image has a sampling of 0.168$^{\prime\prime}$, very similar to the sampling of the MUSE-WFM of 0.2$^{\prime\prime}$ and we resample the HSC image to the MUSE resolution before deriving the SSFR per spaxel.

For the comparison samples we also applied a largely uniform approach to facilitate the comparison. \cite{modjaz20} compares different metallicity calibrations for PTF Ic and Ic-BL SN hosts with a mix of SN site and host nucleus measurements. We do not use their metallicity values directly, but instead derive them consistently using the parameter $\mbox{O3N2} = \log\left({\rm[O\,\textsc{iii}]/H\beta\times H\alpha/[N\textsc{ii}]}\right)$ in the \cite{Marino13} calibration since they also list all needed line fluxes and errors in their paper. Host masses and SFRs are derived from global photometry, also listed in the paper. The SFRs used in the plots are the ones obtained from SED fitting, since their H$\alpha$ fluxes, which we otherwise use to derive the SFR, are (sometimes) not representing the flux in the entire galaxy. 

For long GRB hosts we use global values for line fluxes to derive the metallicity using the O3N2 parameter and masses from different sources in the literature and derive the SFR using the H$\alpha$ line fluxes reported in the literature using the relation \cite{Kennicutt1992}: $\mbox{SFR}/(1\,M_\odot\, \text{yr}^{-1}) = 7.9 \times L({\rm H}\alpha)/(10^{42}\,\rm{erg\,s}^{-1})$. A similar approach was used for the data from the SDSS DR 16. We obtained emission line fluxes to derive the metallicity, the SFR is taken from the database itself, using the median estimate of the total SFR probability distribution function ``SFR\_TOT\_50'', which again was derived from H$\alpha$, corrected for extinction and for aperture effects due to the size of the SDSS fibres. Metallicities for superluminous supernova (SLSN) hosts are taken directly from \cite{Leloudas15, Schulze14, Perley16}, which are partially based on the O3N2 parameter, partially on direct measurements using the electron temperature of oxygen from the [OIII]$\lambda$4363 line. 

To derive the luminosity weighted specific SFR we compare the SFR (derived from H$\alpha$ or the SED fit) to an absolute $B$-band rest-frame magnitude. For the  Ic and Ic-BL sample from \cite{modjaz20} we use the listed host photometry in the $g'$-band and transformed it to an absolute magnitude, the same is done for the SDSS galaxies using the $g'$-band value from the database. The SLSN host samples directly list $M_B$ values. For GRB hosts we use $M_B$ values listed in \cite{Hjorth12, Kruehler11} and measurements from later GRBs from their respective papers.

\subsection{Rate estimates}
The initial trigger of EP suggest a signal to noise ratio of 7, implying that the event would not have been seen far beyond 150 Mpc, we therefore adopt this as $V_{\rm max}$, the observed rate assuming isotropic emission is then simply $1/(\Omega V_{\rm max} t \eta)$ where $\Omega$ is the fraction of the sky observed by the instrument (3600 deg$^2$, corresponding to 0.087), $t$ is the observing time which we set to 2~yr (although the nominal start date was 2024 July 11, several events were detected before this) and $\eta$ is the effective duty cycle of the instrument. This latter number is hard to estimate, in addition to times when the satellite is off-line or slewing, it is also likely that events of this nature within the Galactic plane would be confused or rendered invisible due to high Galactic $N_H$. We therefore set this to 0.5. We therefore obtain a rate of
$R = 0.8^{+1.3}_{-0.6} \times 10^{-6}$ Mpc$^{-3}$ yr$^{-1}$. The ASASSN Ic-BL rate of $0.33^{+0.33}_{-0.22} \times 10^{-6}$ Mpc$^{-3}$ yr$^{-1}$ is well within this rate. We hence conclude that it is possible, although by no means required given the substantial uncertainties, that all Ic-BL SNe create shock-breakouts similar to EP260321a. It is striking in this regard that SN\,2026gzf is a high luminosity SN Ic-BL, similar to those seen in GRBs, and not a lower luminosity event which appear to make up the bulk of the population.

The SN Ic rate from ASAS-SN is an order of magnitude higher than this $4^{+3}_{-2}$ Mpc$^{-3}$ yr$^{-1}$. However, even this value is not substantially higher than the upper bound implied by the detection of EP260321a. If all SN Ic create shock-breakout emission of similar luminosity it may be unlikely that the first such event would be a Ic-BL, but observations to date place few constraints on the possibility of their driving luminous breakouts. We further note that the soft thermal emission associated with breakouts is readily absorbed by gas and dust around the SN. In particular, for a 0.1 keV blackbody, the effective count rate of the EP-WXT is reduced by a factor of two for $N_H = 10^{21}$ cm$^{-2}$, equivalent to $A_V \approx 0.5$\,mag, and similar to the median $N_H$ seen in the {\em Swift} GRB sample. Hence, observations to date do not strongly constrain shock-breakout emission rates in stripped envelope SNe. It should be mentioned, that while WXT's field of view is a huge advantage, the narrow band pass of 0.5\,--\,4\,keV, means that fast evolving $\sim$\,0.1\,keV blackbodies sources, like these shock-breakouts, are more challenging to catch.

\section*{Acknowledgements}
We would like to thank D. B. Malesani for his constant support during the different phases of the observing campaign and for his useful comments that helped improving this paper.

This work is partially based on observations collected by the Stargate collaboration at the European Southern Observatory under ESO programme 114.27PZ and the dedicated GRB/FXT-SN programme 116.297W. 

This paper is partially based on observations obtained with the Las Cumbres Observatory Global Telescope (LCOGT)/Sinistro at Cerro Tololo Inter-American Observatory, Siding Spring Observatory, and Sutherland Observatory (programme CLN2026A-001; Chac\'on PI), with the VLT Survey Telescope (VST)/OmegaCAM at ESO’s Paranal Observatory (ESO programme 116.29H5; Chac\'on PI), and with the Planetary Defense 1m telescope equipped with an Apogee camera at Cerro Tololo Inter-American Observatory (programme CN2026A-67; Chac\'on PI) ; we thank telescope operators Tyler Linder and Lucas Hutton for their support with the latter. 

Data for this paper have been obtained in part under the International Time Programme of the CCI (International Scientific Committee of the Observatorios de Canarias of the IAC) with the NOT and GTC operated on the island of La Palma by the Roque de los Muchachos. 

COLIBR\'I received support from the French government under the France 2030 investment plan, as part of the Initiative d’Excellence d’Aix-Marseille Universit\'e-A*MIDEX through (ANR-11-LABX-0060 - OCEVU) and (AMX-19-IET-008 - IPhU), from LabEx FOCUS (ANR-11-LABX-0013), From Centre National d'Etudes Spatiale (CNES) and from CSAA-INSU-CNRS support program, and in Mexico from UNAM (Secretaria Administrativa, Coordinacion de la Investigacion Cientıfica, Instituto de Astronomıa and PAPIIT grant IN105921), and SECIHTI/CONACyT (277901, Ciencias de Frontera 1046632 and Laboratorios Nacionales). The COLIBR\'I team thanks the staff of the Observatorio Astron\'omico Nacional at Sierra de San Pedro M\'artir, as well as the technical and engineering teams at CEA, CPPM, IRAP, LAM, OHP, OSU Pytheas, and UNAM.

The Gravitational-wave Optical Transient Observer (GOTO) project acknowledges support from the Science and Technology Facilities Council (STFC, grant nos ST T007184/1, ST/T003103/1, ST/T000406/1, ST/X001121/1, and ST/Z000165/1) and the GOTO consortium institutions; the University of Warwick; Monash University; the University of Sheffield; the University of Leicester; Armagh Observatory \& Planetarium; the National Astronomical Research Institute of Thailand (NARIT); the University of Manchester; the University of Birmingham; Instituto de Astrofisica de Canarias (IAC); the University of Portsmouth and the University of Turku.

Data for this paper have been obtained from observations made with the Gran Telescopio Canarias (GTC), installed at the Spanish Observatorio del Roque de los Muchachos of the Instituto de Astrofísica de Canarias, on the island of La Palma, programme GTCMULTIPLE3G-26A (PI Geier).

This paper is partially based on observations made with the Nordic Optical Telescope, owned in collaboration by the University of Turku and Aarhus University, and operated jointly by Aarhus University, the University of Turku and the University of Oslo, representing Denmark, Finland and Norway, the University of Iceland and Stockholm University at the Observatorio del Roque de los Muchachos, La Palma, Spain, of the Instituto de Astrofisica de Canarias. The NOT data were obtained under program ID P71-506 (PI Malesani, Fynbo, Xu).

The Liverpool Telescope is operated on the island of La Palma by Liverpool John Moores University in the Spanish Observatorio del Roque de los Muchachos of the Instituto de Astrofisica de Canarias with financial support from the UK Science and Technology Facilities Council.

The Joan Or\'o Telescope (TJO) of the Montsec Observatory (OdM) is owned by the Generalitat de Catalunya and operated by the Institute for Space Studies of Catalonia (IEEC). We acknowledge the support provided by telescope operators from Observatori del Montsec.

Observations reported here were obtained at the MMT Observatory, a joint facility of the University of Arizona and the Smithsonian Institution. MMT Observatory access was supported by Northwestern University and the Center for Interdisciplinary Exploration and Research in Astrophysics (CIERA).

We acknowledge the scientific and technical staff of the 3.6m Devasthal Optical Telescope (DOT), a national facility operated and managed by the Aryabhatta Research Institute of Observational Sciences (ARIES), an autonomous institute under the Department of Science and Technology, Government of India, for their support in facilitating observations and providing the observational data.

LC and AMC acknowledge the support of the Irish Research Council Postgraduate Scholarship No GOIPG/2022/1008.
BPG acknowledges support from STFC grant no. ST/Y002253/1 and the Leverhulme Trust grant no. RPG-2024-117.
AdUP is supported by the Programme National Astro of CNRS/INSU with INP and IN2P3, co-funded by CEA and CNES through the Thematic Actions PEM, PCMI and CG of INSU Programme National “Astro".
PGJ, JQV, JNDD, JSS and APCH are supported by the European Union (ERC, Starstruck, 101095973, PI Jonker). Views and opinions expressed are, however, those of the author(s) only and do not necessarily reflect those of the European Union or the European Research Council Executive Agency. Neither the European Union nor the granting authority can be held responsible for them.
RLCS acknowledges support from the Leverhulme Trust grant no. RPG-2023-240.
FDC acknowledges support from the DGAPA/PAPIIT grant IN113424.
JDL acknowledges support from a UK Research and Innovation Future Leaders Fellowship (grant references MR/T020784/1 and UKRI1062).
FEB acknowledges support from ANID-Chile BASAL CATA FB210003 and FONDECYT Regular 1241005.
DO acknowledges support from the Leverhulme Trust grant no. RPG-2024-117.
TLK acknowledges support from a Warwick Astrophysics prize post-doctoral fellowship made possible thanks to a generous philanthropic donation.
RB acknowledges funding from the Italian Space Agency, contract ASI/INAF n. I/004/11/6. RB and AM acknowledge support from the INAF grant no. 1.05.24.03.04.
FG acknowledges financial support from the SOXS project (PI S. Campana)
NG and LGG gratefully acknowledge the support of the Simons Foundation (MP-SCMPS-00001470, N.G.).
AMW is grateful for support from UNAM/DGAPA project IN109224.
AR acknowledges financial support from the SOXS project (PI S. Campana) and from the PRIN-INAF 2022 "Shedding light on the nature of gap transients: from the observations to the models".
Based on observations collected at Copernico and Schmidt telescopes (Asiago Mount Ekar, Italy) of the INAF -- Osservatorio Astronomico di Padova.
LR acknowledges funding from the Trottier Space Institute Fellowship, the Natural Sciences and Engineering Research Council of Canada (NSERC) Arthur B. McDonald Fellowship and Discovery Grant programs, the Canada Research Chairs (CRC) program, the Fondes de Recherche Nature et Technologies (FRQNT), the Centre de recherche en astrophysique du Québec (un regroupement stratégique du FRQNT), and the AstroFlash research group. The AstroFlash research group at McGill University, University of Amsterdam, ASTRON, and JIVE is supported by: a Canada Excellence Research Chair in Transient Astrophysics (CERC-2022-00009); an Advanced Grant from the European Research Council (ERC) under the European Union’s Horizon 2020 research and innovation programme (`EuroFlash’; Grant agreement No. 101098079); an NWO-Vici grant (`AstroFlash’; VI.C.192.045); an NSERC Discovery Grant (RGPIN-2025-06681); an ERC Starting Grant (`EnviroFlash’; Grant agreement No. 101223057); and an NWO-Veni grant (VI.Veni.222.295).
JKL acknowledges support from the University of Toronto and Hebrew University of Jerusalem through the University of Toronto -- Hebrew University of Jerusalem Research and Training Alliance program.  The Dunlap Institute is funded through an endowment established by the David Dunlap family and the University of Toronto. 
KM acknowledges support from JSPS KAKENHI grant No. JP24KK0070 and JP24H01810.
JM acknowledges the funding of the National Key R\&D Program of China (2023YFE0101200), the Natural Science Foundation of China 12393813, and the Yunnan Revitalization Talent Support Program (YunLing Scholar Project).
A. Rossi acknowledges from the INAF project ‘Supporto Arizona \& Italia’.
CV acknowledges from the INAF project ‘Supporto Arizona \& Italia’.
BS acknowledges the support of the French Agence Nationale de la Recherche (ANR), under grant ANR-23-CE31-0011 (project PEGaSUS).
MAA acknowledges support from MCIN/AEI/10.13039/501100011033 and the European Union grants PID2021-127495NB-I00 and PID2025-171322NB-C22 funded by, as well as from the Generalitat Valenciana through the Prometeo excellence programme grant CIPROM/2022/13. 
WF gratefully acknowledges support by National Science Foundation under grant Nos. AST-2206494, AST-2308182, AST-2432037, and CAREER grant No. AST-2047919, the David and Lucile Packard Foundation and the Research Corporation for Science Advancement through Cottrell Scholar Award \#28284.

\section*{Data availability}
All spectra presented in this work will be made available without denoising and emission line clipping via \texttt{WISeREP} (\url{https://www.wiserep.org}). The optical/NIR photometry will be made available via \texttt{CDS} portal (\url{https://portal.cds.unistra.fr}) in machine readable format.

\section*{Author contribution}
AMC led this project, including coordination between different telescopes during data acquisition, photometric and spectroscopic data analysis, modelling, and writing this paper. AMC is the PI of a dedicated ESO proposal aimed at the study of SN associated with high-energy prompt emission. CT has led the host galaxy and environment data analysis and wrote those sections in the paper. JKL is the PI of the VLA radio programme and led the scheduling and data analysis of the radio data. GC reduced all the VLT/FORS2 and NOT spectra as well as contributed to photometric analysis. AdUP scheduled and analysed the GTC spectra and has helped to write the paper. He also acted as the coordination link between AMC and the COLIBR\'I collaboration. PJ as the PI of the FXT group has coordinated the photometric observations provided by this team and has provided significant comments throughout the text. LI has provided the data reduction of several spectra, coordinated the inclusion of radio data in this paper and provided suggestions and comments from the early days of the campaign. AL provided valuable suggestions during the concept development of this paper and wrote the sections on rates for the paper. BPG is the coordinator of the GRB working group at the GOTO collaboration and led the investigation, analysis, and writing of the GOTO pre-explosion search. SB, is the PI of COLIBR\'I collaboration that provided multi-band photometry. NS provided valuable conversations and support for the data modelling with \texttt{Redback}. JQV, JAC and FB  provided photometric data via their programmes at different ground-based telescopes (LCO, PD and VST), the data reduction and valualbe feedback throughout the paper. REF was involved in data acquisition with his LT programme, data reduction and provided valuable suggestions throughout the paper. RB coordinated the observations taken by the CIBO collaboration. ALW and JGD contributed to the acquisition, reduction, and analysis of the COLIBR\'I data. NRT and SV are PIs of the Stargate programme and were involved in providing some of the ESO data used in this paper, as well as, comments from the conception of this paper. AM, ARossi, WfF, PK, KM, AG, AR and DX were involved in the acquisition of data taken from different ground-based telescopes as well as their data reduction. MAA contributed to the theoretical interpretation and implications of our observations to the progenitor-system, including the plausibility and limitations of a WR-like stripped-star scenario. All authors contributed to the preparation of this manuscript through contributions to concept development, discussion, and text.

\section*{Competing interests}
The authors declare no competing interests.

\begin{appendices}
\section{Extended Data}

\begin{figure}[!ht]
\centering
\includegraphics[angle=0,width=1\textwidth]{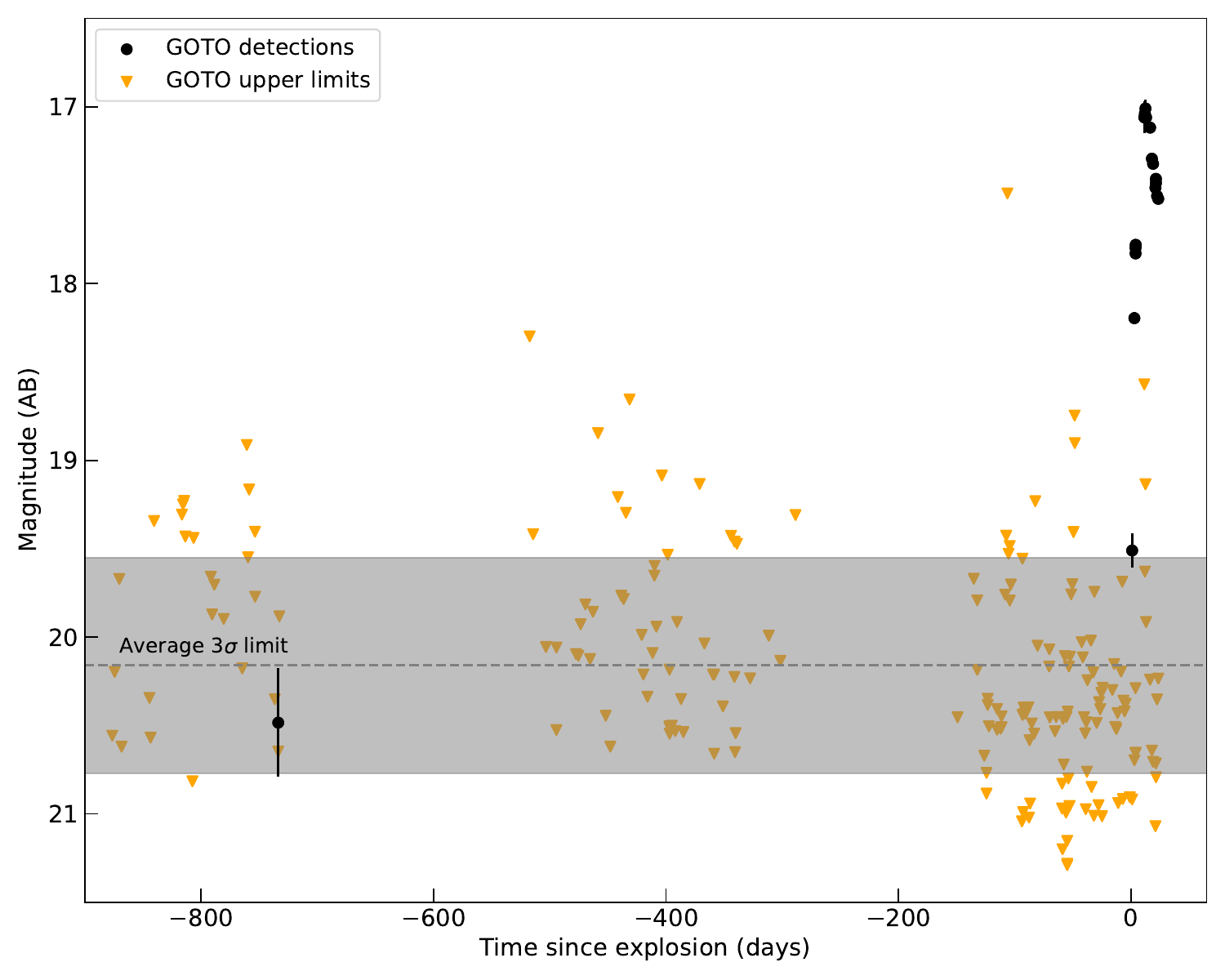}
\caption{GOTO light curve of the SN\,2026gzf location covering from 876 days before the explosion to the peak of the SN The dashed horizontal line shows the average 3\,$\sigma$ limiting magnitude with the shaded gray area representing its 1\,$\sigma$ error. The very early detection falls deep into the average upper limit and thus, does not represent a credible emission detection.}
\label{fig:gotolc}
\end{figure}

\begin{figure}[!ht]
\centering
\includegraphics[angle=0,width=1\textwidth]{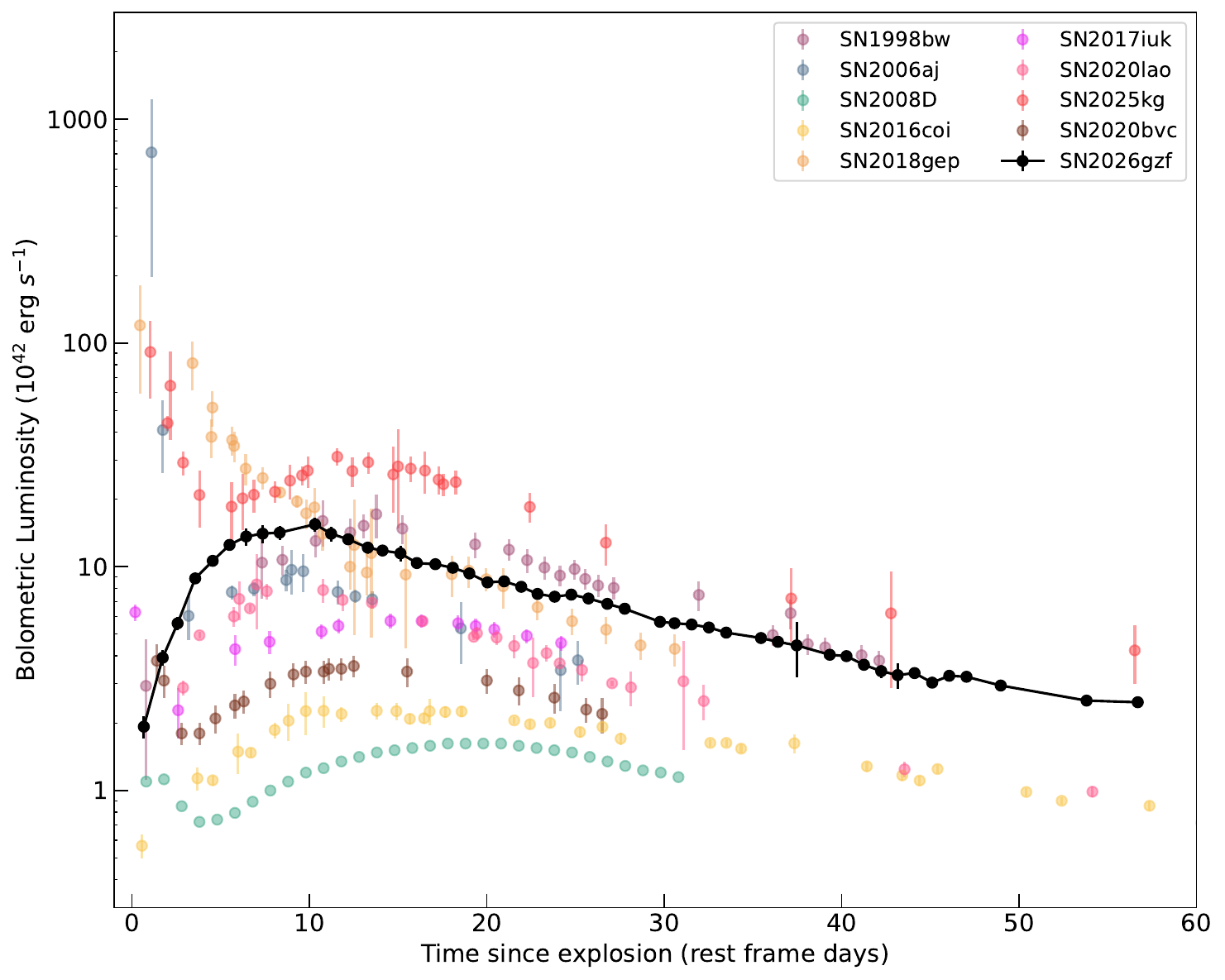}
\caption{Bolometric luminosity obtained through blackbody SED models for SN\,2026gzf and other GRB/FXT-SN and other Ic-BLs with no associated prompt emission.}
\label{fig:sn_luminosity}
\end{figure}

\begin{figure}[!ht]
\centering
\includegraphics[angle=0,width=0.7\textwidth]{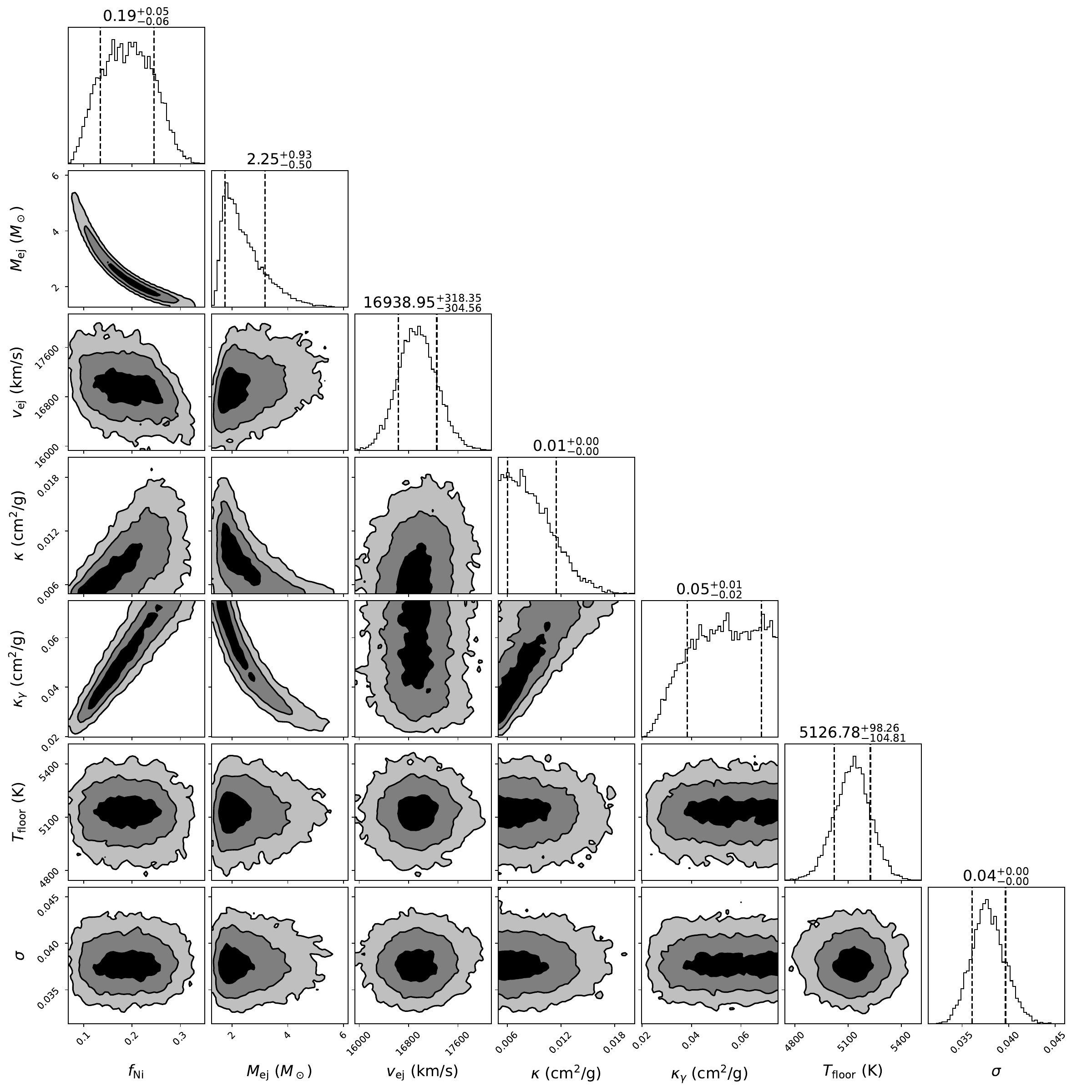}
\includegraphics[angle=0,width=0.7\textwidth]{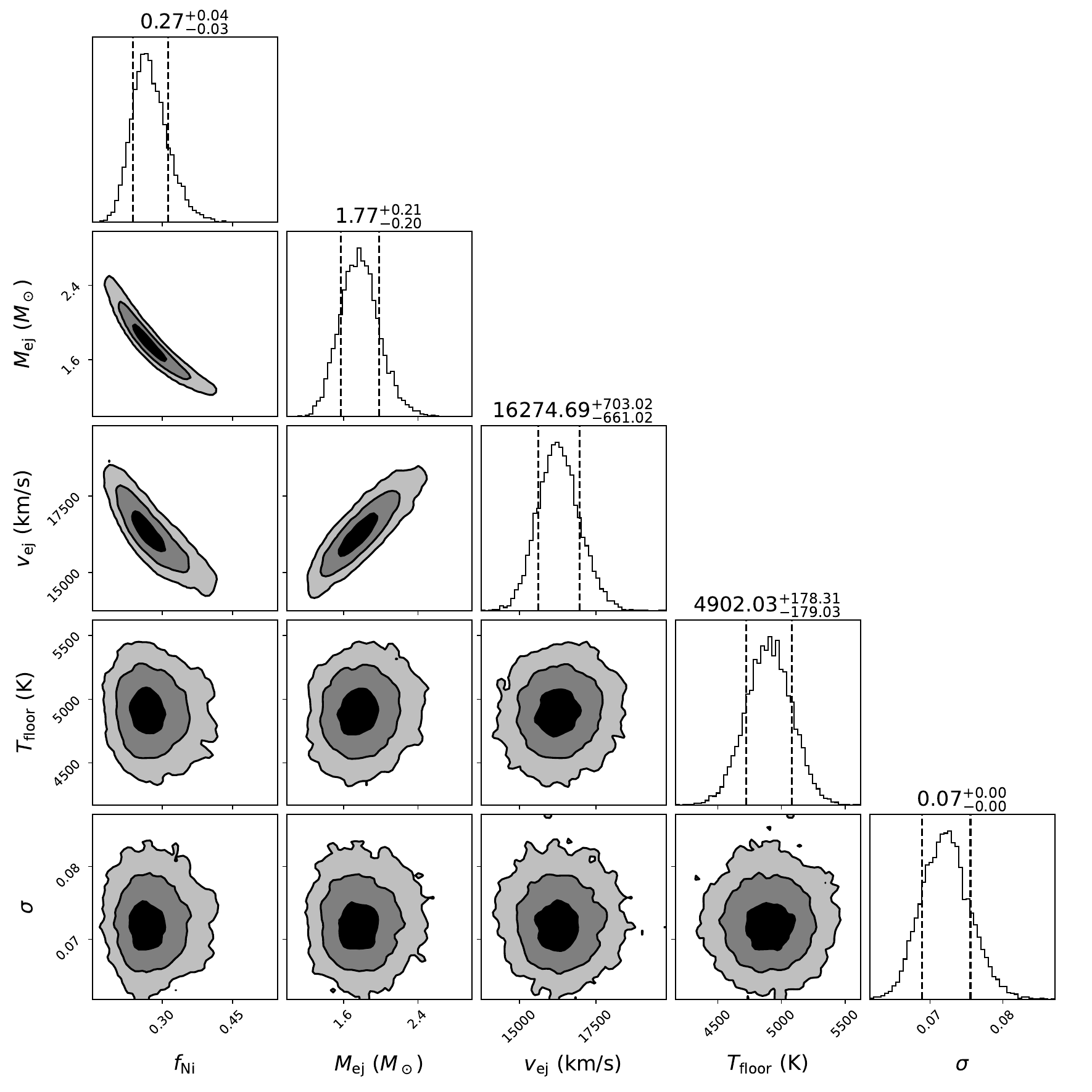}
\caption{\textbf{Top}: Corner plot for the parameters of our Arnett fit model with all parameters set free. \textbf{Bottom}: Corner plot of our Arnett fit model with both opacities frozen ($\kappa$\,=\,0.07 and $\kappa_{\gamma}$\,=\,0.03).}
\label{fig:contours_arnett}
\end{figure}

\begin{figure}[!ht]
\centering
\includegraphics[angle=0,width=0.7\textwidth]{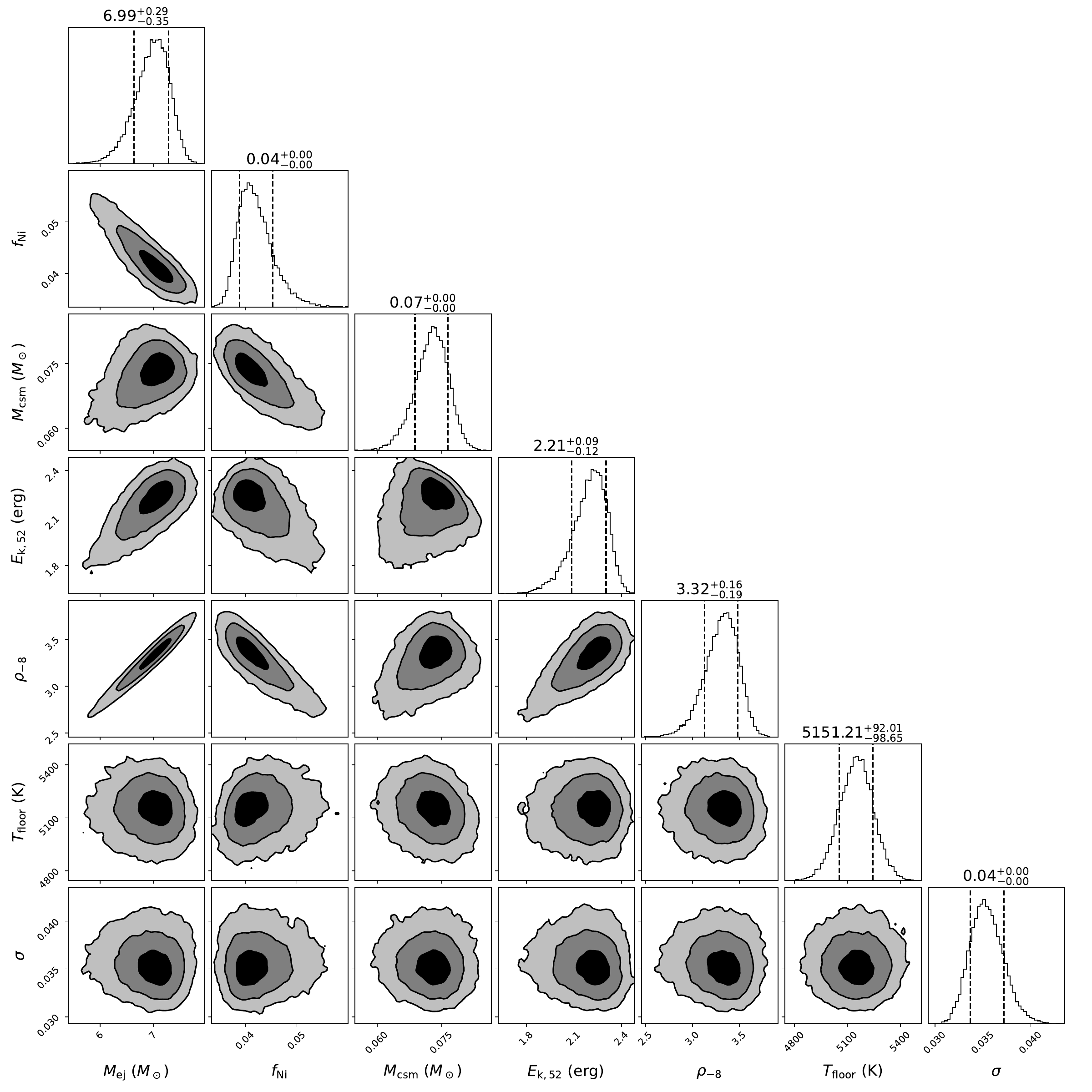}
\includegraphics[angle=0,width=0.7\textwidth]{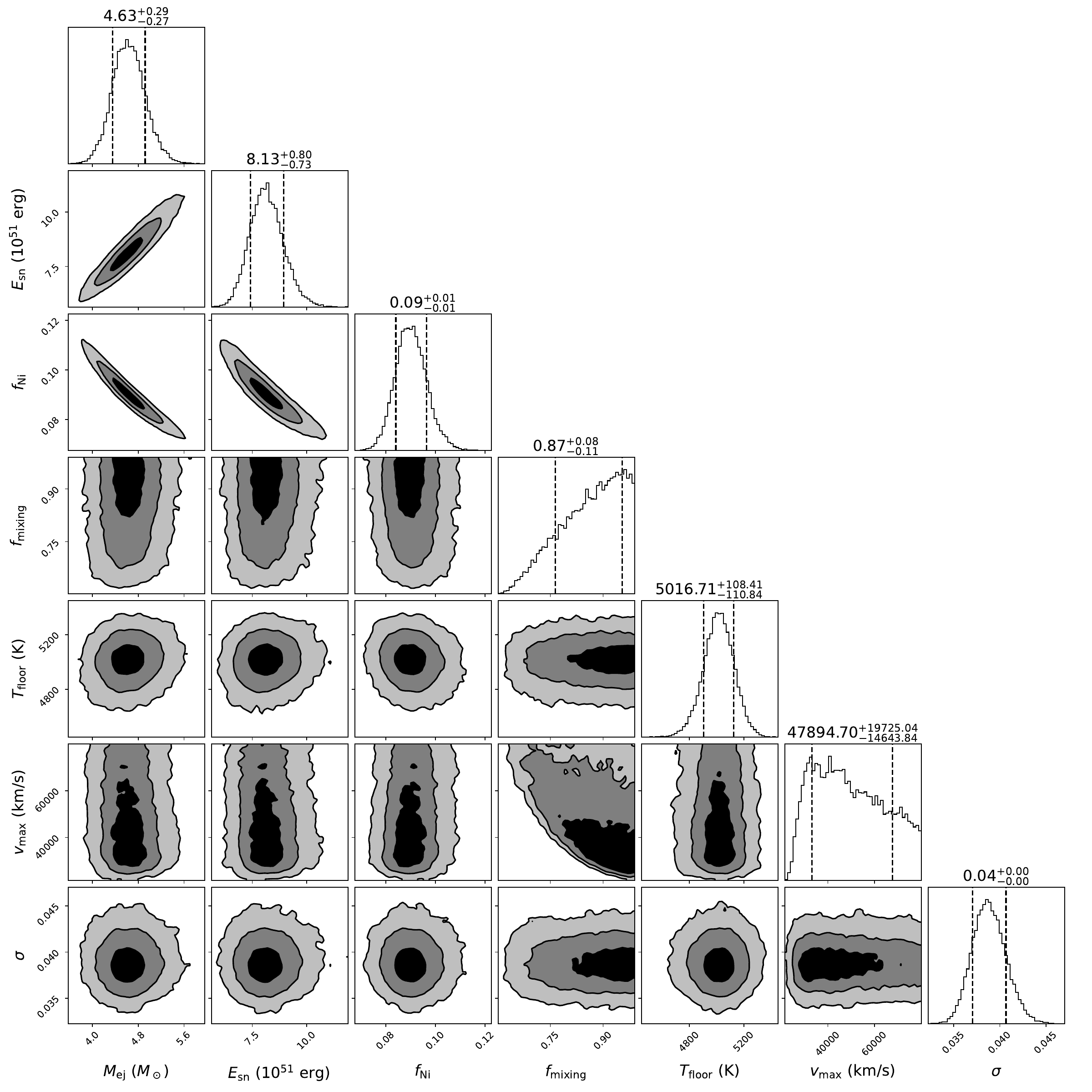}
\caption{\textbf{Top}: Corner plot for the parameters of our CSM\,+\,SN-2 fit model. \textbf{Bottom}: Corner plot for the parameters of our Nickel mixing SN fit model.}
\label{fig:contours_models}
\end{figure}

\begin{figure}[!ht]
\centering
\includegraphics[angle=0,width=0.49\textwidth]{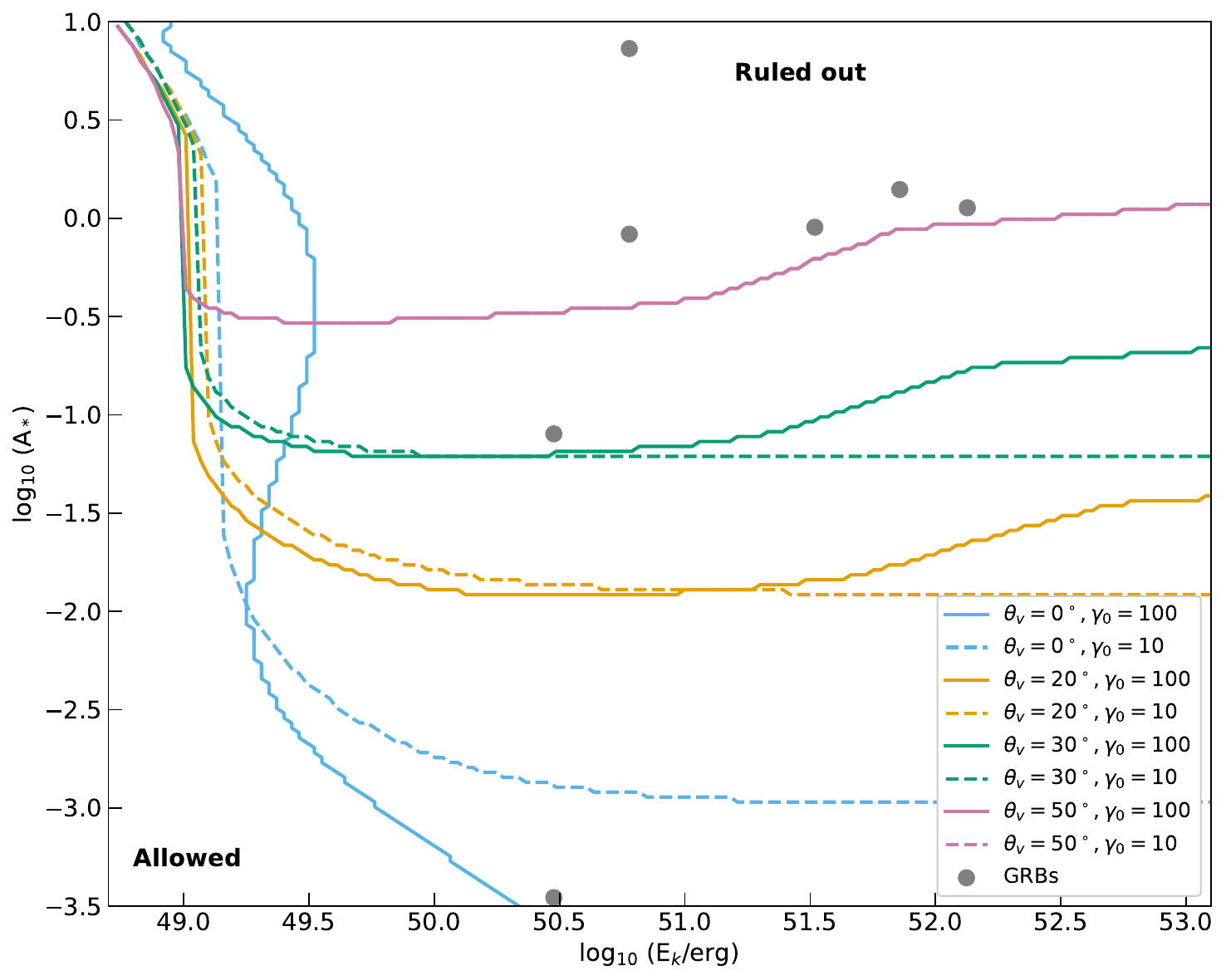}
\includegraphics[angle=0,width=0.49\textwidth]{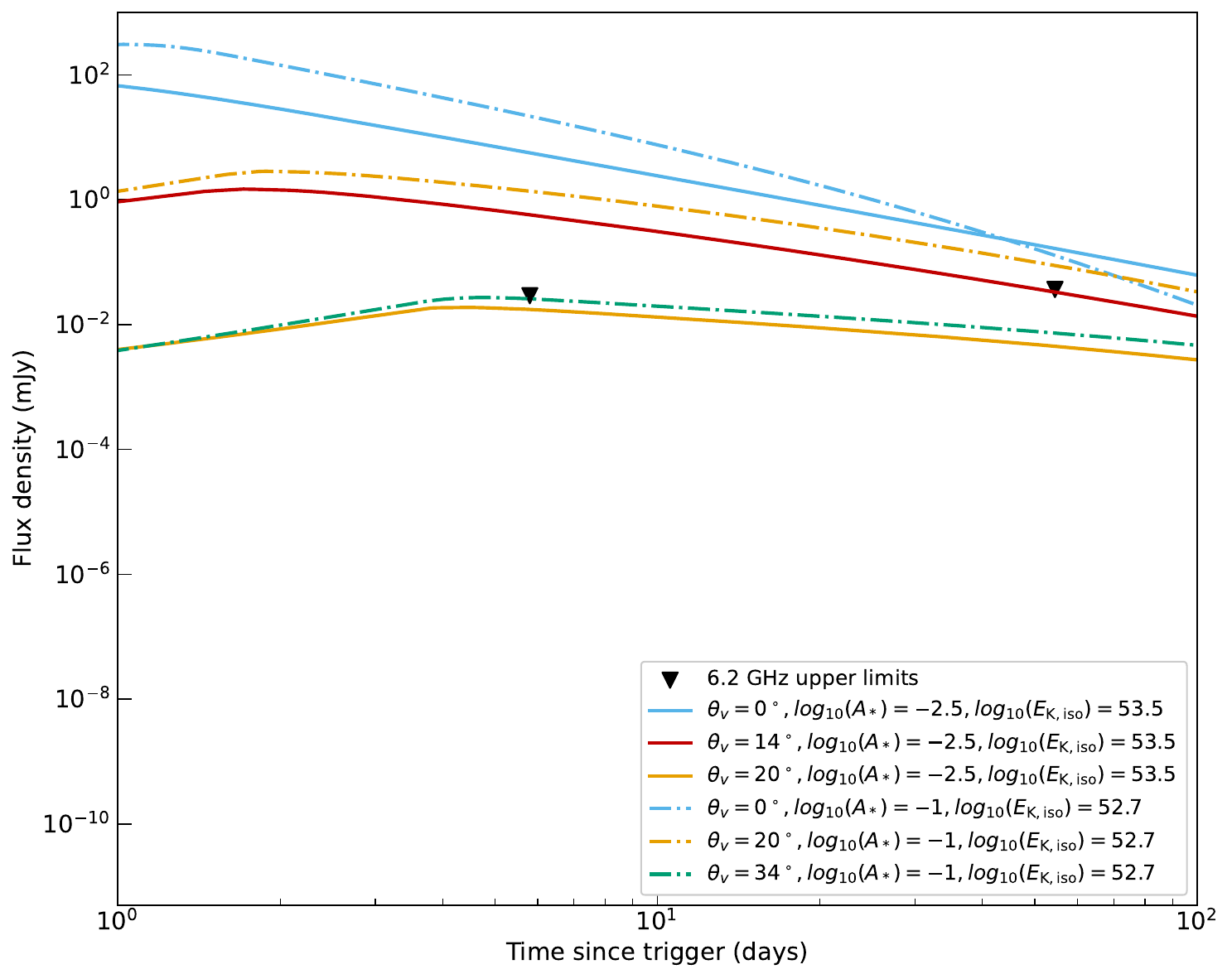}
\caption{\textbf{Left}: The valid parameter space for a relativistic jet in the Wind medium density ($A_{*}$) versus beamed corrected kinetic energy of a jet ($E_{\rm K}$) plane obtained from models generated with the \texttt{Redback} top-hat model for different viewing angles ($\theta_{\rm v}$) and two bulk Lorentz factors ($\gamma_0$). The lines represent the $A_{*}$\,--\,$E_{\rm K}$ boundary limits of \textit{allowed} (below) and \textit{ruled out} (above) from our radio VLA upper limits. The grey circles show values measured for a selection of on-axis Wind GRBs \cite{kangas2021}. \textbf{Right}: Predicted radio afterglow lightcurves at 6.2 GHz from a top-hat model relativistic jet for different viewing angles, wind medium densities and kinetic energies and fixed $\epsilon_e = 0.1$ and $\epsilon_B = 0.001$. Downward triangles mark our VLA observational upper limits. For typical values of $A_{*}$, a jet can only exist for off-axis viewing angles $>14^{\circ}$ in the case of a jet with high $A_{*}$ (typically weak jets), or $>34^{\circ}$, in the case of a more typical GRB afterglow.}
\label{fig:afterglow_wind}
\end{figure}

\begin{figure}[!ht]
\centering
\includegraphics[angle=0,width=\textwidth]{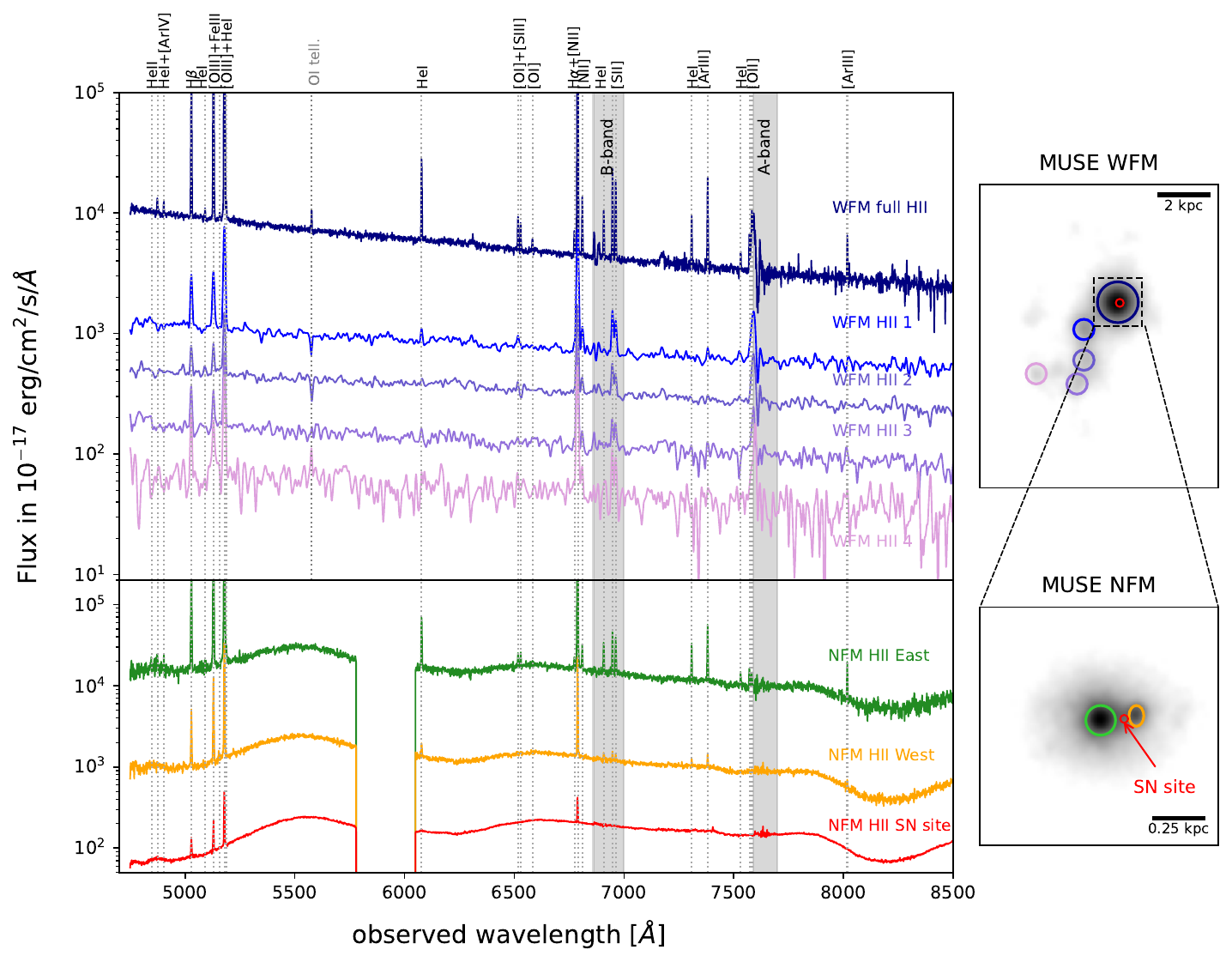}
\caption{Integrated spectra of different H\,\textsc{ii} regions across the galaxy in the MUSE WFM (top panels) and the part of the large H\,\textsc{ii} region covered by the MUSE NFM observations (bottom panels). The images in the right column are H$\alpha$ maps. The colors of the extracted regions in the H$\alpha$ maps correspond to their respective spectra on the left. The emission lines all lie at a common redshift of $z$\,=\,0.0343.}
\label{fig:host_spectra}
\end{figure}

\begin{table*}[!ht]
\centering
\caption{Telescopes and filters, with the number of epochs taken per filter shown in brackets, used for the photometric follow-up of SN\,2026gzf presented in this paper. The optical/NIR photometry will be made available via \texttt{CDS} portal (\url{https://portal.cds.unistra.fr}) in machine readable format.}  
\begin{tabular}{l c }        
\hline \hline
Telescope (acronym)  & Filters (number of epochs) \\
\hline
Las Cumbres Observatory (LCO)      &  $u$ (20), $g$ (64), $r$ (51), $i$ (44), $z$ (48)                          \\
COLIBR\'I                &    $g$ (53), $r$ (51), $i$ (50), $z$ (52), $y$ (50)                         \\
Planetary Defense (PD)      &   $g$ (12), $r$ (13), $i$ (14), $z$ (13)                       \\
VLT Survey Telescope (VST)      &     $u$ (4), $g$ (1), $r$ (1), $i$ (1)                       \\
Liverpool Telescope (LT)      &     $u$ (3), $g$ (12), $r$ (1), $z$ (8)                        \\
Joan Or\'o Telescope (TJO)      &        $g$ (10), $r$ (10), $i$ (10)                     \\
Galileo National Telescope (TNG)      &     $J$ (4), $H$ (5)                       \\
Large Binocular Telescope Observatory (LBT)      &      $g$ (1), $r$ (1),  $i$ (1), $z$ (1), $J$ (2), $H$ (2), $K$ (1)                         \\
Rapid Eye Mount (REM)      &    $g$ (8), $r$ (8), $i$ (8)                        \\
Asiago Schmidt            &    $g$ (2), $r$ (3), $i$ (2)                        \\
Very Large Telescope (VLT)      &   FORS2-R (1)                         \\

 \hline   \hline
    \end{tabular}
    \label{tab:optical_observations}
\end{table*}

\begin{table}[!ht] 
    \centering
        \caption{Observational log of the spectroscopic campaign run to monitor the spectral evolution of SN\,2026gzf.}
    \begin{tabular}{c c c c c }
    \hline \hline
       Start date 	&	$\Delta{t}$	&	Telescope	&	Instrument	&	Wavelength range	\\
       (UT)    &    (days)     &               &               &       (\AA)     \\
       \hline
2026-03-22T05:20:33	&	0.70	&	VLT	&	MUSE	&	4800--9300	\\
2026-03-23T00:51:20	&	1.51	&	VLT	&	X-shooter	&	3000--21000	\\
2026-03-24T00:52:14	&	2.52	&	VLT	&	FORS2	&	3300--11000	\\
2026-03-24T21:15:00 &   3.36    &  Copernico 1.82m   & AFOSC & 3700--8130  \\
2026-03-25T00:41:43	&	3.51	&	VLT	&	FORS2	&	3300--11000	\\
2026-03-26T03:10:50	&	4.61	&	VLT	&	FORS2	&	3300--11000	\\
2026-03-26T21:40:24	&	5.38	&	GTC	&	OSIRIS+	&	3620--9200  \\
2026-03-27T01:21:24	&	5.54	&	VLT	&	FORS2	&	3300--11000	\\
2026-03-28T20:56:54	&	7.35	&	GTC	&	OSIRIS+	&	3620--9200  \\
2026-04-01T00:30:18 &   10.50   &   NOT &   ALFOSC	&	3200--9600 \\
2026-04-02T14:38:17 &   12.08   &   DOT &   ADFOSC  &   4000--8800\\
2026-04-02T22:06:18 &   12.40   &   NOT &   ALFOSC	&	3200--9600 \\
2026-04-07T00:26:52	&	16.50	&	VLT	&	FORS2	&	3300--11000	\\
2026-04-08T00:56:19	&	17.52	&	VLT	&	FORS2	&	3300--11000		\\
2026-04-09T00:03:11	&	18.48	&	VLT	&	UVES	&   3050-10400		\\
2026-04-12T00:47:02	&	21.51	&	VLT	&	MUSE	&	4800--9300	\\
2026-04-13T01:03:47 &	22.52	&	NOT	&	ALFOSC	&	3200--9600	\\
2026-04-15T22:48:27	&	25.43	&	NOT	&	ALFOSC	&	3200--9600	\\
2026-04-17T23:53:01	&	27.47	&	VLT	&	FORS2	&	3300--11000	\\
2026-04-18T23:48:09	&	28.47 	&	VLT	&	X-shooter	&	3000--21000	\\
2026-04-20T00:46:19	&	29.51	&	VLT	&	FORS2	&	3300--11000	\\
2026-04-22T01:10:58	&	31.53	&	VLT	&	FORS2	&	3300--11000	\\
2026-04-23T23:53:58	&	33.47	&	VLT	&	FORS2	&	3300--11000	\\
2026-04-30T22:41:20	&	40.42	&	NOT	&	ALFOSC	&	3200--9600	\\
2026-05-03T23:40:09	&	43.47	&	VLT	&	FORS2	&	3300--11000	\\
2026-05-05T21:13:35	&	45.36	&	LT	&	SPRAT	&	4020--7976 	\\
2026-05-11T23:53:00	&	51.47	&	VLT	&	FORS2	&	4450--8650	\\
2026-05-14T00:22:44	&	53.49	&	VLT	&	FORS2	&	3300--11000	\\
2026-05-15T23:53:53	&	55.47	&	VLT	&	FORS2	&	3300--11000	\\
2026-05-16T01:05:25	&	55.52	&	MMT	&	BINOSPEC	& 4000--9000				\\
2026-05-19T00:46:08	&	58.51	&	MMT	&	BINOSPEC	& 4000--9000				\\
2026-05-25T23:43:01	&	65.47	&	VLT	&	FORS2	&	3300--11000 	\\
\hline \hline
    \end{tabular}
    \label{tab:speclog}
\end{table}

\begin{table}[!ht]     
\centering
\caption{Best fit parameters for the optical multi-band modelling of SN\,2026gzf with the one-zone model using different parameter configurations \texttt{Redback} v1.5.3.}  
\begin{tabular}{l l c c c}        
\hline \hline
Model & & Arnett  & Arnett $\kappa_{\gamma}$ fixed  & Arnett $\kappa$, $\kappa_{\gamma}$ fixed \\
Parameters & Priors & Mean values &   Mean values  &  Mean values\\
\hline
$M_{\mathrm{ej}}~(M_\odot)$ & Uniform(0.1, 10) &2.25$^{+0.93}_{-0.50}$ & 1.27$^{+0.08}_{-0.07}$ & 1.77$^{+0.21}_{-0.20}$\\
$f_{\mathrm{Ni}}$ & LogUniform(0.005, 0.75) & 0.19$^{+0.05}_{-0.06}$& 0.34$^{+0.02}_{-0.02}$ & 0.27$^{+0.04}_{-0.03}$\\
$v_{\mathrm{ejecta}}~(\mathrm{km\,s}^{-1})$ & LogUniform(13000, 30000) &16938.95$^{+318.35}_{-304.56}$ & 17483.91$^{+414.52}_{-369.80}$ &16274.69$^{+703.02}_{-661.02}$\\
$\kappa$& Uniform(0.005, 0.15) &0.01$^{+0.01}_{-0.01}$  & 0.07 & 0.07\\
$T_{\mathrm{floor}}~(\mathrm{K})$ & LogUniform(1000, 10000) &5126.78$^{+98.26}_{-104.81}$ & 4951.77$^{+142.02}_{-141.87}$&4902.03$^{+178.31}_{-179.03}$\\
$\kappa_{\gamma}$& Uniform(0.005, 0.075) & 0.05$^{+0.01}_{-0.02}$ & 0.07$^{+0.01}_{-0.01}$ &0.03\\
\texttt{log\_evidence} & --- & 432.53$^{+1.30}_{-1.93}$ & 352.58$^{+1.51}_{-2.11}$ &284.17$^{+1.16}_{-1.81}$\\
\hline \hline
    \end{tabular}
    \label{tab:arnett_modelling}
\end{table}

\begin{table}[!ht]     
\centering
\caption{Best fit parameters for the optical multi-band modelling of SN\,2026gzf with different models using \texttt{Redback} v1.5.3. Parameters with `---` on any particular column are not part of that model.}  
\begin{tabular}{l l c c c}        
\hline \hline
Model &  &CSM\,+\,SN-1  & CSM\,+\,SN-2  & Nickel mixing \\
Parameters & Priors & Mean values & Mean values  &  Mean values\\
\hline
$M_{\mathrm{ej}}~(M_\odot)$ & Uniform(1, 10) &  7.39$^{+0.26}_{-0.34}$&    6.99$^{+0.29}_{-0.35}$ & 4.63$^{+0.29}_{-0.27}$\\
$f_{\mathrm{Ni}}$ & LogUniform(0.005, 0.75)& 0.04$^{+0.01}_{-0.01}$ & 0.04$^{+0.01}_{-0.01}$& 0.09$^{+0.01}_{-0.01}$\\
$E_{\mathrm{k,52}}~(\mathrm{erg})$ & LogUniform(1e51, 5e52) &2.55$^{+0.16}_{-0.17}$ &  2.21$^{+0.09}_{-0.12}$& 0.81$^{+0.08}_{-0.07}$\\
$\kappa$& frozen & 0.07 &  0.07 & 0.07\\
$M_{\mathrm{CSM}}~(M_\odot)$ & Uniform(0.001, 0.1) &0.07$^{+0.01}_{-0.01}$  & 0.07$^{+0.01}_{-0.01}$ &  --- \\
$R_{0}\,(\rm AU)$& frozen & 2.00 & 2.00 & ---\\
$\eta$& Uniform(1.1, 2) &1.89$^{+0.08}_{-0.09}$ & 1.8 (frozen) &  ---\\
$\rho$ $(10^{-8})$ & LogUniform(0.7, 8) & 2.94$^{+0.50}_{-0.38}$ & 3.32$^{+0.16}_{-0.19}$  &  ---\\
$f_{\mathrm{mixing}}$ & LogUniform(0.01, 0.99) &--- &--- & 0.87$^{+0.08}_{-0.11}$\\
$v_{\rm max}\, ({\rm km\,s}^{-1})$ & LogUniform(15000, 80000) & --- & --- &47894.70$^{+19725.04}_{-14643.84}$\\
$T_{\mathrm{floor}}~(\mathrm{K})$ & LogUniform(4000, 10000) &5191.25$^{+95.73}_{-102.95}$ &5151.21$^{+92.01}_{-98.65}$ & 5016.71$^{+108.41}_{-110.84}$\\
$\kappa_{\gamma}$& frozen & 0.03  & 0.03 & 0.03\\
\texttt{log\_evidence} & --- & 447.24$^{+1.56}_{-2.20}$ &447.36$^{+1.58}_{-2.22}$ & 427.22$^{+1.38}_{-1.94}$\\
 \hline \hline  
    \end{tabular}
    \label{tab:sn_modelling}
\end{table}

\begin{table}[!ht]
\centering
\caption{Log of radio observations presented in this work}  
\begin{tabular}{l c c c c c}        
\hline \hline
Date & $\Delta{T}$  & Telescope & Frequency  & Flux density  & RMS \\
     & (days)       &           & (GHz)     &  (5\,$\sigma$, $\mu$Jy) & ($\mu$Jy\,beam$^{-1}$) \\
\hline
27-03-2026 07:20:05 &5.8 &	VLA	 &	6.2 &	$<$\,29 &	58 \\
27-03-2026 06:55:39	&5.8 &	VLA	 &	9.7 &	$<$\,36 & 	72 \\
27-03-2026 06:28:25	&5.7 &	VLA	& 	14.9 &	$<$\,28 &	56 \\
27-03-2026 05:57:05	&5.7 &	VLA	&	22.0 &	$<$\,39 &	77 \\
15-05-2026 01:38:15	&54.5 &	VLA	&	6.2 &	$<$\,37	&   73  \\
15-05-2026 01:18:49	&54.5 &	VLA	&	9.8 &	$<$\,43	&   85 \\
\hline \hline
    \end{tabular}
    \label{tab:radio_observations}
\end{table}

\begin{table}[!ht]
\caption{Top: Photometry used for the SED fits of the host of EP260321A and two other EP events connected to Ic-BL SNe, EP250827B/ SN\,2025wkm and EP250108A/ SN\,2025kg. The photometry is corrected for Galactic extinction. Second panel: Results from the SED fit performed in CIGALE (see text). Third panel: Emission line measurements. For the host of EP250304A/SN\,2025fhm there is only a spectrum with a few emission lines and no continuum of photometry, hence we only list results from the emission lines. Spectra are from \cite{rastinejad25, cotter26, Corcoran2026GCN} for SN\,2025kg, SN\,2025fhm and SN\,2025wkm, respectively. Line fluxes are in 10$^{-17}$erg\,cm$^{-2}$\,s$^{-1}$ and corrected for Galactic \citep{Schlafly2011} and intrinsic extinction (Balmer decrement).  Photometry sources: $^1$ GALEX \citep{GALEX}, $^2$ Pan-STARRS DR2 \citep{Chambers2016}, $^3$ WISE1 \citep{Cutri2013}, $^4$ Legacy-Survey \citep{LegacySurvey} }
\label{tab:SEDfits}
\begin{tabular}{lcccc}
\hline \hline
 & EP260321a &  EP250827b & EP250108a & EP250304a \\
  & SN\,2026gzf &  SN\,2025wkm & SN\,2025kg & SN\,2025fhm \\
\hline
FUV & 19.52 $\pm$ 0.12$^{1}$& --- & --- & ---\\
NUV & 19.38 $\pm$ 0.07$^{1}$ & --- & --- & ---\\
$g$ & 18.08 $\pm$ 0.03$^{2}$ & 22.29 $\pm$ 0.16$^{2}$ & 23.31 $\pm$ 0.06$^{4}$& ---\\
$r$ & 18.01 $\pm$ 0.03$^{2}$  & 21.93 $\pm$ 0.15$^{2}$ & 22.97 $\pm$ 0.05$^{4}$ & ---\\
$i$ & 18.02 $\pm$ 0.02$^{2}$ & 21.45 $\pm$ 0.09$^{2}$ & 22.93 $\pm$ 0.09$^{4}$ & ---\\
$z$ & 17.98 $\pm$ 0.03$^{2}$ &  22.09 $\pm$ 0.40$^{2}$ & 22.99 $\pm$ 0.17$^{4}$ & ---\\
$y$ & 18.11 $\pm$ 0.20$^{2}$ & 21.41 $\pm$ 0.44$^{2}$ & --- & ---\\
3.4 $\mu$m & 19.33 $\pm$ 0.14$^{3}$ & --- & --- & --- \\
4.6 $\mu$m & 19.74 $\pm$ 0.35$^{3}$ & --- & --- & --- \\
\hline
log$_{10}$ M$_{\star} $[M$_{\odot}$]&$8.45_{-0.18}^{+0.13}$& $<8.63$ & $<8.68$   & ---\\
log$_{10}$SFR [M$_{\odot}$/y] &$< -1.02$  &$-0.10_{-0.51}^{+0.23}$ & $-0.56_{-1.34}^{+0.29}$ & ---\\
A$_V$ [mag]& $< 0.26$ & $1.32 \pm 0.67$ & $<2.28$ & --- \\
\hline
[O~{\sc iii}] 5007& 7052.27$\pm$40.98& 151.20 $\pm$ 29.15	&5.48 $\pm$ 0.31	&2.01 $\pm$ 2.34\\
H$\beta$& 1081.63$\pm$13.80& 39.50 $\pm$ 8.03	&4.79$\pm$  0.31 &0.96 $\pm$1.16\\
H$\alpha$& 2703.44$\pm$ 13.34 &  108.96 $\pm$ 14.78	&10.14 $\pm$ 0.28 &2.62 $\pm$ 2.16\\
$[$N\,\textsc{ii}$]$ 6585  & 24.83$\pm$ 6.83 &2.54 $\pm$ 0.50	&1.07 $\pm$ 0.20	&$<$0.49\\
\hline
redshift& 0.0343 & 0.119 & 0.176 & 0.199\\
g$_\mathrm{abs}$ [mag]& -18.11$\pm$0.03 & -16.58 $\pm$ 0.34	&-16.23 $\pm$ 0.10	&---\\
12+log(O/H) \scriptsize{(O3N2)}& 8.02$\pm$0.01  & 8.06$\pm$0.03 & 8.31$\pm$0.02  & $<$8.3 \\
E(B--V)& 0 & 0.02 $\pm$ 0.05 &0 &0.35 $\pm$ 0.36 \\
SFR [10$^{-2}$M$_{\odot}$/y]& 56.0$\pm$1 & 32.14 $\pm$ 4.36	&7.04 $\pm$ 0.19	&2.39 $\pm$ 1.97\\
sSFR/L [M$_{\odot}$/y/L/L*]&9.91$\pm$0.22  & 18.87 $\pm$ 0.48	&5.69 $\pm$ 0.16	&--- \\
\hline \hline
\end{tabular}
\end{table}

\begin{landscape}
\begin{table}[!ht]
\caption{Emission line fluxes and derived properties from the integrated spectra shown in Fig. \ref{fig:host_spectra} extracted from the MUSE WFM and NFM datacubes. Fluxes are in units of $10^{-17}$ erg s$^{-1}$ cm$^{-2}$ and are corrected for Galactic and intrinsic extinction using the Balmer decrement. The g-band magnitude for each region was obtained from extracting the flux from the HSC g-band image (after resampling to the MUSE WFM spaxel size) in the same region as for the extraction of the spectra, calibrating it to AB magnitudes and applying the same Galactic and intrinsic extinction correction as for the emission lines. Since the giant H\,\textsc{ii} region is not resolved in the HSC images, we cannot do the same for the regions extracted from the NFM cube.} 
\label{tab:integratedspecs}        
\tiny
\begin{tabular}{l|ccc|cccccc}  
\hline\hline                
&\multicolumn{3}{c|}{\scriptsize{NFM}}     & \multicolumn{6}{c}{\scriptsize{WFM}}    \\
\scriptsize{Line $\lambda_\mathrm{rest}$}&\scriptsize{H\,\textsc{ii}-E}&\scriptsize{H\,\textsc{ii}-W}&\scriptsize{SN }&\scriptsize{H\,\textsc{ii} int }&\scriptsize{SE 1      }&\scriptsize{ SE 2    }&\scriptsize{ SE 3 }&\scriptsize{ SE 4 }&\scriptsize{Global } \\
\hline\hline 
[Ar{\sc iv}] 4740	&4.06$\pm$0.62	&$<$1.78 &$<$4.37 &10.74$\pm$1.59&$<$2.39&$<$3.93&$<$1.59&$<$1.57&$<$33.80\\
H$\beta$   4861 & 305.21$\pm$13.10	&40.57$\pm$3.13&	24.86$\pm$5.39&	735.54$\pm$5.81	&25.25$\pm$0.72&	14.65$\pm$3.82	&7.98$\pm$0.58	&6.58$\pm$0.53	&1081.63$\pm$13.80\\

[Fe{\sc iii}] 4986 &$<$3.13 &3.06$\pm$0.44 &$<$4.37&3.81$\pm$0.95	&$<$1.71&$<$2.10 	&$<$1.22&$<$1.90 &$<$26.66\\

[O~{\sc iii}] 5007 & 2373.04$\pm$98.27	&299.99$\pm$22.17	&185.84$\pm$38.29	&5368.78$\pm$32.82&	83.28$\pm$  0.66&38.97$\pm$9.59	&20.11$\pm$0.53	&14.38$\pm$0.65&	7052.27$\pm$40.98 \\

He{\sc ii}  4847 & 2.97$\pm$0.62 &$<$2.73&	$<$3.25 &3.01$\pm$1.40	&$<$2.41&$<$2.92 &$<$2.41&$<$1.06&$<$44.16 \\

He{\sc i}   5876 & 31.30$\pm$1.17&4.37$\pm$0.43	&$<$4.03 &	74.21$\pm$0.84&3.04$\pm$0.51 &$<$2.35&	1.21 $\pm$ 0.34&$<$1.40&114.36$\pm$  5.57 \\

[O{\sc i}] 6300	&3.30$\pm$0.53	&$<$1.27 &	$<$2.58  &12.40$\pm$2.90	&1.85$\pm$0.48	&1.89$\pm$0.47	&$<$1.03&$<$0.85&41.55$\pm$8.05\\

[S{\sc iii}] 6312	&4.07$\pm$0.56	&$<$1.36 &	$<$2.57 &	9.65$\pm$3.22& $<$1.54 &$<$1.05 &	$<$0.63  &$<$0.84  &16.25$\pm$5.15\\

H$\alpha$  6563 & 834.89$\pm$24.73	&111.22$\pm$5.86	&68.33$\pm$9.98	&1804.38$\pm$13.65	&68.72$\pm$0.52	&40.31$\pm$6.99	&21.56$\pm$0.25	&15.25$\pm$0.31	&2703.44$\pm$ 13.34 \\

[N{\sc ii}]   6583 & 7.05  0.26& $<$1.04 & 1.26$\pm$0.56&25.50$\pm$0.57&	4.38$\pm$ 0.47	&29.76$\pm$5.14	&1.52  $\pm$0.23	&1.05$\pm$0.30	&69.23$\pm$3.40\\
He{\sc i} 6680	&	8.97$\pm$0.40	&1.70$\pm$0.38&	7.18$\pm$1.19	&18.31$\pm$0.62	&0.78$\pm$0.31&	$<$1.15 &$<$0.73 &$<$0.62&24.83$\pm$ 6.83\\

[S{\sc ii}]   6716 & 13.79$\pm$0.48	&1.77$\pm$0.22	&$<$3.28&	153.62$\pm$6.87&	8.81$\pm$0.48	&6.70$\pm$1.18	&2.89$\pm$0.39	&2.26$\pm$0.32	&152.55$\pm$6.10\\

[S{\sc ii}]  6731 & 11.76$\pm$0.42	&1.07$\pm$0.13	&$<$3.78 &114.54$\pm$3.10	&6.18$\pm$0.39	&3.94$\pm$0.73	&2.04$\pm$0.23	&1.37$\pm$0.25	&114.88$\pm$3.71\\

He{\sc i} 7065	&	10.11$\pm$0.40	&1.03$\pm$0.16	&$<$3.66 &17.72$\pm$1.04	&$<$1.49 &$<$1.21 &$<$1.43 &$<$1.49 &	24.57  $\pm$7.15\\

[Ar{\sc iii}] 7136	&	17.63$\pm$0.45	&1.93$\pm$0.17	&$<$2.69&42.30$\pm$0.87	&1.49$\pm$0.43&	0.82$\pm$0.31	&$<$0.59&$<$1.04&66.26$\pm$5.92\\

\hline
size [kpc$^2$] &0.018 & 0.014 & 0.005  &1.98 &0.74 &0.63 &0.52& 0.66 &---\\
$g$ [mag] &--- & --- & ---  &19.313 &21.400 &21.091 &22.380&23.34&17.748\\
M$_g$ [mag] &--- & --- & ---  &-16.545  &-14.458 &-14.767&	-13.478&-12.514&-18.110 \\
12+log(O/H) & 7.90$\pm$0.01 & 7.91$\pm$0.01 & 7.97$\pm$0.06 & 7.95$\pm$0.01 & 8.17$\pm$0.01 & 8.41$\pm$0.01 & 8.20$\pm$0.02  & 8.21$\pm$0.03 &8.02$\pm$0.01 \\
E(B--V)& 0.311$\pm$0.013	&0.235$\pm$0.023	&0.153$\pm$0.064	&0.000  &0.000	&0.111$\pm$0.076	&0.000 &0.000 &0.000\\
SFR \tiny{[10$^{-2}$M$_{\odot}$/y]}& 17.994$\pm$0.533	&2.397$\pm$0.126	&1.473$\pm$0.215&38.889$\pm$0.294	&1.481 $\pm$0.011	&0.869$\pm$0.151	&0.465$\pm$0.005	&0.329$\pm$0.007	&58.265$\pm$0.287\\
sSFR/L \tiny{[M$_{\odot}$/y/L/L*]}& --- & --- & --- &23.532$\pm$0.178	&6.126$\pm$0.046&	2.705$\pm$0.469	&4.740$\pm$0.055	&8.147$\pm$0.168	&8.343 $\pm$0.041\\


\end{tabular}
\end{table}
\end{landscape}
\end{appendices}

\bibliography{sn-bibliography}

@ARTICLE{SDSS_DR16,
       author = {{Ahumada}, Romina and {Allende Prieto}, Carlos and {Almeida}, Andr{\'e}s and {Anders}, Friedrich and {Anderson}, Scott F. and {Andrews}, Brett H. and {Anguiano}, Borja and {Arcodia}, Riccardo and {Armengaud}, Eric and {Aubert}, Marie and {Avila}, Santiago and {Avila-Reese}, Vladimir and {Badenes}, Carles and {Balland}, Christophe and {Barger}, Kat and {Barrera-Ballesteros}, Jorge K. and {Basu}, Sarbani and {Bautista}, Julian and {Beaton}, Rachael L. and {Beers}, Timothy C. and {Benavides}, B. Izamar T. and {Bender}, Chad F. and {Bernardi}, Mariangela and {Bershady}, Matthew and {Beutler}, Florian and {Bidin}, Christian Moni and {Bird}, Jonathan and {Bizyaev}, Dmitry and {Blanc}, Guillermo A. and {Blanton}, Michael R. and {Boquien}, M{\'e}d{\'e}ric and {Borissova}, Jura and {Bovy}, Jo and {Brandt}, W.~N. and {Brinkmann}, Jonathan and {Brownstein}, Joel R. and {Bundy}, Kevin and {Bureau}, Martin and {Burgasser}, Adam and {Burtin}, Etienne and {Cano-D{\'\i}az}, Mariana and {Capasso}, Raffaella and {Cappellari}, Michele and {Carrera}, Ricardo and {Chabanier}, Sol{\`e}ne and {Chaplin}, William and {Chapman}, Michael and {Cherinka}, Brian and {Chiappini}, Cristina and {Doohyun Choi}, Peter and {Chojnowski}, S. Drew and {Chung}, Haeun and {Clerc}, Nicolas and {Coffey}, Damien and {Comerford}, Julia M. and {Comparat}, Johan and {da Costa}, Luiz and {Cousinou}, Marie-Claude and {Covey}, Kevin and {Crane}, Jeffrey D. and {Cunha}, Katia and {Ilha}, Gabriele da Silva and {Dai}, Yu Sophia and {Damsted}, Sanna B. and {Darling}, Jeremy and {Davidson}, Jr., James W. and {Davies}, Roger and {Dawson}, Kyle and {De}, Nikhil and {de la Macorra}, Axel and {De Lee}, Nathan and {Queiroz}, Anna B{\'a}rbara de Andrade and {Deconto Machado}, Alice and {de la Torre}, Sylvain and {Dell'Agli}, Flavia and {du Mas des Bourboux}, H{\'e}lion and {Diamond-Stanic}, Aleksandar M. and {Dillon}, Sean and {Donor}, John and {Drory}, Niv and {Duckworth}, Chris and {Dwelly}, Tom and {Ebelke}, Garrett and {Eftekharzadeh}, Sarah and {Davis Eigenbrot}, Arthur and {Elsworth}, Yvonne P. and {Eracleous}, Mike and {Erfanianfar}, Ghazaleh and {Escoffier}, Stephanie and {Fan}, Xiaohui and {Farr}, Emily and {Fern{\'a}ndez-Trincado}, Jos{\'e} G. and {Feuillet}, Diane and {Finoguenov}, Alexis and {Fofie}, Patricia and {Fraser-McKelvie}, Amelia and {Frinchaboy}, Peter M. and {Fromenteau}, Sebastien and {Fu}, Hai and {Galbany}, Llu{\'\i}s and {Garcia}, Rafael A. and {Garc{\'\i}a-Hern{\'a}ndez}, D.~A. and {Garma Oehmichen}, Luis Alberto and {Ge}, Junqiang and {Geimba Maia}, Marcio Antonio and {Geisler}, Doug and {Gelfand}, Joseph and {Goddy}, Julian and {Gonzalez-Perez}, Violeta and {Grabowski}, Kathleen and {Green}, Paul and {Grier}, Catherine J. and {Guo}, Hong and {Guy}, Julien and {Harding}, Paul and {Hasselquist}, Sten and {Hawken}, Adam James and {Hayes}, Christian R. and {Hearty}, Fred and {Hekker}, S. and {Hogg}, David W. and {Holtzman}, Jon A. and {Horta}, Danny and {Hou}, Jiamin and {Hsieh}, Bau-Ching and {Huber}, Daniel and {Hunt}, Jason A.~S. and {Ider Chitham}, J. and {Imig}, Julie and {Jaber}, Mariana and {Jimenez Angel}, Camilo Eduardo and {Johnson}, Jennifer A. and {Jones}, Amy M. and {J{\"o}nsson}, Henrik and {Jullo}, Eric and {Kim}, Yerim and {Kinemuchi}, Karen and {Kirkpatrick}, IV, Charles C. and {Kite}, George W. and {Klaene}, Mark and {Kneib}, Jean-Paul and {Kollmeier}, Juna A. and {Kong}, Hui and {Kounkel}, Marina and {Krishnarao}, Dhanesh and {Lacerna}, Ivan and {Lan}, Ting-Wen and {Lane}, Richard R. and {Law}, David R. and {Le Goff}, Jean-Marc and {Leung}, Henry W. and {Lewis}, Hannah and {Li}, Cheng and {Lian}, Jianhui and {Lin}, Lihwai and {Long}, Dan and {Longa-Pe{\~n}a}, Pen{\'e}lope and {Lundgren}, Britt and {Lyke}, Brad W. and {Mackereth}, J. Ted and {MacLeod}, Chelsea L. and {Majewski}, Steven R. and {Manchado}, Arturo and {Maraston}, Claudia and {Martini}, Paul and {Masseron}, Thomas and {Masters}, Karen L. and {Mathur}, Savita and {McDermid}, Richard M. and {Merloni}, Andrea and {Merrifield}, Michael and {M{\'e}sz{\'a}ros}, Szabolcs and {Miglio}, Andrea and {Minniti}, Dante and {Minsley}, Rebecca and {Miyaji}, Takamitsu and {Mohammad}, Faizan Gohar and {Mosser}, Benoit and {Mueller}, Eva-Maria and {Muna}, Demitri and {Mu{\~n}oz-Guti{\'e}rrez}, Andrea and {Myers}, Adam D. and {Nadathur}, Seshadri and {Nair}, Preethi and {Nandra}, Kirpal and {Correa do Nascimento}, Janaina and {Nevin}, Rebecca Jean and {Newman}, Jeffrey A. and {Nidever}, David L. and {Nitschelm}, Christian and {Noterdaeme}, Pasquier and {O'Connell}, Julia E. and {Olmstead}, Matthew D. and {Oravetz}, Daniel and {Oravetz}, Audrey and {Osorio}, Yeisson and {Pace}, Zachary J. and {Padilla}, Nelson and {Palanque-Delabrouille}, Nathalie and {Palicio}, Pedro A.},
        title = "{The 16th Data Release of the Sloan Digital Sky Surveys: First Release from the APOGEE-2 Southern Survey and Full Release of eBOSS Spectra}",
      journal = {\apjs},
         year = 2020,
        month = jul,
       volume = {249},
       number = {1},
          eid = {3},
        pages = {3},
          doi = {10.3847/1538-4365/ab929e},
archivePrefix = {arXiv},
       eprint = {1912.02905},
 primaryClass = {astro-ph.GA},
       adsurl = {https://ui.adsabs.harvard.edu/abs/2020ApJS..249....3A}
}

@ARTICLE{gcn44084,
       author = {{Ahumada}, T. and {Hall}, X.~J. and {Perley}, D.~A. and {Zwicky Transient Facility}},
        title = "{EP260321a: ZTF and Rubin detections of the candidate optical counterpart to EP260321a}",
      journal = {GRB Coordinates Network},
         year = 2026,
        month = mar,
       volume = {44084},
        pages = {1},
       adsurl = {https://ui.adsabs.harvard.edu/abs/2026GCN.44084....1A}
}

@ARTICLE{GALEX,
       author = {{Bianchi}, Luciana and {Shiao}, Bernie and {Thilker}, David},
        title = "{Revised Catalog of GALEX Ultraviolet Sources. I. The All-Sky Survey: GUVcat\_AIS}",
      journal = {\apjs},
         year = 2017,
        month = jun,
       volume = {230},
       number = {2},
          eid = {24},
        pages = {24},
          doi = {10.3847/1538-4365/aa7053},
archivePrefix = {arXiv},
       eprint = {1704.05903},
 primaryClass = {astro-ph.GA},
       adsurl = {https://ui.adsabs.harvard.edu/abs/2017ApJS..230...24B}
}

@ARTICLE{LegacySurvey,
       author = {{Dey}, Arjun and {Schlegel}, David J. and {Lang}, Dustin and {Blum}, Robert and {Burleigh}, Kaylan and {Fan}, Xiaohui and {Findlay}, Joseph R. and {Finkbeiner}, Doug and {Herrera}, David and {Juneau}, St{\'e}phanie and {Landriau}, Martin and {Levi}, Michael and {McGreer}, Ian and {Meisner}, Aaron and {Myers}, Adam D. and {Moustakas}, John and {Nugent}, Peter and {Patej}, Anna and {Schlafly}, Edward F. and {Walker}, Alistair R. and {Valdes}, Francisco and {Weaver}, Benjamin A. and {Y{\`e}che}, Christophe and {Zou}, Hu and {Zhou}, Xu and {Abareshi}, Behzad and {Abbott}, T.~M.~C. and {Abolfathi}, Bela and {Aguilera}, C. and {Alam}, Shadab and {Allen}, Lori and {Alvarez}, A. and {Annis}, James and {Ansarinejad}, Behzad and {Aubert}, Marie and {Beechert}, Jacqueline and {Bell}, Eric F. and {BenZvi}, Segev Y. and {Beutler}, Florian and {Bielby}, Richard M. and {Bolton}, Adam S. and {Brice{\~n}o}, C{\'e}sar and {Buckley-Geer}, Elizabeth J. and {Butler}, Karen and {Calamida}, Annalisa and {Carlberg}, Raymond G. and {Carter}, Paul and {Casas}, Ricard and {Castander}, Francisco J. and {Choi}, Yumi and {Comparat}, Johan and {Cukanovaite}, Elena and {Delubac}, Timoth{\'e}e and {DeVries}, Kaitlin and {Dey}, Sharmila and {Dhungana}, Govinda and {Dickinson}, Mark and {Ding}, Zhejie and {Donaldson}, John B. and {Duan}, Yutong and {Duckworth}, Christopher J. and {Eftekharzadeh}, Sarah and {Eisenstein}, Daniel J. and {Etourneau}, Thomas and {Fagrelius}, Parker A. and {Farihi}, Jay and {Fitzpatrick}, Mike and {Font-Ribera}, Andreu and {Fulmer}, Leah and {G{\"a}nsicke}, Boris T. and {Gaztanaga}, Enrique and {George}, Koshy and {Gerdes}, David W. and {Gontcho}, Satya Gontcho A. and {Gorgoni}, Claudio and {Green}, Gregory and {Guy}, Julien and {Harmer}, Diane and {Hernandez}, M. and {Honscheid}, Klaus and {Huang}, Lijuan Wendy and {James}, David J. and {Jannuzi}, Buell T. and {Jiang}, Linhua and {Joyce}, Richard and {Karcher}, Armin and {Karkar}, Sonia and {Kehoe}, Robert and {Kneib}, Jean-Paul and {Kueter-Young}, Andrea and {Lan}, Ting-Wen and {Lauer}, Tod R. and {Le Guillou}, Laurent and {Le Van Suu}, Auguste and {Lee}, Jae Hyeon and {Lesser}, Michael and {Perreault Levasseur}, Laurence and {Li}, Ting S. and {Mann}, Justin L. and {Marshall}, Robert and {Mart{\'\i}nez-V{\'a}zquez}, C.~E. and {Martini}, Paul and {du Mas des Bourboux}, H{\'e}lion and {McManus}, Sean and {Meier}, Tobias Gabriel and {M{\'e}nard}, Brice and {Metcalfe}, Nigel and {Mu{\~n}oz-Guti{\'e}rrez}, Andrea and {Najita}, Joan and {Napier}, Kevin and {Narayan}, Gautham and {Newman}, Jeffrey A. and {Nie}, Jundan and {Nord}, Brian and {Norman}, Dara J. and {Olsen}, Knut A.~G. and {Paat}, Anthony and {Palanque-Delabrouille}, Nathalie and {Peng}, Xiyan and {Poppett}, Claire L. and {Poremba}, Megan R. and {Prakash}, Abhishek and {Rabinowitz}, David and {Raichoor}, Anand and {Rezaie}, Mehdi and {Robertson}, A.~N. and {Roe}, Natalie A. and {Ross}, Ashley J. and {Ross}, Nicholas P. and {Rudnick}, Gregory and {Safonova}, Sasha and {Saha}, Abhijit and {S{\'a}nchez}, F. Javier and {Savary}, Elodie and {Schweiker}, Heidi and {Scott}, Adam and {Seo}, Hee-Jong and {Shan}, Huanyuan and {Silva}, David R. and {Slepian}, Zachary and {Soto}, Christian and {Sprayberry}, David and {Staten}, Ryan and {Stillman}, Coley M. and {Stupak}, Robert J. and {Summers}, David L. and {Sien Tie}, Suk and {Tirado}, H. and {Vargas-Maga{\~n}a}, Mariana and {Vivas}, A. Katherina and {Wechsler}, Risa H. and {Williams}, Doug and {Yang}, Jinyi and {Yang}, Qian and {Yapici}, Tolga and {Zaritsky}, Dennis and {Zenteno}, A. and {Zhang}, Kai and {Zhang}, Tianmeng and {Zhou}, Rongpu and {Zhou}, Zhimin},
        title = "{Overview of the DESI Legacy Imaging Surveys}",
      journal = {\aj},
         year = 2019,
        month = may,
       volume = {157},
       number = {5},
          eid = {168},
        pages = {168},
          doi = {10.3847/1538-3881/ab089d},
archivePrefix = {arXiv},
       eprint = {1804.08657},
 primaryClass = {astro-ph.IM},
       adsurl = {https://ui.adsabs.harvard.edu/abs/2019AJ....157..168D}
}

@misc{Cutri2013,
  author = {{Cutri}, R. M. and others},
  title = {Explanatory Supplement to the AllWISE Data Release Products},
  year = {2013},
  publisher = {IPAC/Caltech},
  note = {\url{https://wise2.ipac.caltech.edu/docs/release/allwise/expsup/}}
}

@ARTICLE{arnett82,
       author = {{Arnett}, W.~D.},
        title = "{Type I supernovae. I - Analytic solutions for the early part of the light curve}",
      journal = {\apj},
         year = 1982,
        month = feb,
       volume = {253},
        pages = {785-797},
          doi = {10.1086/159681},
       adsurl = {https://ui.adsabs.harvard.edu/abs/1982ApJ...253..785A}
}

@misc{mpdaf,
       author = {{Bacon}, Roland and {Piqueras}, Laure and {Conseil}, Simon and {Richard}, Johan and {Shepherd}, Martin},
        title = "{MPDAF: MUSE Python Data Analysis Framework}",
 howpublished = {Astrophysics Source Code Library, record ascl:1611.003},
         year = 2016,
        month = nov,
          eid = {ascl:1611.003},
archivePrefix = {ascl},
       eprint = {1611.003},
       adsurl = {https://ui.adsabs.harvard.edu/abs/2016ascl.soft11003B}
}

@INPROCEEDINGS{Basa2022,
       author = {{Basa}, St{\'e}phane and {Lee}, William H. and {Dolon}, Fran{\c{c}}ois and {Watson}, Alan M. and {Floriot}, Johan and {Atteia}, Jean-Luc and {Butler}, Nathaniel R. and {Dornic}, Damien and {Lombardo}, Simona and {Ronayette}, Samuel and {Ageron}, Michel and {Agneray}, Fran{\c{c}}ois and {{\'A}ngeles}, Fernando and {Bautista}, Ludovik and {Benamar-Aissa}, Hafid and {Blanpain}, Cyril and {Boulade}, Olivier and {Boy}, J{\'e}r{\'e}mie and {Buat}, Veronique and {Cadena}, Edgar and {Cuevas}, Salvador and {Farah}, Alejandro and {Figueroa}, Liliana and {Fuentes}, Jorge and {Ga{\"\i}ti}, Carole and {Gallais}, Pascal and {Kajfasz}, Eric and {Langarica}, Rosal{\'\i}a. and {Langlois}, Arthur and {Larrieu}, Marie and {Le Van Suu}, Auguste and {Lecubin}, Julien and {L{\'o}pez {\'A}ngeles}, Eduardo and {Lugo}, Erica and {Malgoyre}, Adrien and {Mathon}, Romain and {Moreau}, Chrystel and {Nouvel-De-La-Fl{\`e}che}, Alix and {Ochoa}, Jos{\'e} Luis and {Pedrayes-L{\'o}pez}, Maria and {Ramon}, Pascale and {Ru{\'\i}z-D{\'\i}az-Soto}, Jaime and {Tinoco}, Silvio and {Valentin}, Herv{\'e}},
        title = "{COLIBRI, a wide-field 1.3 m robotic telescope dedicated to the transient sky}",
    booktitle = {Ground-based and Airborne Telescopes IX},
         year = 2022,
       editor = {{Marshall}, Heather K. and {Spyromilio}, Jason and {Usuda}, Tomonori},
       series = {Society of Photo-Optical Instrumentation Engineers (SPIE) Conference Series},
       volume = {12182},
        month = aug,
          eid = {121821S},
        pages = {121821S},
          doi = {10.1117/12.2627139},
       adsurl = {https://ui.adsabs.harvard.edu/abs/2022SPIE12182E..1SB}
}

@ARTICLE{Belkin24,
       author = {{Belkin}, S. and {Gompertz}, B.~P. and {Kumar}, A. and {Ackley}, K. and {Galloway}, D.~K. and {Jim{\'e}nez-Ibarra}, F. and {Killestein}, T.~L. and {O'Neill}, D. and {Wiersema}, K. and {Malesani}, D.~B. and {Levan}, A.~J. and {Lyman}, J. and {Dyer}, M.~J. and {Ulaczyk}, K. and {Steeghs}, D. and {Dhillon}, V.~S. and {O'Brien}, P. and {Ramsay}, G. and {Noysena}, K. and {Kotak}, R. and {Breton}, R.~P. and {Nuttall}, L.~K. and {Pall{\'e}}, E. and {Pollacco}, D. and {Awiphan}, S. and {Burhanudin}, U. and {Chote}, P. and {Chrimes}, A. and {Daw}, E. and {Duffy}, C. and {Eyles-Ferris}, R. and {Godson}, B. and {Heikkil{\"a}}, T. and {Irawati}, P. and {Kelsey}, L. and {Kennedy}, M.~R. and {Littlefair}, S. and {Makrygianni}, L. and {Marsh}, T. and {Mata S{\'a}nchez}, D. and {Mattila}, S. and {Maund}, J. and {McCormac}, J. and {Mkrtichian}, D. and {Mullaney}, J. and {Patel}, M. and {Rana}, J. and {Rol}, E. and {Sawangwit}, U. and {Stanway}, E. and {Starling}, R. and {Str{\o}m}, P.~A. and {Warwick}, B.},
        title = "{GRB 230911A: The First Discovery of a Fermi GRB Optical Counterpart with the Gravitational-wave Optical Transient Observer (GOTO)}",
      journal = {Research Notes of the American Astronomical Society},
         year = 2024,
        month = jan,
       volume = {8},
       number = {1},
          eid = {6},
        pages = {6},
          doi = {10.3847/2515-5172/ad1876},
       adsurl = {https://ui.adsabs.harvard.edu/abs/2024RNAAS...8....6B}
}

@ARTICLE{Boquien2019,
       author = {{Boquien}, M. and {Burgarella}, D. and {Roehlly}, Y. and {Buat}, V. and {Ciesla}, L. and {Corre}, D. and {Inoue}, A.~K. and {Salas}, H.},
        title = "{CIGALE: a python Code Investigating GALaxy Emission}",
      journal = {\aap},
         year = 2019,
        month = feb,
       volume = {622},
          eid = {A103},
        pages = {A103},
          doi = {10.1051/0004-6361/201834156},
archivePrefix = {arXiv},
       eprint = {1811.03094},
 primaryClass = {astro-ph.GA},
       adsurl = {https://ui.adsabs.harvard.edu/abs/2019A&A...622A.103B}
}

@ARTICLE{Campana2006,
       author = {{Campana}, S. and {Mangano}, V. and {Blustin}, A.~J. and {Brown}, P. and {Burrows}, D.~N. and {Chincarini}, G. and {Cummings}, J.~R. and {Cusumano}, G. and {Della Valle}, M. and {Malesani}, D. and {M{\'e}sz{\'a}ros}, P. and {Nousek}, J.~A. and {Page}, M. and {Sakamoto}, T. and {Waxman}, E. and {Zhang}, B. and {Dai}, Z.~G. and {Gehrels}, N. and {Immler}, S. and {Marshall}, F.~E. and {Mason}, K.~O. and {Moretti}, A. and {O'Brien}, P.~T. and {Osborne}, J.~P. and {Page}, K.~L. and {Romano}, P. and {Roming}, P.~W.~A. and {Tagliaferri}, G. and {Cominsky}, L.~R. and {Giommi}, P. and {Godet}, O. and {Kennea}, J.~A. and {Krimm}, H. and {Angelini}, L. and {Barthelmy}, S.~D. and {Boyd}, P.~T. and {Palmer}, D.~M. and {Wells}, A.~A. and {White}, N.~E.},
        title = "{The association of GRB 060218 with a supernova and the evolution of the shock wave}",
      journal = {\nat},
         year = 2006,
        month = aug,
       volume = {442},
       number = {7106},
        pages = {1008-1010},
          doi = {10.1038/nature04892},
archivePrefix = {arXiv},
       eprint = {astro-ph/0603279},
 primaryClass = {astro-ph},
       adsurl = {https://ui.adsabs.harvard.edu/abs/2006Natur.442.1008C}
}

@ARTICLE{Cano17b,
       author = {{Cano}, Z. and {Izzo}, L. and {de Ugarte Postigo}, A. and {Th{\"o}ne}, C.~C. and {Kr{\"u}hler}, T. and {Heintz}, K.~E. and {Malesani}, D. and {Geier}, S. and {Fuentes}, C. and {Chen}, T. -W. and {Covino}, S. and {D'Elia}, V. and {Fynbo}, J.~P.~U. and {Goldoni}, P. and {Gomboc}, A. and {Hjorth}, J. and {Jakobsson}, P. and {Kann}, D.~A. and {Milvang-Jensen}, B. and {Pugliese}, G. and {S{\'a}nchez-Ram{\'\i}rez}, R. and {Schulze}, S. and {Sollerman}, J. and {Tanvir}, N.~R. and {Wiersema}, K.},
        title = "{GRB 161219B/SN 2016jca: A low-redshift gamma-ray burst supernova powered by radioactive heating}",
      journal = {\aap},
         year = 2017,
        month = sep,
       volume = {605},
          eid = {A107},
        pages = {A107},
          doi = {10.1051/0004-6361/201731005},
archivePrefix = {arXiv},
       eprint = {1704.05401},
 primaryClass = {astro-ph.HE},
       adsurl = {https://ui.adsabs.harvard.edu/abs/2017A&A...605A.107C}
}

@ARTICLE{cano17,
       author = {{Cano}, Zach and {Wang}, Shan-Qin and {Dai}, Zi-Gao and {Wu}, Xue-Feng},
        title = "{The Observer's Guide to the Gamma-Ray Burst Supernova Connection}",
      journal = {Advances in Astronomy},
         year = 2017,
        month = jan,
       volume = {2017},
          eid = {8929054},
        pages = {8929054},
          doi = {10.1155/2017/8929054},
archivePrefix = {arXiv},
       eprint = {1604.03549},
 primaryClass = {astro-ph.HE},
       adsurl = {https://ui.adsabs.harvard.edu/abs/2017AdAst2017E...5C}
}

@ARTICLE{astronote18,
       author = {{Chambers}, K.~C. and {Boer}, T.~D. and {Fairlamb}, J. and {Huber}, M. and {Lin}, C.~C. and {Lowe}, T. and {Magnier}, E. and {Minguez}, P. and {Paek}, G. and {Schultz}, A. and {Smith}, I.~A. and {Wainscoat}, R.~J. and {Smartt}, S.~J. and {Smith}, K.~W. and {Srivastav}, S. and {Young}, D.~R. and {Fulton}, M. and {Nicholl}, M. and {Moore}, T. and {Weston}, J. and {Angus}, C.~R.},
        title = "{Pan-STARRS Transient Discovery Report for 2026-03-23}",
      journal = {Transient Name Server Discovery Report},
         year = 2026,
        month = mar,
       volume = {2026-1247},
        pages = {1},
       adsurl = {https://ui.adsabs.harvard.edu/abs/2026TNSTR1247....1C}
}

@ARTICLE{chatzopoulos13,
       author = {{Chatzopoulos}, E. and {Wheeler}, J. Craig and {Vinko}, J. and {Horvath}, Z.~L. and {Nagy}, A.},
        title = "{Analytical Light Curve Models of Superluminous Supernovae: {\ensuremath{\chi}}$^{2}$-minimization of Parameter Fits}",
      journal = {\apj},
         year = 2013,
        month = aug,
       volume = {773},
       number = {1},
          eid = {76},
        pages = {76},
          doi = {10.1088/0004-637X/773/1/76},
archivePrefix = {arXiv},
       eprint = {1306.3447},
 primaryClass = {astro-ph.HE},
       adsurl = {https://ui.adsabs.harvard.edu/abs/2013ApJ...773...76C}
}

@ARTICLE{janet26,
       author = {{Chen}, T-W. and {et al}.},
        title = "{Precursor activity and early circumstellar interaction in an Ic-BL supernova}",
      journal = {submitted},
         year = 2026,
        month = jun,
          eid = {},
        pages = {},
          doi = {},
archivePrefix = {},
       eprint = {},
 primaryClass = {},
       adsurl = {}
}

@ARTICLE{Christensen08,
   author = {{Christensen}, L. and {Vreeswijk}, P.~M. and {Sollerman}, J. and 
	{Th{\"o}ne}, C.~C. and {Le Floc'h}, E. and {Wiersema}, K.},
    title = "{IFU observations of the GRB 980425/SN 1998bw host galaxy: emission line ratios in GRB regions}",
  journal = {\aap},
archivePrefix = "arXiv",
   eprint = {0807.3554},
     year = 2008,
    month = oct,
   volume = 490,
    pages = {45-59},
      doi = {10.1051/0004-6361:200809896},
   adsurl = {http://adsabs.harvard.edu/abs/2008A%26A...490...45C}
}

@ARTICLE{Corcoran2026GCN,
       author = {{Corcoran}, G. and {Martin-Carrillo}, A. and {Izzo}, L. and {de Ugarte Postigo}, A. and {Malesani}, D.~B. and {De Pasquale}, M. and {Tanvir}, N.~R. and {Eyles-Ferris}, R.~A.~J. and {Levan}, A.~J. and {Stargate Collaboration}},
        title = "{EP260321a: VLT/FORS2 spectroscopy confirmation of an associated type Ic-BL supernova SN 2026gzf}",
      journal = {GRB Coordinates Network},
         year = 2026,
        month = mar,
       volume = {44105},
        pages = {1},
       adsurl = {https://ui.adsabs.harvard.edu/abs/2026GCN.44105....1C}
}

@article{corsi23,
doi = {10.3847/1538-4357/acd3f2},
url = {https://doi.org/10.3847/1538-4357/acd3f2},
year = {2023},
month = {aug},
publisher = {The American Astronomical Society},
volume = {953},
number = {2},
pages = {179},
author = {Corsi, Alessandra and Ho, Anna Y. Q. and Cenko, S. Bradley and Kulkarni, Shrinivas R. and Anand, Shreya and Yang, Sheng and Sollerman, Jesper and Srinivasaragavan, Gokul P. and Omand, Conor M. B. and Balasubramanian, Arvind and Frail, Dale A. and Fremling, Christoffer and Perley, Daniel A. and Yao, Yuhan and Dahiwale, Aishwarya S. and De, Kishalay and Dugas, Alison and Hankins, Matthew and Jencson, Jacob and Kasliwal, Mansi M. and Tzanidakis, Anastasios and Bellm, Eric C. and Laher, Russ R. and Masci, Frank J. and Purdum, Josiah N. and Regnault, Nicolas},
title = {A Search for Relativistic Ejecta in a Sample of ZTF Broad-lined Type Ic Supernovae},
journal = {The Astrophysical Journal}
}

@ARTICLE{Corre2018,
       author = {{Corre}, D. and {Buat}, V. and {Basa}, S. and {Boissier}, S. and {Japelj}, J. and {Palmerio}, J. and {Salvaterra}, R. and {Vergani}, S.~D. and {Zafar}, T.},
        title = "{Investigation of dust attenuation and star formation activity in galaxies hosting GRBs}",
      journal = {\aap},
         year = 2018,
        month = sep,
       volume = {617},
          eid = {A141},
        pages = {A141},
          doi = {10.1051/0004-6361/201832926},
archivePrefix = {arXiv},
       eprint = {1807.00635},
 primaryClass = {astro-ph.GA},
       adsurl = {https://ui.adsabs.harvard.edu/abs/2018A&A...617A.141C}
}

@ARTICLE{Chabrier03,
   author = {{Chabrier}, G.},
    title = "{Galactic Stellar and Substellar Initial Mass Function}",
  journal = {\pasp},
     year = 2003,
   volume = 115,
    pages = {763-795},
      doi = {10.1086/376392}
}

@misc{cotter26,
      title={Probing a new subclass of llGRB-SN transients: Insights from EP250304a and its associated supernova}, 
      author={L. Cotter and A. Martin-Carrillo and R. A. J. Eyles-Ferris and L. Izzo and D. B. Malesani and Y. Julakanti and G. Corcoran and A. Saccardi and P. G. Jonker and A. J. Levan and F. Carotenuto and P. T. O'Brien and J. H. Gillanders and J. N. D. van Dalen and M. E. Ravasio and S. Schulze and N. Sarin and F. E. Bauer and M. Fraser and J. Quirola-Vasquez and A. P. C. van Hoof and S. J. Smartt and C. Gall and A. Rest and C. T. Murphey and N. Tanvir and T. -W. Chen and S. Campana and C. Ashall and J. P. Anderson and J. A. Chacon and F. J. Cowie and V. D'Elia and L. Galbany and C. P. Gutierrez and D. H. Hartmann and P. Jakobsson and S. Kobayashi and A. H. Kong and P. Mazalli and T. E. Muller-Bravo and M. De Pasquale and L. Rhodes and A. Rossi and J. Sanchez-Sierras and J. Sollerman and A. Andersson and A. Aryan and T. de Boer and J. S. Bright and K. C. Chambers and M. Gromadzki and M. E. Huber and C. Inserra and T. Lowe and P. Minguez and G. S. Narayan and M. Nicholl and G. S. H. Paek and A. Sedgewick and K. W. Smith and J. W. Tweddle and S. Yang},
      year={2026},
      eprint={2606.06213},
      archivePrefix={arXiv},
      primaryClass={astro-ph.HE},
      url={https://arxiv.org/abs/2606.06213}, 
}

@ARTICLE{decolle18,
       author = {{De Colle}, Fabio and {Kumar}, Pawan and {Aguilera-Dena}, David R.},
        title = "{Radio Emission from the Cocoon of a GRB Jet: Implications for Relativistic Supernovae and Off-axis GRB Emission}",
      journal = {\apj},
         year = 2018,
        month = aug,
       volume = {863},
       number = {1},
          eid = {32},
        pages = {32},
          doi = {10.3847/1538-4357/aad04d},
archivePrefix = {arXiv},
       eprint = {1803.00602},
 primaryClass = {astro-ph.HE},
       adsurl = {https://ui.adsabs.harvard.edu/abs/2018ApJ...863...32D}
}

@ARTICLE{DellaValle06,
       author = {{Della Valle}, M. and {Chincarini}, G. and {Panagia}, N. and {Tagliaferri}, G. and {Malesani}, D. and {Testa}, V. and {Fugazza}, D. and {Campana}, S. and {Covino}, S. and {Mangano}, V. and {Antonelli}, L.~A. and {D'Avanzo}, P. and {Hurley}, K. and {Mirabel}, I.~F. and {Pellizza}, L.~J. and {Piranomonte}, S. and {Stella}, L.},
        title = "{An enigmatic long-lasting {\ensuremath{\gamma}}-ray burst not accompanied by a bright supernova}",
      journal = {\nat},
         year = 2006,
        month = dec,
       volume = {444},
       number = {7122},
        pages = {1050-1052},
          doi = {10.1038/nature05374},
archivePrefix = {arXiv},
       eprint = {astro-ph/0608322},
 primaryClass = {astro-ph},
       adsurl = {https://ui.adsabs.harvard.edu/abs/2006Natur.444.1050D}
}

@ARTICLE{deUgarte20,
       author = {{de Ugarte Postigo}, A. and {Th{\"o}ne}, C.~C. and {Mart{\'\i}n}, S. and {Japelj}, J. and {Levan}, A.~J. and {Micha{\l}owski}, M.~J. and {Selsing}, J. and {Kann}, D.~A. and {Schulze}, S. and {Palmerio}, J.~T. and {Vergani}, S.~D. and {Tanvir}, N.~R. and {Bensch}, K. and {Covino}, S. and {D'Elia}, V. and {De Pasquale}, M. and {Fruchter}, A.~S. and {Fynbo}, J.~P.~U. and {Hartmann}, D. and {Heintz}, K.~E. and {van der Horst}, A.~J. and {Izzo}, L. and {Jakobsson}, P. and {Ng}, K.~C.~Y. and {Perley}, D.~A. and {Rossi}, A. and {Sbarufatti}, B. and {Salvaterra}, R. and {S{\'a}nchez-Ram{\'\i}rez}, R. and {Watson}, D. and {Xu}, D.},
        title = "{GRB 190114C in the nuclear region of an interacting galaxy. A detailed host analysis using ALMA, the HST, and the VLT}",
      journal = {\aap},
         year = 2020,
        month = jan,
       volume = {633},
          eid = {A68},
        pages = {A68},
          doi = {10.1051/0004-6361/201936668},
archivePrefix = {arXiv},
       eprint = {1911.07876},
 primaryClass = {astro-ph.HE},
       adsurl = {https://ui.adsabs.harvard.edu/abs/2020A&A...633A..68D}
}

@ARTICLE{deUgarte18,
       author = {{de Ugarte Postigo}, A. and {Th{\"o}ne}, C.~C. and {Bensch}, K. and {van der Horst}, A.~J. and {Kann}, D.~A. and {Cano}, Z. and {Izzo}, L. and {Goldoni}, P. and {Mart{\'\i}n}, S. and {Filgas}, R. and {Schady}, P. and {Gorosabel}, J. and {Bikmaev}, I. and {Bremer}, M. and {Burenin}, R. and {Castro-Tirado}, A.~J. and {Covino}, S. and {Fynbo}, J.~P.~U. and {Garcia-Appadoo}, D. and {de Gregorio-Monsalvo}, I. and {Jel{\'\i}nek}, M. and {Khamitov}, I. and {Kamble}, A. and {Kouveliotou}, C. and {Kr{\"u}hler}, T. and {Leloudas}, G. and {Melnikov}, S. and {Nardini}, M. and {Perley}, D.~A. and {Petitpas}, G. and {Pooley}, G. and {Rau}, A. and {Rol}, E. and {S{\'a}nchez-Ram{\'\i}rez}, R. and {Starling}, R.~L.~C. and {Tanvir}, N.~R. and {Wiersema}, K. and {Wijers}, R.~A.~M.~J. and {Zafar}, T.},
        title = "{The luminous host galaxy, faint supernova and rapid afterglow rebrightening of GRB 100418A}",
      journal = {\aap},
         year = 2018,
        month = dec,
       volume = {620},
          eid = {A190},
        pages = {A190},
          doi = {10.1051/0004-6361/201833636},
archivePrefix = {arXiv},
       eprint = {1807.04281},
 primaryClass = {astro-ph.HE},
       adsurl = {https://ui.adsabs.harvard.edu/abs/2018A&A...620A.190D}
}

@INPROCEEDINGS{Dyer24,
       author = {{Dyer}, Martin J. and {Ackley}, Kendall and {Jim{\'e}nez-Ibarra}, Felipe and {Lyman}, Joseph and {Ulaczyk}, Krzysztof and {Steeghs}, Danny and {Galloway}, Duncan K. and {Dhillon}, Vik S. and {O'Brien}, Paul and {Ramsay}, Gavin and {Noysena}, Kanthanakorn and {Kotak}, Rubina and {Breton}, Rene and {Nuttall}, Laura and {Pall{\'e}}, Enric and {Pollacco}, Don and {Killestein}, Tom and {Kumar}, Amit and {O'Neill}, David and {Kelsey}, Lisa and {Godson}, Ben and {Jarvis}, Dan},
        title = "{The Gravitational-wave Optical Transient Observer (GOTO)}",
    booktitle = {Ground-based and Airborne Telescopes X},
         year = 2024,
       editor = {{Marshall}, Heather K. and {Spyromilio}, Jason and {Usuda}, Tomonori},
       series = {Society of Photo-Optical Instrumentation Engineers (SPIE) Conference Series},
       volume = {13094},
        month = aug,
          eid = {130941X},
        pages = {130941X},
          doi = {10.1117/12.3018305},
archivePrefix = {arXiv},
       eprint = {2407.17176},
 primaryClass = {astro-ph.IM},
       adsurl = {https://ui.adsabs.harvard.edu/abs/2024SPIE13094E..1XD}
}

@ARTICLE{eylesferris25,
       author = {{Eyles-Ferris}, Rob A.~J. and {Jonker}, Peter G. and {Levan}, Andrew J. and {Malesani}, Daniele Bj{\o}rn and {Sarin}, Nikhil and {Fryer}, Christopher L. and {Rastinejad}, Jillian C. and {Burns}, Eric and {Tanvir}, Nial R. and {O'Brien}, Paul T. and {Fong}, Wen-fai and {Mandel}, Ilya and {Gompertz}, Benjamin P. and {Kilpatrick}, Charles D. and {Bloemen}, Steven and {Bright}, Joe S. and {Carotenuto}, Francesco and {Corcoran}, Gregory and {Cotter}, Laura and {Groot}, Paul J. and {Izzo}, Luca and {Laskar}, Tanmoy and {Martin-Carrillo}, Antonio and {Palmerio}, Jesse and {Ravasio}, Maria E. and {van Roestel}, Jan and {Saccardi}, Andrea and {Starling}, Rhaana L.~C. and {Thakur}, Aishwarya Linesh and {Vergani}, Susanna D. and {Vreeswijk}, Paul M. and {Bauer}, Franz E. and {Campana}, Sergio and {Chac{\'o}n}, Jennifer A. and {Chrimes}, Ashley A. and {Covino}, Stefano and {van Dalen}, Joyce N.~D. and {D'Elia}, Valerio and {De Pasquale}, Massimiliano and {Habeeb}, Nusrin and {Hartmann}, Dieter H. and {van Hoof}, Agnes P.~C. and {Jakobsson}, P{\'a}ll and {Julakanti}, Yashaswi and {Leloudas}, Giorgos and {Mata S{\'a}nchez}, Daniel and {Nixon}, Christopher J. and {Pieterse}, Dani{\"e}lle L.~A. and {Pugliese}, Giovanna and {Quirola-V{\'a}squez}, Jonathan and {Rayson}, Ben C. and {Salvaterra}, Ruben and {Schneider}, Ben and {Torres}, Manuel A.~P. and {Zafar}, Tayyaba},
        title = "{The Kangaroo's First Hop: The Early Fast Cooling Phase of EP250108a/SN 2025kg}",
      journal = {\apjl},
         year = 2025,
        month = jul,
       volume = {988},
       number = {1},
          eid = {L14},
        pages = {L14},
          doi = {10.3847/2041-8213/ade1d9},
archivePrefix = {arXiv},
       eprint = {2504.08886},
 primaryClass = {astro-ph.HE},
       adsurl = {https://ui.adsabs.harvard.edu/abs/2025ApJ...988L..14E}
}

@ARTICLE{finneran25a,
       author = {{Finneran}, Gabriel and {Cotter}, Laura and {Martin-Carrillo}, Antonio},
        title = "{The GRBSN webtool: An open-source repository for gamma-ray burst-supernova associations}",
      journal = {Astronomy and Computing},
         year = 2025,
        month = jul,
       volume = {52},
          eid = {100954},
        pages = {100954},
          doi = {10.1016/j.ascom.2025.100954},
archivePrefix = {arXiv},
       eprint = {2411.08866},
 primaryClass = {astro-ph.HE},
       adsurl = {https://ui.adsabs.harvard.edu/abs/2025A&C....5200954F}
}

@ARTICLE{finneran2025b,
       author = {{Finneran}, G. and {Cotter}, L. and {Martin-Carrillo}, A.},
        title = "{Velocity evolution of broad-lined type-Ic supernovae with and without gamma-ray bursts}",
      journal = {\aap},
         year = 2025,
        month = aug,
       volume = {700},
          eid = {A200},
        pages = {A200},
          doi = {10.1051/0004-6361/202453047},
archivePrefix = {arXiv},
       eprint = {2411.11503},
 primaryClass = {astro-ph.HE},
       adsurl = {https://ui.adsabs.harvard.edu/abs/2025A&A...700A.200F}
}

@ARTICLE{EDPS,
       author = {{Freudling}, W. and {Zampieri}, S. and {Coccato}, L. and {Podgorski}, S. and {Romaniello}, M. and {Modigliani}, A. and {Pritchard}, J.},
        title = "{Adaptive data reduction workflows for astronomy: The ESO Data Processing System (EDPS)}",
      journal = {\aap},
         year = 2024,
        month = jan,
       volume = {681},
          eid = {A93},
        pages = {A93},
          doi = {10.1051/0004-6361/202347651},
archivePrefix = {arXiv},
       eprint = {2311.03822},
 primaryClass = {astro-ph.IM},
       adsurl = {https://ui.adsabs.harvard.edu/abs/2024A&A...681A..93F}
}

@ARTICLE{fryer99,
       author = {{Fryer}, Chris L.},
        title = "{Mass Limits For Black Hole Formation}",
      journal = {\apj},
         year = 1999,
        month = sep,
       volume = {522},
       number = {1},
        pages = {413-418},
          doi = {10.1086/307647},
archivePrefix = {arXiv},
       eprint = {astro-ph/9902315},
 primaryClass = {astro-ph},
       adsurl = {https://ui.adsabs.harvard.edu/abs/1999ApJ...522..413F}
}

@ARTICLE{Galama1998,
       author = {{Galama}, T.~J. and {Vreeswijk}, P.~M. and {van Paradijs}, J. and {Kouveliotou}, C. and {Augusteijn}, T. and {B{\"o}hnhardt}, H. and {Brewer}, J.~P. and {Doublier}, V. and {Gonzalez}, J.-F. and {Leibundgut}, B. and {Lidman}, C. and {Hainaut}, O.~R. and {Patat}, F. and {Heise}, J. and {in't Zand}, J. and {Hurley}, K. and {Groot}, P.~J. and {Strom}, R.~G. and {Mazzali}, P.~A. and {Iwamoto}, K. and {Nomoto}, K. and {Umeda}, H. and {Nakamura}, T. and {Young}, T.~R. and {Suzuki}, T. and {Shigeyama}, T. and {Koshut}, T. and {Kippen}, M. and {Robinson}, C. and {de Wildt}, P. and {Wijers}, R.~A.~M.~J. and {Tanvir}, N. and {Greiner}, J. and {Pian}, E. and {Palazzi}, E. and {Frontera}, F. and {Masetti}, N. and {Nicastro}, L. and {Feroci}, M. and {Costa}, E. and {Piro}, L. and {Peterson}, B.~A. and {Tinney}, C. and {Boyle}, B. and {Cannon}, R. and {Stathakis}, R. and {Sadler}, E. and {Begam}, M.~C. and {Ianna}, P.},
        title = "{An unusual supernova in the error box of the {\ensuremath{\gamma}}-ray burst of 25 April 1998}",
      journal = {\nat},
         year = 1998,
        month = oct,
       volume = {395},
       number = {6703},
        pages = {670-672},
          doi = {10.1038/27150},
archivePrefix = {arXiv},
       eprint = {astro-ph/9806175},
 primaryClass = {astro-ph},
       adsurl = {https://ui.adsabs.harvard.edu/abs/1998Natur.395..670G}
}

@ARTICLE{grafener17,
       author = {{Gr{\"a}fener}, G. and {Owocki}, S.~P. and {Grassitelli}, L. and {Langer}, N.},
        title = "{On the optically thick winds of Wolf-Rayet stars}",
      journal = {\aap},
         year = 2017,
        month = dec,
       volume = {608},
          eid = {A34},
        pages = {A34},
          doi = {10.1051/0004-6361/201731590},
archivePrefix = {arXiv},
       eprint = {1710.04543},
 primaryClass = {astro-ph.SR},
       adsurl = {https://ui.adsabs.harvard.edu/abs/2017A&A...608A..34G}
}

@ARTICLE{Gompertz20,
       author = {{Gompertz}, B.~P. and {Cutter}, R. and {Steeghs}, D. and {Galloway}, D.~K. and {Lyman}, J. and {Ulaczyk}, K. and {Dyer}, M.~J. and {Ackley}, K. and {Dhillon}, V.~S. and {O'Brien}, P.~T. and {Ramsay}, G. and {Poshyachinda}, S. and {Kotak}, R. and {Nuttall}, L. and {Breton}, R.~P. and {Pall{\'e}}, E. and {Pollacco}, D. and {Thrane}, E. and {Aukkaravittayapun}, S. and {Awiphan}, S. and {Brown}, M.~J.~I. and {Burhanudin}, U. and {Chote}, P. and {Chrimes}, A.~A. and {Daw}, E. and {Duffy}, C. and {Eyles-Ferris}, R.~A.~J. and {Heikkil{\"a}}, T. and {Irawati}, P. and {Kennedy}, M.~R. and {Killestein}, T. and {Levan}, A.~J. and {Littlefair}, S. and {Makrygianni}, L. and {Marsh}, T. and {Mata S{\'a}nchez}, D. and {Mattila}, S. and {Maund}, J. and {McCormac}, J. and {Mkrtichian}, D. and {Mong}, Y.-L. and {Mullaney}, J. and {M{\"u}ller}, B. and {Obradovic}, A. and {Rol}, E. and {Sawangwit}, U. and {Stanway}, E.~R. and {Starling}, R.~L.~C. and {Str{\o}m}, P.~A. and {Tooke}, S. and {West}, R. and {Wiersema}, K.},
        title = "{Searching for electromagnetic counterparts to gravitational-wave merger events with the prototype Gravitational-Wave Optical Transient Observer (GOTO-4)}",
      journal = {\mnras},
         year = 2020,
        month = sep,
       volume = {497},
       number = {1},
        pages = {726-738},
          doi = {10.1093/mnras/staa1845},
archivePrefix = {arXiv},
       eprint = {2004.00025},
 primaryClass = {astro-ph.HE},
       adsurl = {https://ui.adsabs.harvard.edu/abs/2020MNRAS.497..726G}
}

@ARTICLE{Hamidani25a,
       author = {{Hamidani}, Hamid and {Sato}, Yuri and {Kashiyama}, Kazumi and {Tanaka}, Masaomi and {Ioka}, Kunihito and {Kimura}, Shigeo S.},
        title = "{EP240414a: A Gamma-Ray Burst Jet Weakened by an Extended Circumstellar Material}",
      journal = {\apjl},
         year = 2025,
        month = jun,
       volume = {986},
       number = {1},
          eid = {L4},
        pages = {L4},
          doi = {10.3847/2041-8213/add99d},
archivePrefix = {arXiv},
       eprint = {2503.16243},
 primaryClass = {astro-ph.HE},
       adsurl = {https://ui.adsabs.harvard.edu/abs/2025ApJ...986L...4H}
}

@ARTICLE{Hamidani25,
       author = {{Hamidani}, Hamid and {Ioka}, Kunihito and {Kashiyama}, Kazumi and {Tanaka}, Masaomi},
        title = "{Gamma-Ray Burst Jets in Circumstellar Material: Dynamics, Breakout, and Diversity of Transients}",
      journal = {\apj},
         year = 2025,
        month = jul,
       volume = {988},
       number = {1},
          eid = {30},
        pages = {30},
          doi = {10.3847/1538-4357/addd13},
archivePrefix = {arXiv},
       eprint = {2503.16242},
 primaryClass = {astro-ph.HE},
       adsurl = {https://ui.adsabs.harvard.edu/abs/2025ApJ...988...30H}
}

@ARTICLE{Han10,
   author = {{Han}, X.~H. and {Hammer}, F. and {Liang}, Y.~C. and {Flores}, H. and 
	{Rodrigues}, M. and {Hou}, J.~L. and {Wei}, J.~Y.},
    title = "{The Wolf-Rayet features and mass-metallicity relation of long-duration gamma-ray burst host galaxies}",
  journal = {\aap},
archivePrefix = "arXiv",
   eprint = {1001.2476},
     year = 2010,
    month = may,
   volume = 514,
      eid = {A24},
    pages = {A24},
      doi = {10.1051/0004-6361/200912475},
   adsurl = {http://adsabs.harvard.edu/abs/2010A%26A...514A..24H}
}

@ARTICLE{Heintz18,
       author = {{Heintz}, K.~E. and {Malesani}, D. and {Wiersema}, K. and {Jakobsson}, P. and {Fynbo}, J.~P.~U. and {Savaglio}, S. and {Cano}, Z. and {Covino}, S. and {D'Elia}, V. and {Gomboc}, A. and {Hammer}, F. and {Kaper}, L. and {Milvang-Jensen}, B. and {M{\o}ller}, P. and {Piranomonte}, S. and {Selsing}, J. and {Rhodin}, N.~H.~P. and {Tanvir}, N.~R. and {Th{\"o}ne}, C.~C. and {de Ugarte Postigo}, A. and {Vergani}, S.~D. and {Watson}, D.},
        title = "{The luminous, massive and solar metallicity galaxy hosting the Swift {\ensuremath{\gamma}}-ray burst GRB 160804A at z = 0.737}",
      journal = {\mnras},
         year = 2018,
        month = feb,
       volume = {474},
       number = {2},
        pages = {2738-2749},
          doi = {10.1093/mnras/stx2895},
archivePrefix = {arXiv},
       eprint = {1711.02706},
 primaryClass = {astro-ph.GA},
       adsurl = {https://ui.adsabs.harvard.edu/abs/2018MNRAS.474.2738H}
}

@ARTICLE{Hjorth12,
       author = {{Hjorth}, Jens and {Malesani}, Daniele and {Jakobsson}, P{\'a}ll and {Jaunsen}, Andreas O. and {Fynbo}, Johan P.~U. and {Gorosabel}, Javier and {Kr{\"u}hler}, Thomas and {Levan}, Andrew J. and {Micha{\l}owski}, Micha{\l} J. and {Milvang-Jensen}, Bo and {M{\o}ller}, Palle and {Schulze}, Steve and {Tanvir}, Nial R. and {Watson}, Darach},
        title = "{The Optically Unbiased Gamma-Ray Burst Host (TOUGH) Survey. I. Survey Design and Catalogs}",
      journal = {\apj},
         year = 2012,
        month = sep,
       volume = {756},
       number = {2},
          eid = {187},
        pages = {187},
          doi = {10.1088/0004-637X/756/2/187},
archivePrefix = {arXiv},
       eprint = {1205.3162},
 primaryClass = {astro-ph.CO},
       adsurl = {https://ui.adsabs.harvard.edu/abs/2012ApJ...756..187H}
}

@ARTICLE{ho19,
       author = {{Ho}, Anna Y.~Q. and {Goldstein}, Daniel A. and {Schulze}, Steve and {Khatami}, David K. and {Perley}, Daniel A. and {Ergon}, Mattias and {Gal-Yam}, Avishay and {Corsi}, Alessandra and {Andreoni}, Igor and {Barbarino}, Cristina and {Bellm}, Eric C. and {Blagorodnova}, Nadia and {Bright}, Joe S. and {Burns}, E. and {Cenko}, S. Bradley and {Cunningham}, Virginia and {De}, Kishalay and {Dekany}, Richard and {Dugas}, Alison and {Fender}, Rob P. and {Fransson}, Claes and {Fremling}, Christoffer and {Goldstein}, Adam and {Graham}, Matthew J. and {Hale}, David and {Horesh}, Assaf and {Hung}, Tiara and {Kasliwal}, Mansi M. and {Kuin}, N. Paul M. and {Kulkarni}, S.~R. and {Kupfer}, Thomas and {Lunnan}, Ragnhild and {Masci}, Frank J. and {Ngeow}, Chow-Choong and {Nugent}, Peter E. and {Ofek}, Eran O. and {Patterson}, Maria T. and {Petitpas}, Glen and {Rusholme}, Ben and {Sai}, Hanna and {Sfaradi}, Itai and {Shupe}, David L. and {Sollerman}, Jesper and {Soumagnac}, Maayane T. and {Tachibana}, Yutaro and {Taddia}, Francesco and {Walters}, Richard and {Wang}, Xiaofeng and {Yao}, Yuhan and {Zhang}, Xinhan},
        title = "{Evidence for Late-stage Eruptive Mass Loss in the Progenitor to SN2018gep, a Broad-lined Ic Supernova: Pre-explosion Emission and a Rapidly Rising Luminous Transient}",
      journal = {\apj},
         year = 2019,
        month = dec,
       volume = {887},
       number = {2},
          eid = {169},
        pages = {169},
          doi = {10.3847/1538-4357/ab55ec},
archivePrefix = {arXiv},
       eprint = {1904.11009},
 primaryClass = {astro-ph.HE},
       adsurl = {https://ui.adsabs.harvard.edu/abs/2019ApJ...887..169H}
}

@ARTICLE{HSC_SSP,
       author = {{Aihara}, Hiroaki and {Arimoto}, Nobuo and {Armstrong}, Robert and {Arnouts}, St{\'e}phane and {Bahcall}, Neta A. and {Bickerton}, Steven and {Bosch}, James and {Bundy}, Kevin and {Capak}, Peter L. and {Chan}, James H.~H. and {Chiba}, Masashi and {Coupon}, Jean and {Egami}, Eiichi and {Enoki}, Motohiro and {Finet}, Francois and {Fujimori}, Hiroki and {Fujimoto}, Seiji and {Furusawa}, Hisanori and {Furusawa}, Junko and {Goto}, Tomotsugu and {Goulding}, Andy and {Greco}, Johnny P. and {Greene}, Jenny E. and {Gunn}, James E. and {Hamana}, Takashi and {Harikane}, Yuichi and {Hashimoto}, Yasuhiro and {Hattori}, Takashi and {Hayashi}, Masao and {Hayashi}, Yusuke and {He{\l}miniak}, Krzysztof G. and {Higuchi}, Ryo and {Hikage}, Chiaki and {Ho}, Paul T.~P. and {Hsieh}, Bau-Ching and {Huang}, Kuiyun and {Huang}, Song and {Ikeda}, Hiroyuki and {Imanishi}, Masatoshi and {Inoue}, Akio K. and {Iwasawa}, Kazushi and {Iwata}, Ikuru and {Jaelani}, Anton T. and {Jian}, Hung-Yu and {Kamata}, Yukiko and {Karoji}, Hiroshi and {Kashikawa}, Nobunari and {Katayama}, Nobuhiko and {Kawanomoto}, Satoshi and {Kayo}, Issha and {Koda}, Jin and {Koike}, Michitaro and {Kojima}, Takashi and {Komiyama}, Yutaka and {Konno}, Akira and {Koshida}, Shintaro and {Koyama}, Yusei and {Kusakabe}, Haruka and {Leauthaud}, Alexie and {Lee}, Chien-Hsiu and {Lin}, Lihwai and {Lin}, Yen-Ting and {Lupton}, Robert H. and {Mandelbaum}, Rachel and {Matsuoka}, Yoshiki and {Medezinski}, Elinor and {Mineo}, Sogo and {Miyama}, Shoken and {Miyatake}, Hironao and {Miyazaki}, Satoshi and {Momose}, Rieko and {More}, Anupreeta and {More}, Surhud and {Moritani}, Yuki and {Moriya}, Takashi J. and {Morokuma}, Tomoki and {Mukae}, Shiro and {Murata}, Ryoma and {Murayama}, Hitoshi and {Nagao}, Tohru and {Nakata}, Fumiaki and {Niida}, Mana and {Niikura}, Hiroko and {Nishizawa}, Atsushi J. and {Obuchi}, Yoshiyuki and {Oguri}, Masamune and {Oishi}, Yukie and {Okabe}, Nobuhiro and {Okamoto}, Sakurako and {Okura}, Yuki and {Ono}, Yoshiaki and {Onodera}, Masato and {Onoue}, Masafusa and {Osato}, Ken and {Ouchi}, Masami and {Price}, Paul A. and {Pyo}, Tae-Soo and {Sako}, Masao and {Sawicki}, Marcin and {Shibuya}, Takatoshi and {Shimasaku}, Kazuhiro and {Shimono}, Atsushi and {Shirasaki}, Masato and {Silverman}, John D. and {Simet}, Melanie and {Speagle}, Joshua and {Spergel}, David N. and {Strauss}, Michael A. and {Sugahara}, Yuma and {Sugiyama}, Naoshi and {Suto}, Yasushi and {Suyu}, Sherry H. and {Suzuki}, Nao and {Tait}, Philip J. and {Takada}, Masahiro and {Takata}, Tadafumi and {Tamura}, Naoyuki and {Tanaka}, Manobu M. and {Tanaka}, Masaomi and {Tanaka}, Masayuki and {Tanaka}, Yoko and {Terai}, Tsuyoshi and {Terashima}, Yuichi and {Toba}, Yoshiki and {Tominaga}, Nozomu and {Toshikawa}, Jun and {Turner}, Edwin L. and {Uchida}, Tomohisa and {Uchiyama}, Hisakazu and {Umetsu}, Keiichi and {Uraguchi}, Fumihiro and {Urata}, Yuji and {Usuda}, Tomonori and {Utsumi}, Yousuke and {Wang}, Shiang-Yu and {Wang}, Wei-Hao and {Wong}, Kenneth C. and {Yabe}, Kiyoto and {Yamada}, Yoshihiko and {Yamanoi}, Hitomi and {Yasuda}, Naoki and {Yeh}, Sherry and {Yonehara}, Atsunori and {Yuma}, Suraphong},
        title = "{The Hyper Suprime-Cam SSP Survey: Overview and survey design}",
      journal = {\pasj},
         year = 2018,
        month = jan,
       volume = {70},
          eid = {S4},
        pages = {S4},
          doi = {10.1093/pasj/psx066},
archivePrefix = {arXiv},
       eprint = {1704.05858},
 primaryClass = {astro-ph.IM},
       adsurl = {https://ui.adsabs.harvard.edu/abs/2018PASJ...70S...4A}
}

@ARTICLE{huang02,
       author = {{Huang}, Y.~F. and {Dai}, Z.~G. and {Lu}, T.},
        title = "{Failed gamma-ray bursts and orphan afterglows}",
      journal = {\mnras},
         year = 2002,
        month = may,
       volume = {332},
       number = {3},
        pages = {735-740},
          doi = {10.1046/j.1365-8711.2002.05334.x},
archivePrefix = {arXiv},
       eprint = {astro-ph/0112469},
 primaryClass = {astro-ph},
       adsurl = {https://ui.adsabs.harvard.edu/abs/2002MNRAS.332..735H}
}

@ARTICLE{gcn44068,
       author = {{Huang}, Q.~J. and {Zou}, Z.~C. and {Mao}, X. and {Li}, D.~Y. and {Pan}, H.~W. and {Einstein Probe Team}},
        title = "{EP260321a: Einstein Probe detection of an X-ray transient}",
      journal = {GRB Coordinates Network},
         year = 2026,
        month = mar,
       volume = {44068},
        pages = {1},
       adsurl = {https://ui.adsabs.harvard.edu/abs/2026GCN.44068....1H}
}

@ARTICLE{gcn44075,
       author = {{Huang}, Q.~J. and {Zou}, Z.-C. and {Li}, D.~Y. and {Mao}, X. and {Pan}, H.~W. and {Einstein Probe Team}},
        title = "{EP260321a: refined analysis of the EP-WXT and EP-FXT observations, implying a possible supernova shock breakout candidate}",
      journal = {GRB Coordinates Network},
         year = 2026,
        month = mar,
       volume = {44075},
        pages = {1},
       adsurl = {https://ui.adsabs.harvard.edu/abs/2026GCN.44075....1H}
}

@INPROCEEDINGS{fors_recipe,
       author = {{Izzo}, C. and {de Bilbao}, L. and {Larsen}, J. and {Bagnulo}, S. and {Freudling}, W. and {Moehler}, S. and {Ballester}, P.},
        title = "{The new FORS pipeline}",
    booktitle = {Observatory Operations: Strategies, Processes, and Systems III},
         year = 2010,
       editor = {{Silva}, David R. and {Peck}, Alison B. and {Soifer}, B. Thomas},
       series = {Society of Photo-Optical Instrumentation Engineers (SPIE) Conference Series},
       volume = {7737},
        month = jul,
          eid = {773729},
        pages = {773729},
          doi = {10.1117/12.858043},
       adsurl = {https://ui.adsabs.harvard.edu/abs/2010SPIE.7737E..29I}
}

@ARTICLE{Izzo17,
       author = {{Izzo}, L. and {Th{\"o}ne}, C.~C. and {Schulze}, S. and {Mehner}, A. and {Flores}, H. and {Cano}, Z. and {de Ugarte Postigo}, A. and {Kann}, D.~A. and {Amor{\'\i}n}, R. and {Anderson}, J.~P. and {Bauer}, F.~E. and {Bensch}, K. and {Christensen}, L. and {Covino}, S. and {Della Valle}, M. and {Fynbo}, J.~P.~U. and {Jakobsson}, P. and {Klose}, S. and {Kuncarayakti}, H. and {Leloudas}, G. and {Milvang-Jensen}, B. and {M{\o}ller}, P. and {Puech}, M. and {Rossi}, A. and {S{\'a}nchez-Ram{\'\i}rez}, R. and {Vergani}, S.~D.},
        title = "{The MUSE view of the host galaxy of GRB 100316D}",
      journal = {\mnras},
         year = 2017,
        month = dec,
       volume = {472},
       number = {4},
        pages = {4480-4496},
          doi = {10.1093/mnras/stx2244},
archivePrefix = {arXiv},
       eprint = {1704.05509},
 primaryClass = {astro-ph.GA},
       adsurl = {https://ui.adsabs.harvard.edu/abs/2017MNRAS.472.4480I}
}

@ARTICLE{Izzo2019,
       author = {{Izzo}, L. and {de Ugarte Postigo}, A. and {Maeda}, K. and {Th{\"o}ne}, C.~C. and {Kann}, D.~A. and {Della Valle}, M. and {Sagues Carracedo}, A. and {Micha{\l}owski}, M.~J. and {Schady}, P. and {Schmidl}, S. and {Selsing}, J. and {Starling}, R.~L.~C. and {Suzuki}, A. and {Bensch}, K. and {Bolmer}, J. and {Campana}, S. and {Cano}, Z. and {Covino}, S. and {Fynbo}, J.~P.~U. and {Hartmann}, D.~H. and {Heintz}, K.~E. and {Hjorth}, J. and {Japelj}, J. and {Kami{\'n}ski}, K. and {Kaper}, L. and {Kouveliotou}, C. and {Kru{\.Z}y{\'n}ski}, M. and {Kwiatkowski}, T. and {Leloudas}, G. and {Levan}, A.~J. and {Malesani}, D.~B. and {Micha{\l}owski}, T. and {Piranomonte}, S. and {Pugliese}, G. and {Rossi}, A. and {S{\'a}nchez-Ram{\'\i}rez}, R. and {Schulze}, S. and {Steeghs}, D. and {Tanvir}, N.~R. and {Ulaczyk}, K. and {Vergani}, S.~D. and {Wiersema}, K.},
        title = "{Signatures of a jet cocoon in early spectra of a supernova associated with a {\ensuremath{\gamma}}-ray burst}",
      journal = {\nat},
         year = 2019,
        month = jan,
       volume = {565},
       number = {7739},
        pages = {324-327},
          doi = {10.1038/s41586-018-0826-3},
archivePrefix = {arXiv},
       eprint = {1901.05500},
 primaryClass = {astro-ph.HE},
       adsurl = {https://ui.adsabs.harvard.edu/abs/2019Natur.565..324I}
}

@ARTICLE{Izzo2020,
       author = {{Izzo}, L. and {Auchettl}, K. and {Hjorth}, J. and {De Colle}, F. and {Gall}, C. and {Angus}, C.~R. and {Raimundo}, S.~I. and {Ramirez-Ruiz}, E.},
        title = "{Broad-line type Ic SN 2020bvc. Signatures of an off-axis gamma-ray burst afterglow}",
      journal = {\aap},
         year = 2020,
        month = jul,
       volume = {639},
          eid = {L11},
        pages = {L11},
          doi = {10.1051/0004-6361/202038152},
archivePrefix = {arXiv},
       eprint = {2004.05941},
 primaryClass = {astro-ph.HE},
       adsurl = {https://ui.adsabs.harvard.edu/abs/2020A&A...639L..11I}
}

@ARTICLE{jiang20,
       author = {{Jiang}, Brighten and {Jiang}, Shuai and {Ashley Villar}, V.},
        title = "{Extended Self-similar Solution for Circumstellar Material-supernova Ejecta Interaction}",
      journal = {Research Notes of the American Astronomical Society},
         year = 2020,
        month = jan,
       volume = {4},
       number = {1},
          eid = {16},
        pages = {16},
          doi = {10.3847/2515-5172/ab7128},
archivePrefix = {arXiv},
       eprint = {2008.10397},
 primaryClass = {astro-ph.HE},
       adsurl = {https://ui.adsabs.harvard.edu/abs/2020RNAAS...4...16J}
}

@ARTICLE{kangas2021,
       author = {{Kangas}, Tuomas and {Fruchter}, Andrew S.},
        title = "{The Late-time Radio Behavior of Gamma-ray Burst Afterglows: Testing the Standard Model}",
      journal = {\apj},
         year = 2021,
        month = apr,
       volume = {911},
       number = {1},
          eid = {14},
        pages = {14},
          doi = {10.3847/1538-4357/abe76b},
archivePrefix = {arXiv},
       eprint = {1911.01938},
 primaryClass = {astro-ph.HE},
       adsurl = {https://ui.adsabs.harvard.edu/abs/2021ApJ...911...14K}
}

@ARTICLE{Kelly13,
       author = {{Kelly}, Patrick L. and {Filippenko}, Alexei V. and {Fox}, Ori D. and {Zheng}, Weikang and {Clubb}, Kelsey I.},
        title = "{Evidence that Gamma-Ray Burst 130702A Exploded in a Dwarf Satellite of a Massive Galaxy}",
      journal = {\apjl},
         year = 2013,
        month = sep,
       volume = {775},
       number = {1},
          eid = {L5},
        pages = {L5},
          doi = {10.1088/2041-8205/775/1/L5},
archivePrefix = {arXiv},
       eprint = {1307.5103},
 primaryClass = {astro-ph.CO},
       adsurl = {https://ui.adsabs.harvard.edu/abs/2013ApJ...775L...5K}
}

@ARTICLE{Kennicutt1992,
       author = {{Kennicutt}, Jr., Robert C.},
        title = "{The Integrated Spectra of Nearby Galaxies: General Properties and Emission-Line Spectra}",
      journal = {\apj},
         year = 1992,
        month = apr,
       volume = {388},
        pages = {310},
          doi = {10.1086/171154},
       adsurl = {https://ui.adsabs.harvard.edu/abs/1992ApJ...388..310K}
}

@ARTICLE{Kruehler11,
       author = {{Kr{\"u}hler}, T. and {Greiner}, J. and {Schady}, P. and {Savaglio}, S. and {Afonso}, P.~M.~J. and {Clemens}, C. and {Elliott}, J. and {Filgas}, R. and {Gruber}, D. and {Kann}, D.~A. and {Klose}, S. and {K{\"u}pc{\"u}-Yolda{\textcommabelow s}}, A. and {McBreen}, S. and {Olivares}, F. and {Pierini}, D. and {Rau}, A. and {Rossi}, A. and {Nardini}, M. and {Nicuesa Guelbenzu}, A. and {Sudilovsky}, V. and {Updike}, A.~C.},
        title = "{The SEDs and host galaxies of the dustiest GRB afterglows}",
      journal = {\aap},
         year = 2011,
        month = oct,
       volume = {534},
          eid = {A108},
        pages = {A108},
          doi = {10.1051/0004-6361/201117428},
archivePrefix = {arXiv},
       eprint = {1108.0674},
 primaryClass = {astro-ph.CO},
       adsurl = {https://ui.adsabs.harvard.edu/abs/2011A&A...534A.108K}
}

@ARTICLE{Kruhler15,
       author = {{Kr{\"u}hler}, T. and {Malesani}, D. and {Fynbo}, J.~P.~U. and
         {Hartoog}, O.~E. and {Hjorth}, J. and {Jakobsson}, P. and
         {Perley}, D.~A. and {Rossi}, A. and {Schady}, P. and {Schulze}, S. and
         {Tanvir}, N.~R. and {Vergani}, S.~D. and {Wiersema}, K. and
         {Afonso}, P.~M.~J. and {Bolmer}, J. and {Cano}, Z. and {Covino}, S. and
         {D'Elia}, V. and {de Ugarte Postigo}, A. and {Filgas}, R. and
         {Friis}, M. and {Graham}, J.~F. and {Greiner}, J. and {Goldoni}, P. and
         {Gomboc}, A. and {Hammer}, F. and {Japelj}, J. and {Kann}, D.~A. and
         {Kaper}, L. and {Klose}, S. and {Levan}, A.~J. and {Leloudas}, G. and
         {Milvang-Jensen}, B. and {Nicuesa Guelbenzu}, A. and {Palazzi}, E. and
         {Pian}, E. and {Piranomonte}, S. and {S{\'a}nchez-Ram{\'\i}rez}, R. and
         {Savaglio}, S. and {Selsing}, J. and {Tagliaferri}, G. and
         {Vreeswijk}, P.~M. and {Watson}, D.~J. and {Xu}, D.},
        title = "{GRB hosts through cosmic time. VLT/X-Shooter emission-line spectroscopy of 96 {\ensuremath{\gamma}}-ray-burst-selected galaxies at 0.1 <z < 3.6}",
      journal = {\aap},
         year = 2015,
        month = sep,
       volume = {581},
          eid = {A125},
        pages = {A125},
          doi = {10.1051/0004-6361/201425561},
archivePrefix = {arXiv},
       eprint = {1505.06743},
 primaryClass = {astro-ph.GA},
       adsurl = {https://ui.adsabs.harvard.edu/abs/2015A&A...581A.125K}
}

@ARTICLE{Kumar25,
       author = {{Kumar}, Amit and {Gompertz}, B.~P. and {Schneider}, B. and {Belkin}, S. and {Wortley}, M.~E. and {Saccardi}, A. and {O'Neill}, D. and {Ackley}, K. and {Rayson}, B. and {Postigo}, A. de Ugarte and {Gulati}, A. and {Steeghs}, D. and {Malesani}, D.~B. and {Maund}, J.~R. and {Dyer}, M.~J. and {Giarratana}, S. and {Serino}, M. and {Julakanti}, Y. and {Kumar}, B. and {Xu}, D. and {Eyles-Ferris}, R.~A.~J. and {Zhu}, Z.-P. and {Warwick}, B. and {Hu}, Y.-D. and {Allen}, I. and {Ramsay}, G. and {Starling}, R.~L.~C. and {Lyman}, J. and {Ulaczyk}, K. and {Godson}, B. and {Galloway}, D.~K. and {Dhillon}, V.~S. and {O'Brien}, P. and {Noysena}, K. and {Kotak}, R. and {Breton}, R.~P. and {Nuttall}, L.~K. and {Pollacco}, D. and {Casares}, J. and {Killestein}, T.~L. and {Kennedy}, M.~R. and {Habeeb}, N. and {Moran}, S. and {Wiersema}, K. and {Worssam}, I. and {Coppejans}, D.~L. and {Phillips}, C.~A. and {Martin-Carrillo}, A. and {Pankov}, N.~S. and {Ag{\"u}{\'\i} Fern{\'a}ndez}, J.~F. and {Aloy}, M.~A. and {An}, J. and {Anderson}, G.~E. and {Bochenek}, A. and {Castro-Tirado}, A.~J. and {Chen}, X. and {Cotter}, L. and {Dastidar}, R. and {De Pasquale}, M. and {D'Elia}, V. and {Fang}, Y. and {Fu}, S.~Y. and {Fynbo}, J.~P.~U. and {Hartmann}, D.~H. and {He}, L.~B. and {Izzo}, L. and {Jiang}, S.~Q. and {Kawakubo}, Y. and {Klunko}, E.~V. and {Levan}, A.~J. and {Liu}, X.-W. and {Liu}, X. and {Lombardi}, G. and {Maiorano}, E. and {Palmerio}, J.~T. and {Perley}, D.~A. and {Pieterse}, D.~L.~A. and {Pozanenko}, A.~S. and {Pugliese}, G. and {Rossi}, A. and {Sbarufatti}, B. and {Seshashayana}, S. Bijavara and {Tanvir}, N.~R. and {Th{\"o}ne}, C.~C. and {van der Horst}, A.~J. and {Vergani}, S.~D. and {Volnova}, A.~A. and {Wijers}, R.~A.~M.~J. and {Wise}, J.~L.},
        title = "{Discovery and analysis of afterglows from poorly localized GRBs with the Gravitational-wave Optical Transient Observer (GOTO) All-sky Survey}",
      journal = {\mnras},
         year = 2025,
        month = dec,
       volume = {544},
       number = {2},
        pages = {1541-1587},
          doi = {10.1093/mnras/staf1689},
archivePrefix = {arXiv},
       eprint = {2509.09827},
 primaryClass = {astro-ph.HE},
       adsurl = {https://ui.adsabs.harvard.edu/abs/2025MNRAS.544.1541K}
}

@INPROCEEDINGS{Langarica2024,
       author = {{Langarica}, Rosal{\'\i}a. and {Watson}, Alan M. and {{\'A}lvarez N{\'u}{\~n}ez}, Luis Carlos and {Angeles}, Fernando and {Atteia}, Jean-Luc and {Basa}, St{\'e}phane and {Cuevas}, Salvador and {Dolon}, Fran{\c{c}}ois and {Farah}, Alejandro and {Dornic}, Damien and {Floriot}, Johan and {Fuentes-Fern{\'a}ndez}, Jorge and {Langlois}, Arthur and {Lee}, William H. and {Lombardo}, Simona and {Pereyra}, Margarita and {Ru{\'\i}z D{\'\i}az-Soto}, Jaime and {Ronayette}, Samuel and {Tinoco}, Silvio and {Valent{\'\i}n}, Herv{\'e}},
        title = "{The DDRAGO wide-field imager for the COLIBR{\'I} telescope}",
    booktitle = {Ground-based and Airborne Instrumentation for Astronomy X},
         year = 2024,
       editor = {{Bryant}, Julia J. and {Motohara}, Kentaro and {Vernet}, Jo{\"e}l. R.~D.},
       series = {Society of Photo-Optical Instrumentation Engineers (SPIE) Conference Series},
       volume = {13096},
        month = jul,
          eid = {130963D},
        pages = {130963D},
          doi = {10.1117/12.3020545},
       adsurl = {https://ui.adsabs.harvard.edu/abs/2024SPIE13096E..3DL}
}

@ARTICLE{laskar15,
       author = {{Laskar}, Tanmoy and {Berger}, Edo and {Margutti}, Raffaella and {Perley}, Daniel and {Zauderer}, B. Ashley and {Sari}, Re'em and {Fong}, Wen-fai},
        title = "{Energy Injection in Gamma-Ray Burst Afterglows}",
      journal = {\apj},
         year = 2015,
        month = nov,
       volume = {814},
       number = {1},
          eid = {1},
        pages = {1},
          doi = {10.1088/0004-637X/814/1/1},
archivePrefix = {arXiv},
       eprint = {1504.03702},
 primaryClass = {astro-ph.HE},
       adsurl = {https://ui.adsabs.harvard.edu/abs/2015ApJ...814....1L}
}

@ARTICLE{gcn44070,
       author = {{Lee}, M.-H. and {Aryan}, A. and {Chen}, T.-W. and {Hou}, W.-J. and {Yang}, S. and {Smartt}, S.~J. and {Gillanders}, J. and {Kong}, A.~K.~H. and {Yang}, Y.~J. and {Lee}, Y.-H. and {Sankar. K}, A. and {Pan}, Y.-C. and {Ngeow}, C.-C. and {Lai}, C.-H. and {Lin}, C.-S. Lin H.-C. and {Hsiao}, H.-Y. and {Guo}, J.-K. and {Wang}, Z.~N. and {Qiang}, D.~C. and {Fan}, L.~L. and {Lin}, H.-W. and {Stevance}, H.~F. and {Srivastav}, S. and {Rhodes}, L. and {Nicholl}, M. and {Fulton}, M. and {Moore}, T. and {Smith}, K.~W. and {Angus}, C. and {Aamer}, A. and {Schultz}, A. and {Huber}, M.},
        title = "{EP260321a: Kinder observations detect a blue variable star and set limits on a source from the z =0.034 galaxy within the error circle}",
      journal = {GRB Coordinates Network},
         year = 2026,
        month = mar,
       volume = {44070},
        pages = {1},
       adsurl = {https://ui.adsabs.harvard.edu/abs/2026GCN.44070....1L}
}

@ARTICLE{Leloudas15,
       author = {{Leloudas}, G. and {Schulze}, S. and {Kr{\"u}hler}, T. and {Gorosabel}, J. and {Christensen}, L. and {Mehner}, A. and {de Ugarte Postigo}, A. and {Amor{\'\i}n}, R. and {Th{\"o}ne}, C.~C. and {Anderson}, J.~P. and {Bauer}, F.~E. and {Gallazzi}, A. and {He{\l}miniak}, K.~G. and {Hjorth}, J. and {Ibar}, E. and {Malesani}, D. and {Morell}, N. and {Vinko}, J. and {Wheeler}, J.~C.},
        title = "{Spectroscopy of superluminous supernova host galaxies. A preference of hydrogen-poor events for extreme emission line galaxies}",
      journal = {\mnras},
         year = 2015,
        month = may,
       volume = {449},
       number = {1},
        pages = {917-932},
          doi = {10.1093/mnras/stv320},
archivePrefix = {arXiv},
       eprint = {1409.8331},
 primaryClass = {astro-ph.GA},
       adsurl = {https://ui.adsabs.harvard.edu/abs/2015MNRAS.449..917L}
}

@ARTICLE{Lyman16,
       author = {{Lyman}, J.~D. and {Bersier}, D. and {James}, P.~A. and {Mazzali}, P.~A. and {Eldridge}, J.~J. and {Fraser}, M. and {Pian}, E.},
        title = "{Bolometric light curves and explosion parameters of 38 stripped-envelope core-collapse supernovae}",
      journal = {\mnras},
         year = 2016,
        month = mar,
       volume = {457},
       number = {1},
        pages = {328-350},
          doi = {10.1093/mnras/stv2983},
archivePrefix = {arXiv},
       eprint = {1406.3667},
 primaryClass = {astro-ph.SR},
       adsurl = {https://ui.adsabs.harvard.edu/abs/2016MNRAS.457..328L}
}

@ARTICLE{Lyman26,
       author = {{Lyman}, J.~D. and {O'Neill}, D. and {Killestein}, T. and {Jarvis}, D. and {Kumar}, A. and {Ulaczyk}, K. and {Ackley}, K. and {Chote}, P. and {Dyer}, M.~J. and {Pursiainen}, M. and {Steeghs}, D. and {Godson}, B. and {Magee}, M. and {Mullaney}, J.~R. and {Warwick}, B. and {Belkin}, S. and {Galloway}, D.~K. and {Ramsay}, G. and {Dhillon}, V.~S. and {O'Brien}, P. and {Noysena}, K. and {Kotak}, R. and {Breton}, R.~P. and {Nuttall}, L.~K. and {Gompertz}, B. and {Pollacco}, D. and {Casares}, J. and {Coppejans}, D.~L. and {Eyles-Ferris}, R.~A.~J. and {Graur}, O. and {Kelsey}, L. and {Kennedy}, M.~R. and {Levan}, A. and {Littlefair}, S. and {Mandhai}, S. and {Mata S{\'a}nchez}, D. and {Mattila}, S. and {McCormac}, J. and {Moran}, S. and {Phillips}, C. and {Pu}, K. and {Sahu}, A. and {Shrestha}, M. and {Stanway}, E. and {Starling}, R.~L.~C. and {Vincetti}, L. and {Wickens}, E. and {Wiersema}, K.},
        title = "{The Gravitational-wave Optical Transient Observer (GOTO) data pipeline and workflow for transient discovery}",
      journal = {arXiv e-prints},
         year = 2026,
        month = mar,
          eid = {arXiv:2603.02330},
        pages = {arXiv:2603.02330},
          doi = {10.48550/arXiv.2603.02330},
archivePrefix = {arXiv},
       eprint = {2603.02330},
 primaryClass = {astro-ph.IM},
       adsurl = {https://ui.adsabs.harvard.edu/abs/2026arXiv260302330L}
}

@ARTICLE{Marino13,
       author = {{Marino}, R.~A. and {Rosales-Ortega}, F.~F. and {S{\'a}nchez}, S.~F. and {Gil de Paz}, A. and {V{\'\i}lchez}, J. and {Miralles-Caballero}, D. and {Kehrig}, C. and {P{\'e}rez-Montero}, E. and {Stanishev}, V. and {Iglesias-P{\'a}ramo}, J. and {D{\'\i}az}, A.~I. and {Castillo-Morales}, A. and {Kennicutt}, R. and {L{\'o}pez-S{\'a}nchez}, A.~R. and {Galbany}, L. and {Garc{\'\i}a-Benito}, R. and {Mast}, D. and {Mendez-Abreu}, J. and {Monreal-Ibero}, A. and {Husemann}, B. and {Walcher}, C.~J. and {Garc{\'\i}a-Lorenzo}, B. and {Masegosa}, J. and {Del Olmo Orozco}, A. and {Mour{\~a}o}, A.~M. and {Ziegler}, B. and {Moll{\'a}}, M. and {Papaderos}, P. and {S{\'a}nchez-Bl{\'a}zquez}, P. and {Gonz{\'a}lez Delgado}, R.~M. and {Falc{\'o}n-Barroso}, J. and {Roth}, M.~M. and {van de Ven}, G. and {CALIFA Team}},
        title = "{The O3N2 and N2 abundance indicators revisited: improved calibrations based on CALIFA and T$_{e}$-based literature data}",
      journal = {\aap},
         year = 2013,
        month = nov,
       volume = {559},
          eid = {A114},
        pages = {A114},
          doi = {10.1051/0004-6361/201321956},
archivePrefix = {arXiv},
       eprint = {1307.5316},
 primaryClass = {astro-ph.CO},
       adsurl = {https://ui.adsabs.harvard.edu/abs/2013A&A...559A.114M}
}

@ARTICLE{Melandri19,
       author = {{Melandri}, A. and {Malesani}, D.~B. and {Izzo}, L. and {Japelj}, J. and {Vergani}, S.~D. and {Schady}, P. and {Sagu{\'e}s Carracedo}, A. and {de Ugarte Postigo}, A. and {Anderson}, J.~P. and {Barbarino}, C. and {Bolmer}, J. and {Breeveld}, A. and {Calissendorff}, P. and {Campana}, S. and {Cano}, Z. and {Carini}, R. and {Covino}, S. and {D'Avanzo}, P. and {D'Elia}, V. and {della Valle}, M. and {De Pasquale}, M. and {Fynbo}, J.~P.~U. and {Gromadzki}, M. and {Hammer}, F. and {Hartmann}, D.~H. and {Heintz}, K.~E. and {Inserra}, C. and {Jakobsson}, P. and {Kann}, D.~A. and {Kotilainen}, J. and {Maguire}, K. and {Masetti}, N. and {Nicholl}, M. and {Olivares E}, F. and {Pugliese}, G. and {Rossi}, A. and {Salvaterra}, R. and {Sollerman}, J. and {Stone}, M.~B. and {Tagliaferri}, G. and {Tomasella}, L. and {Th{\"o}ne}, C.~C. and {Xu}, D. and {Young}, D.~R.},
        title = "{GRB 171010A/SN 2017htp: a GRB-SN at z = 0.33}",
      journal = {\mnras},
         year = 2019,
        month = dec,
       volume = {490},
       number = {4},
        pages = {5366-5374},
          doi = {10.1093/mnras/stz2900},
archivePrefix = {arXiv},
       eprint = {1910.14160},
 primaryClass = {astro-ph.HE},
       adsurl = {https://ui.adsabs.harvard.edu/abs/2019MNRAS.490.5366M}
}

@ARTICLE{modjaz09,
       author = {{Modjaz}, M. and {Li}, W. and {Butler}, N. and {Chornock}, R. and {Perley}, D. and {Blondin}, S. and {Bloom}, J.~S. and {Filippenko}, A.~V. and {Kirshner}, R.~P. and {Kocevski}, D. and {Poznanski}, D. and {Hicken}, M. and {Foley}, R.~J. and {Stringfellow}, G.~S. and {Berlind}, P. and {Barrado y Navascues}, D. and {Blake}, C.~H. and {Bouy}, H. and {Brown}, W.~R. and {Challis}, P. and {Chen}, H. and {de Vries}, W.~H. and {Dufour}, P. and {Falco}, E. and {Friedman}, A. and {Ganeshalingam}, M. and {Garnavich}, P. and {Holden}, B. and {Illingworth}, G. and {Lee}, N. and {Liebert}, J. and {Marion}, G.~H. and {Olivier}, S.~S. and {Prochaska}, J.~X. and {Silverman}, J.~M. and {Smith}, N. and {Starr}, D. and {Steele}, T.~N. and {Stockton}, A. and {Williams}, G.~G. and {Wood-Vasey}, W.~M.},
        title = "{From Shock Breakout to Peak and Beyond: Extensive Panchromatic Observations of the Type Ib Supernova 2008D Associated with Swift X-ray Transient 080109}",
      journal = {\apj},
         year = 2009,
        month = sep,
       volume = {702},
       number = {1},
        pages = {226-248},
          doi = {10.1088/0004-637X/702/1/226},
archivePrefix = {arXiv},
       eprint = {0805.2201},
 primaryClass = {astro-ph},
       adsurl = {https://ui.adsabs.harvard.edu/abs/2009ApJ...702..226M}
}

@ARTICLE{modjaz16,
       author = {{Modjaz}, Maryam and {Liu}, Yuqian Q. and {Bianco}, Federica B. and {Graur}, Or},
        title = "{The Spectral SN-GRB Connection: Systematic Spectral Comparisons between Type Ic Supernovae and Broad-lined Type Ic Supernovae with and without Gamma-Ray Bursts}",
      journal = {\apj},
         year = 2016,
        month = dec,
       volume = {832},
       number = {2},
          eid = {108},
        pages = {108},
          doi = {10.3847/0004-637X/832/2/108},
archivePrefix = {arXiv},
       eprint = {1509.07124},
 primaryClass = {astro-ph.HE},
       adsurl = {https://ui.adsabs.harvard.edu/abs/2016ApJ...832..108M}
}

@ARTICLE{modjaz20,
       author = {{Modjaz}, Maryam and {Bianco}, Federica B. and {Siwek}, Magdalena and {Huang}, Shan and {Perley}, Daniel A. and {Fierroz}, David and {Liu}, Yu-Qian and {Arcavi}, Iair and {Gal-Yam}, Avishay and {Filippenko}, Alexei V. and {Blagorodnova}, Nadia and {Cenko}, Bradley S. and {Kasliwal}, Mansi and {Kulkarni}, Shri and {Schulze}, Steve and {Taggart}, Kirsty and {Zheng}, Weikang},
        title = "{Host Galaxies of Type Ic and Broad-lined Type Ic Supernovae from the Palomar Transient Factory: Implications for Jet Production}",
      journal = {\apj},
         year = 2020,
        month = apr,
       volume = {892},
       number = {2},
          eid = {153},
        pages = {153},
          doi = {10.3847/1538-4357/ab4185},
archivePrefix = {arXiv},
       eprint = {1901.00872},
 primaryClass = {astro-ph.HE},
       adsurl = {https://ui.adsabs.harvard.edu/abs/2020ApJ...892..153M}
}

@manual{nessai,
   author       = {Michael J. Williams},
   title        = {nessai: Nested Sampling with Artificial Intelligence},
   month        = feb,
   year         = 2021,
   publisher    = {Zenodo},
   version      = {latest},
   doi          = {10.5281/zenodo.4550693},
   url          = {https://doi.org/10.5281/zenodo.4550693}
 }

@ARTICLE{Palmerio19,
       author = {{Palmerio}, J.~T. and {Vergani}, S.~D. and {Salvaterra}, R. and {Sanders}, R.~L. and {Japelj}, J. and {Vidal-Garc{\'\i}a}, A. and {D'Avanzo}, P. and {Corre}, D. and {Perley}, D.~A. and {Shapley}, A.~E. and {Boissier}, S. and {Greiner}, J. and {Le Floc'h}, E. and {Wiseman}, P.},
        title = "{Are long gamma-ray bursts biased tracers of star formation? Clues from the host galaxies of the Swift/BAT6 complete sample of bright LGRBs. III. Stellar masses, star formation rates, and metallicities at z > 1}",
      journal = {\aap},
         year = 2019,
        month = mar,
       volume = {623},
          eid = {A26},
        pages = {A26},
          doi = {10.1051/0004-6361/201834179},
archivePrefix = {arXiv},
       eprint = {1901.02457},
 primaryClass = {astro-ph.HE},
       adsurl = {https://ui.adsabs.harvard.edu/abs/2019A&A...623A..26P}
}

@ARTICLE{Perley16,
       author = {{Perley}, D.~A. and {Quimby}, R.~M. and {Yan}, L. and {Vreeswijk}, P.~M. and {De Cia}, A. and {Lunnan}, R. and {Gal-Yam}, A. and {Yaron}, O. and {Filippenko}, A.~V. and {Graham}, M.~L. and {Laher}, R. and {Nugent}, P.~E.},
        title = "{Host-galaxy Properties of 32 Low-redshift Superluminous Supernovae from the Palomar Transient Factory}",
      journal = {\apj},
         year = 2016,
        month = oct,
       volume = {830},
       number = {1},
          eid = {13},
        pages = {13},
          doi = {10.3847/0004-637X/830/1/13},
archivePrefix = {arXiv},
       eprint = {1604.08207},
 primaryClass = {astro-ph.HE},
       adsurl = {https://ui.adsabs.harvard.edu/abs/2016ApJ...830...13P}
}

@ARTICLE{Pignata11,
       author = {{Pignata}, Giuliano and {Stritzinger}, Maximilian and {Soderberg}, Alicia and {Mazzali}, Paolo and {Phillips}, M.~M. and {Morrell}, Nidia and {Anderson}, J.~P. and {Boldt}, Luis and {Campillay}, Abdo and {Contreras}, Carlos and {Folatelli}, Gast{\'o}n and {F{\"o}rster}, Francisco and {Gonz{\'a}lez}, Sergio and {Hamuy}, Mario and {Krzeminski}, Wojtek and {Maza}, Jos{\'e} and {Roth}, Miguel and {Salgado}, Francisco and {Levesque}, Emily M. and {Rest}, Armin and {Crain}, J. Adam and {Foster}, Andrew C. and {Haislip}, Joshua B. and {Ivarsen}, Kevin M. and {LaCluyze}, Aaron P. and {Nysewander}, Melissa C. and {Reichart}, Daniel E.},
        title = "{SN 2009bb: A Peculiar Broad-lined Type Ic Supernova}",
      journal = {\apj},
         year = 2011,
        month = feb,
       volume = {728},
       number = {1},
          eid = {14},
        pages = {14},
          doi = {10.1088/0004-637X/728/1/14},
archivePrefix = {arXiv},
       eprint = {1011.6126},
 primaryClass = {astro-ph.SR},
       adsurl = {https://ui.adsabs.harvard.edu/abs/2011ApJ...728...14P}
}

@ARTICLE{piran19,
       author = {{Piran}, Tsvi and {Nakar}, Ehud and {Mazzali}, Paolo and {Pian}, Elena},
        title = "{Relativistic Jets in Core-collapse Supernovae}",
      journal = {\apjl},
         year = 2019,
        month = feb,
       volume = {871},
       number = {2},
          eid = {L25},
        pages = {L25},
          doi = {10.3847/2041-8213/aaffce},
       adsurl = {https://ui.adsabs.harvard.edu/abs/2019ApJ...871L..25P}
}

@article{piro13, 
  year     = {2013}, 
  title    = {{WHAT} {CAN} {WE} {LEARN} {FROM} {THE} {RISING} {LIGHT} {CURVES} {OF} {RADIOACTIVELY} {POWERED} {SUPERNOVAE}?}, 
  author   = {Piro, Anthony L. and Nakar, Ehud}, 
  journal  = {The Astrophysical Journal}, 
  issn     = {0004-637X}, 
  doi      = {10.1088/0004-637x/769/1/67}, 
  eprint   = {1210.3032}, 
  pages    = {67}, 
  number   = {1}, 
  volume   = {769}
}

@ARTICLE{Piro18,
       author = {{Piro}, Anthony L. and {Kollmeier}, Juna A.},
        title = "{Evidence for Cocoon Emission from the Early Light Curve of SSS17a}",
      journal = {\apj},
         year = 2018,
        month = mar,
       volume = {855},
       number = {2},
          eid = {103},
        pages = {103},
          doi = {10.3847/1538-4357/aaaab3},
archivePrefix = {arXiv},
       eprint = {1710.05822},
 primaryClass = {astro-ph.HE},
       adsurl = {https://ui.adsabs.harvard.edu/abs/2018ApJ...855..103P}
}

@ARTICLE{planck18,
       author = {{Planck Collaboration} and {Aghanim}, N. and {Akrami}, Y. and {Ashdown}, M. and {Aumont}, J. and {Baccigalupi}, C. and {Ballardini}, M. and {Banday}, A.~J. and {Barreiro}, R.~B. and {Bartolo}, N. and {Basak}, S. and {Battye}, R. and {Benabed}, K. and {Bernard}, J. -P. and {Bersanelli}, M. and {Bielewicz}, P. and {Bock}, J.~J. and {Bond}, J.~R. and {Borrill}, J. and {Bouchet}, F.~R. and {Boulanger}, F. and {Bucher}, M. and {Burigana}, C. and {Butler}, R.~C. and {Calabrese}, E. and {Cardoso}, J. -F. and {Carron}, J. and {Challinor}, A. and {Chiang}, H.~C. and {Chluba}, J. and {Colombo}, L.~P.~L. and {Combet}, C. and {Contreras}, D. and {Crill}, B.~P. and {Cuttaia}, F. and {de Bernardis}, P. and {de Zotti}, G. and {Delabrouille}, J. and {Delouis}, J. -M. and {Di Valentino}, E. and {Diego}, J.~M. and {Dor{\'e}}, O. and {Douspis}, M. and {Ducout}, A. and {Dupac}, X. and {Dusini}, S. and {Efstathiou}, G. and {Elsner}, F. and {En{\ss}lin}, T.~A. and {Eriksen}, H.~K. and {Fantaye}, Y. and {Farhang}, M. and {Fergusson}, J. and {Fernandez-Cobos}, R. and {Finelli}, F. and {Forastieri}, F. and {Frailis}, M. and {Fraisse}, A.~A. and {Franceschi}, E. and {Frolov}, A. and {Galeotta}, S. and {Galli}, S. and {Ganga}, K. and {G{\'e}nova-Santos}, R.~T. and {Gerbino}, M. and {Ghosh}, T. and {Gonz{\'a}lez-Nuevo}, J. and {G{\'o}rski}, K.~M. and {Gratton}, S. and {Gruppuso}, A. and {Gudmundsson}, J.~E. and {Hamann}, J. and {Handley}, W. and {Hansen}, F.~K. and {Herranz}, D. and {Hildebrandt}, S.~R. and {Hivon}, E. and {Huang}, Z. and {Jaffe}, A.~H. and {Jones}, W.~C. and {Karakci}, A. and {Keih{\"a}nen}, E. and {Keskitalo}, R. and {Kiiveri}, K. and {Kim}, J. and {Kisner}, T.~S. and {Knox}, L. and {Krachmalnicoff}, N. and {Kunz}, M. and {Kurki-Suonio}, H. and {Lagache}, G. and {Lamarre}, J. -M. and {Lasenby}, A. and {Lattanzi}, M. and {Lawrence}, C.~R. and {Le Jeune}, M. and {Lemos}, P. and {Lesgourgues}, J. and {Levrier}, F. and {Lewis}, A. and {Liguori}, M. and {Lilje}, P.~B. and {Lilley}, M. and {Lindholm}, V. and {L{\'o}pez-Caniego}, M. and {Lubin}, P.~M. and {Ma}, Y. -Z. and {Mac{\'\i}as-P{\'e}rez}, J.~F. and {Maggio}, G. and {Maino}, D. and {Mandolesi}, N. and {Mangilli}, A. and {Marcos-Caballero}, A. and {Maris}, M. and {Martin}, P.~G. and {Martinelli}, M. and {Mart{\'\i}nez-Gonz{\'a}lez}, E. and {Matarrese}, S. and {Mauri}, N. and {McEwen}, J.~D. and {Meinhold}, P.~R. and {Melchiorri}, A. and {Mennella}, A. and {Migliaccio}, M. and {Millea}, M. and {Mitra}, S. and {Miville-Desch{\^e}nes}, M. -A. and {Molinari}, D. and {Montier}, L. and {Morgante}, G. and {Moss}, A. and {Natoli}, P. and {N{\o}rgaard-Nielsen}, H.~U. and {Pagano}, L. and {Paoletti}, D. and {Partridge}, B. and {Patanchon}, G. and {Peiris}, H.~V. and {Perrotta}, F. and {Pettorino}, V. and {Piacentini}, F. and {Polastri}, L. and {Polenta}, G. and {Puget}, J. -L. and {Rachen}, J.~P. and {Reinecke}, M. and {Remazeilles}, M. and {Renzi}, A. and {Rocha}, G. and {Rosset}, C. and {Roudier}, G. and {Rubi{\~n}o-Mart{\'\i}n}, J.~A. and {Ruiz-Granados}, B. and {Salvati}, L. and {Sandri}, M. and {Savelainen}, M. and {Scott}, D. and {Shellard}, E.~P.~S. and {Sirignano}, C. and {Sirri}, G. and {Spencer}, L.~D. and {Sunyaev}, R. and {Suur-Uski}, A. -S. and {Tauber}, J.~A. and {Tavagnacco}, D. and {Tenti}, M. and {Toffolatti}, L. and {Tomasi}, M. and {Trombetti}, T. and {Valenziano}, L. and {Valiviita}, J. and {Van Tent}, B. and {Vibert}, L. and {Vielva}, P. and {Villa}, F. and {Vittorio}, N. and {Wandelt}, B.~D. and {Wehus}, I.~K. and {White}, M. and {White}, S.~D.~M. and {Zacchei}, A. and {Zonca}, A.},
        title = "{Planck 2018 results. VI. Cosmological parameters}",
      journal = {\aap},
         year = 2020,
        month = sep,
       volume = {641},
          eid = {A6},
        pages = {A6},
          doi = {10.1051/0004-6361/201833910},
archivePrefix = {arXiv},
       eprint = {1807.06209},
 primaryClass = {astro-ph.CO},
       adsurl = {https://ui.adsabs.harvard.edu/abs/2020A&A...641A...6P}
}

@ARTICLE{rastinejad25,
       author = {{Rastinejad}, Jillian C. and {Levan}, Andrew J. and {Jonker}, Peter G. and {Kilpatrick}, Charles D. and {Fryer}, Christopher L. and {Sarin}, Nikhil and {Gompertz}, Benjamin P. and {Liu}, Chang and {Eyles-Ferris}, Rob A.~J. and {Fong}, Wen-fai and {Burns}, Eric and {Gillanders}, James H. and {Mandel}, Ilya and {Malesani}, Daniele Bj{\o}rn and {O'Brien}, Paul T. and {Tanvir}, Nial R. and {Ackley}, Kendall and {Aryan}, Amar and {Bauer}, Franz E. and {Bloemen}, Steven and {de Boer}, Thomas and {Bom}, Cl{\'e}cio R. and {Chac{\'o}n}, Jennifer A. and {Chambers}, Ken and {Chen}, Ting-Wan and {Chrimes}, Ashley A. and {van Dalen}, Joyce N.~D. and {D'Elia}, Valerio and {De Pasquale}, Massimiliano and {Fulton}, Michael D. and {Groot}, Paul J. and {Gupta}, Rahul and {Hartmann}, Dieter H. and {van Hoof}, Agnes P.~C. and {Huber}, Mark E. and {Izzo}, Luca and {Jacobson-Galan}, Wynn and {Jakobsson}, P{\'a}ll and {Kong}, Albert and {Laskar}, Tanmoy and {Lowe}, Thomas B. and {Magnier}, Eugene A. and {Maiorano}, Elisabetta and {Martin-Carrillo}, Antonio and {Mas-Ribas}, Lluis and {Mata S{\'a}nchez}, Daniel and {Nicholl}, Matt and {Nixon}, Christopher J. and {Oates}, Samantha R. and {Paek}, Gregory and {Palmerio}, Jesse and {Paris}, Diego and {Pieterse}, Dani{\"e}lle L.~A. and {Pugliese}, Giovanna and {Quirola Vasquez}, Jonathan A. and {van Roestel}, Jan and {Rossi}, Andrea and {Rouco Escorial}, Alicia and {Salvaterra}, Ruben and {Schneider}, Benjamin and {Smartt}, Stephen J. and {Smith}, Ken and {Smith}, Ian A. and {Srivastav}, Shubham and {Torres}, Manuel A.~P. and {Ventura}, Chiara and {Vreeswijk}, Paul and {Wainscoat}, Richard and {Yang}, Yi-Jung and {Yang}, Sheng},
        title = "{EP 250108a/SN 2025kg: Observations of the Most Nearby Broad-line Type Ic Supernova Following an Einstein Probe Fast X-Ray Transient}",
      journal = {\apjl},
         year = 2025,
        month = jul,
       volume = {988},
       number = {1},
          eid = {L13},
        pages = {L13},
          doi = {10.3847/2041-8213/ade7f9},
archivePrefix = {arXiv},
       eprint = {2504.08889},
 primaryClass = {astro-ph.HE},
       adsurl = {https://ui.adsabs.harvard.edu/abs/2025ApJ...988L..13R}
}

@ARTICLE{Sanchez19,
       author = {{S{\'a}nchez}, S.~F. and {Barrera-Ballesteros}, J.~K. and {L{\'o}pez-Cob{\'a}}, C. and {Brough}, S. and {Bryant}, J.~J. and {Bland-Hawthorn}, J. and {Croom}, S.~M. and {van de Sande}, J. and {Cortese}, L. and {Goodwin}, M. and {Lawrence}, J.~S. and {L{\'o}pez-S{\'a}nchez}, A.~R. and {Sweet}, S.~M. and {Owers}, M.~S. and {Richards}, S.~N. and {Walcher}, C.~J.},
        title = "{The SAMI galaxy survey: exploring the gas-phase mass-metallicity relation}",
      journal = {\mnras},
         year = 2019,
        month = apr,
       volume = {484},
       number = {3},
        pages = {3042-3070},
          doi = {10.1093/mnras/stz019},
archivePrefix = {arXiv},
       eprint = {1812.11263},
 primaryClass = {astro-ph.GA},
       adsurl = {https://ui.adsabs.harvard.edu/abs/2019MNRAS.484.3042S}
}

@ARTICLE{sarin24,
       author = {{Sarin}, Nikhil and {H{\"u}bner}, Moritz and {Omand}, Conor M.~B. and {Setzer}, Christian N. and {Schulze}, Steve and {Adhikari}, Naresh and {Sagu{\'e}s-Carracedo}, Ana and {Galaudage}, Shanika and {Wallace}, Wendy F. and {Lamb}, Gavin P. and {Lin}, En-Tzu},
        title = "{REDBACK: a Bayesian inference software package for electromagnetic transients}",
      journal = {\mnras},
         year = 2024,
        month = jun,
       volume = {531},
       number = {1},
        pages = {1203-1227},
          doi = {10.1093/mnras/stae1238},
archivePrefix = {arXiv},
       eprint = {2308.12806},
 primaryClass = {astro-ph.HE},
       adsurl = {https://ui.adsabs.harvard.edu/abs/2024MNRAS.531.1203S}
}

@ARTICLE{Schlafly2011,
       author = {{Schlafly}, Edward F. and {Finkbeiner}, Douglas P.},
        title = "{Measuring Reddening with Sloan Digital Sky Survey Stellar Spectra and Recalibrating SFD}",
      journal = {\apj},
         year = 2011,
        month = aug,
       volume = {737},
       number = {2},
          eid = {103},
        pages = {103},
          doi = {10.1088/0004-637X/737/2/103},
archivePrefix = {arXiv},
       eprint = {1012.4804},
 primaryClass = {astro-ph.GA},
       adsurl = {https://ui.adsabs.harvard.edu/abs/2011ApJ...737..103S}
}

@ARTICLE{schroeder25,
       author = {{Schroeder}, Genevieve and {Ho}, Anna Y.~Q. and {Dastidar}, Ranadeep G. and {Modjaz}, Maryam and {Corsi}, Alessandra and {Duffell}, Paul C.},
        title = "{A Late-time Radio Search for Highly Off-axis Jets from PTF Broad-lined Ic Supernovae in GRB-like Host Galaxy Environments}",
      journal = {\apj},
         year = 2025,
        month = dec,
       volume = {995},
       number = {1},
          eid = {61},
        pages = {61},
          doi = {10.3847/1538-4357/ae129b},
archivePrefix = {arXiv},
       eprint = {2507.15928},
 primaryClass = {astro-ph.HE},
       adsurl = {https://ui.adsabs.harvard.edu/abs/2025ApJ...995...61S}
}

@ARTICLE{Schulze14,
   author = {{Schulze}, S. and {Malesani}, D. and {Cucchiara}, A. and {Tanvir}, N.~R. and 
	{Kr{\"u}hler}, T. and {de Ugarte Postigo}, A. and {Leloudas}, G. and 
	{Lyman}, J. and {Bersier}, D. and {Wiersema}, K. and {Perley}, D.~A. and 
	{Schady}, P. and {Gorosabel}, J. and {Anderson}, J.~P. and {Castro-Tirado}, A.~J. and 
	{Cenko}, S.~B. and {De Cia}, A. and {Ellerbroek}, L.~E. and 
	{Fynbo}, J.~P.~U. and {Greiner}, J. and {Hjorth}, J. and {Kann}, D.~A. and 
	{Kaper}, L. and {Klose}, S. and {Levan}, A.~J. and {Mart{\'{\i}}n}, S. and 
	{O'Brien}, P.~T. and {Page}, K.~L. and {Pignata}, G. and {Rapaport}, S. and 
	{S{\'a}nchez-Ram{\'{\i}}rez}, R. and {Sollerman}, J. and {Smith}, I.~A. and 
	{Sparre}, M. and {Th{\"o}ne}, C.~C. and {Watson}, D.~J. and 
	{Xu}, D. and {Bauer}, F.~E. and {Bayliss}, M. and {Bj{\"o}rnsson}, G. and 
	{Bremer}, M. and {Cano}, Z. and {Covino}, S. and {D'Elia}, V. and 
	{Frail}, D.~A. and {Geier}, S. and {Goldoni}, P. and {Hartoog}, O.~E. and 
	{Jakobsson}, P. and {Korhonen}, H. and {Lee}, K.~Y. and {Milvang-Jensen}, B. and 
	{Nardini}, M. and {Nicuesa Guelbenzu}, A. and {Oguri}, M. and 
	{Pandey}, S.~B. and {Petitpas}, G. and {Rossi}, A. and {Sandberg}, A. and 
	{Schmidl}, S. and {Tagliaferri}, G. and {Tilanus}, R.~P.~J. and 
	{Winters}, J.~M. and {Wright}, D. and {Wuyts}, E.},
    title = "{GRB 120422A/SN 2012bz: Bridging the gap between low- and high-luminosity gamma-ray bursts}",
  journal = {\aap},
archivePrefix = "arXiv",
   eprint = {1401.3774},
 primaryClass = "astro-ph.HE",
     year = 2014,
    month = jun,
   volume = 566,
      eid = {A102},
    pages = {A102},
      doi = {10.1051/0004-6361/201423387},
   adsurl = {http://adsabs.harvard.edu/abs/2014A%26A...566A.102S}
}

@ARTICLE{Schulze18,
       author = {{Schulze}, S. and {Kr{\"u}hler}, T. and {Leloudas}, G. and {Gorosabel}, J. and {Mehner}, A. and {Buchner}, J. and {Kim}, S. and {Ibar}, E. and {Amor{\'\i}n}, R. and {Herrero-Illana}, R. and {Anderson}, J.~P. and {Bauer}, F.~E. and {Christensen}, L. and {de Pasquale}, M. and {de Ugarte Postigo}, A. and {Gallazzi}, A. and {Hjorth}, J. and {Morrell}, N. and {Malesani}, D. and {Sparre}, M. and {Stalder}, B. and {Stark}, A.~A. and {Th{\"o}ne}, C.~C. and {Wheeler}, J.~C.},
        title = "{Cosmic evolution and metal aversion in superluminous supernova host galaxies}",
      journal = {\mnras},
         year = 2018,
        month = jan,
       volume = {473},
       number = {1},
        pages = {1258-1285},
          doi = {10.1093/mnras/stx2352},
archivePrefix = {arXiv},
       eprint = {1612.05978},
 primaryClass = {astro-ph.GA},
       adsurl = {https://ui.adsabs.harvard.edu/abs/2018MNRAS.473.1258S}
}

@ARTICLE{smith14,
       author = {{Smith}, Nathan},
        title = "{Mass Loss: Its Effect on the Evolution and Fate of High-Mass Stars}",
      journal = {\araa},
         year = 2014,
        month = aug,
       volume = {52},
        pages = {487-528},
          doi = {10.1146/annurev-astro-081913-040025},
archivePrefix = {arXiv},
       eprint = {1402.1237},
 primaryClass = {astro-ph.SR},
       adsurl = {https://ui.adsabs.harvard.edu/abs/2014ARA&A..52..487S}
}

@ARTICLE{Soderberg2006,
       author = {{Soderberg}, A.~M. and {Kulkarni}, S.~R. and {Nakar}, E. and {Berger}, E. and {Cameron}, P.~B. and {Fox}, D.~B. and {Frail}, D. and {Gal-Yam}, A. and {Sari}, R. and {Cenko}, S.~B. and {Kasliwal}, M. and {Chevalier}, R.~A. and {Piran}, T. and {Price}, P.~A. and {Schmidt}, B.~P. and {Pooley}, G. and {Moon}, D.-S. and {Penprase}, B.~E. and {Ofek}, E. and {Rau}, A. and {Gehrels}, N. and {Nousek}, J.~A. and {Burrows}, D.~N. and {Persson}, S.~E. and {McCarthy}, P.~J.},
        title = "{Relativistic ejecta from X-ray flash XRF 060218 and the rate of cosmic explosions}",
      journal = {\nat},
         year = 2006,
        month = aug,
       volume = {442},
       number = {7106},
        pages = {1014-1017},
          doi = {10.1038/nature05087},
archivePrefix = {arXiv},
       eprint = {astro-ph/0604389},
 primaryClass = {astro-ph},
       adsurl = {https://ui.adsabs.harvard.edu/abs/2006Natur.442.1014S}
}

@ARTICLE{srinivasaragavan24,
       author = {{Srinivasaragavan}, Gokul P. and {Yang}, Sheng and {Anand}, Shreya and {Sollerman}, Jesper and {Ho}, Anna Y.~Q. and {Corsi}, Alessandra and {Cenko}, S. Bradley and {Perley}, Daniel and {Schulze}, Steve and {Sanchez-Fleming}, Marquice and {Pope}, Jack and {Sarin}, Nikhil and {Omand}, Conor and {Das}, Kaustav K. and {Fremling}, Christoffer and {Andreoni}, Igor and {Bruch}, Rachel and {Burdge}, Kevin B. and {De}, Kishalay and {Gal-Yam}, Avishay and {Gangopadhyay}, Anjasha and {Graham}, Matthew J. and {Jencson}, Jacob E. and {Karambelkar}, Viraj and {Kasliwal}, Mansi M. and {Kulkarni}, S.~R. and {Martikainen}, Julia and {Sharma}, Yashvi S. and {Tzanidakis}, Anastasios and {Yan}, Lin and {Yao}, Yuhan and {Bellm}, Eric C. and {Groom}, Steven L. and {Masci}, Frank J. and {Nir}, Guy and {Purdum}, Josiah and {Smith}, Roger and {Sravan}, Niharika},
        title = "{Optical and Radio Analysis of Systematically Classified Broad-lined Type Ic Supernovae from the Zwicky Transient Facility}",
      journal = {\apj},
         year = 2024,
        month = nov,
       volume = {976},
       number = {1},
          eid = {71},
        pages = {71},
          doi = {10.3847/1538-4357/ad7fde},
archivePrefix = {arXiv},
       eprint = {2408.14586},
 primaryClass = {astro-ph.HE},
       adsurl = {https://ui.adsabs.harvard.edu/abs/2024ApJ...976...71S}
}

@ARTICLE{srinivasaragavan25a,
       author = {{Srinivasaragavan}, Gokul P. and {Hamidani}, Hamid and {Schroeder}, Genevieve and {Sarin}, Nikhil and {Ho}, Anna Y.~Q. and {Piro}, Anthony L. and {Cenko}, S. Bradley and {Anand}, Shreya and {Sollerman}, Jesper and {Perley}, Daniel A. and {Maeda}, Keiichi and {O'Connor}, Brendan and {Kuncarayakti}, Hanindyo and {Miller}, M. Coleman and {Ahumada}, Tom{\'a}s and {Vail}, Jada L. and {Duffell}, Paul and {Dastidar}, Ranadeep and {Andreoni}, Igor and {Bochenek}, Aleksandra and {Brennan}, Se{\'a}n. J. and {Carney}, Jonathan and {Chen}, Ping and {Freeburn}, James and {Gal-Yam}, Avishay and {Jacobson-Gal{\'a}n}, Wynn and {Kasliwal}, Mansi M. and {Li}, Jiaxuan and {Li}, Maggie L. and {Sravan}, Niharika and {Warshofsky}, Daniel E.},
        title = "{EP250108a/SN 2025kg: A Jet-driven Stellar Explosion Interacting with Circumstellar Material}",
      journal = {\apjl},
         year = 2025,
        month = aug,
       volume = {988},
       number = {2},
          eid = {L60},
        pages = {L60},
          doi = {10.3847/2041-8213/ade870},
archivePrefix = {arXiv},
       eprint = {2504.17516},
 primaryClass = {astro-ph.HE},
       adsurl = {https://ui.adsabs.harvard.edu/abs/2025ApJ...988L..60S}
}

@ARTICLE{srinivasaragavan25b,
       author = {{Srinivasaragavan}, Gokul P. and {Li}, Dongyue and {Hall}, Xander J. and {Gottlieb}, Ore and {Schroeder}, Genevieve and {Liu}, Heyang and {O'Connor}, Brendan and {Jin}, Chichuan and {Kasliwal}, Mansi and {Ahumada}, Tom{\'a}s and {Wu}, Qinyu and {Fryer}, Christopher L. and {Niblett}, Annabelle E. and {Xu}, Dong and {Edvige Ravasio}, Maria and {Daja}, Grace and {Li}, Wenxiong and {Anand}, Shreya and {Ho}, Anna Y.~Q. and {Sun}, Hui and {Perley}, Daniel A. and {Yan}, Lin and {Burns}, Eric and {Cenko}, S. Bradley and {Sollerman}, Jesper and {Sarin}, Nikhil and {Piro}, Anthony L. and {Aryan}, Amar and {Miller}, M. Coleman and {An}, Jie and {An}, Tao and {Andrews}, Moira and {Augustin}, Jule and {Bellm}, Eric C. and {Bochenek}, Aleksandra and {Busmann}, Malte and {Chanchaiworawit}, Krittapas and {Chen}, Huaqing and {Caballero-Garc{\'\i}a}, Maria D. and {Castro-Tirado}, Alberto J. and {Esamdin}, Ali and {Faba-Moreno}, Jennifer and {Farah}, Joseph and {Fern{\'a}ndez-Garc{\'\i}a}, Emilio and {Fu}, Shaoyu and {Fynbo}, Johan P.~U. and {Gassert}, Julius and {Padilla Gonzalez}, Estefania and {P{\'e}rez-Garc{\'\i}a}, Ignacio and {Graham}, Matthew and {Gritsevich}, Maria and {Gruen}, Daniel and {Guziy}, Sergiy and {Howell}, D. Andrew and {He}, Linbo and {Hu}, Jingwei and {Hu}, You-Dong and {Iskandar}, Abdusamatjan and {Castaneda Jaims}, Joahan and {Jiang}, Ji-An and {Jiang}, Ning and {Jiang}, Shuaijiao and {Liang}, Runduo and {Ling}, Zhixing and {Liu}, Jialian and {Liu}, Xing and {Liu}, Yuan and {Masci}, Frank J. and {McCully}, Curtis and {Newsome}, Megan and {Noysena}, Kanthanakorn and {Pandey}, Shashi B. and {Ni}, Kangrui and {Palmese}, Antonella and {Peng}, Han-Long and {Purdum}, Josiah and {Qin}, Yu-Jing and {Rose}, Sam and {Rusholme}, Ben and {S{\'a}nchez-Ram{\'\i}rez}, Rub{\'e}n and {Sevilla}, Cassie and {Smith}, Roger and {Song}, Yujia and {Sravan}, Niharika and {Stein}, Robert and {Tabor}, Constantin and {Terreran}, Giacomo and {Tinyanont}, Samaporn and {Vega}, Pablo and {Wang}, Letian and {Wang}, Tinggu and {Wang}, Xiaofeng and {Wu}, Siyu and {Wu}, Xuefeng and {Wynn}, Kathryn and {Xu}, Yunfei and {Yan}, Shengyu and {Yuan}, Weimin and {Zhang}, Binbin and {Zhang}, Chen and {Zhu}, Zipei and {Zuo}, Xiaoxiong and {Bhullar}, Gursimran},
        title = "{EP250827b/SN 2025wkm: An X-ray Flash-Supernova Powered by a Central Engine and Circumstellar Interaction}",
      journal = {arXiv e-prints},
         year = 2025,
        month = dec,
          eid = {arXiv:2512.10239},
        pages = {arXiv:2512.10239},
          doi = {10.48550/arXiv.2512.10239},
archivePrefix = {arXiv},
       eprint = {2512.10239},
 primaryClass = {astro-ph.HE},
       adsurl = {https://ui.adsabs.harvard.edu/abs/2025arXiv251210239S}
}

@ARTICLE{Srivastav25,
       author = {{Srivastav}, S. and {Chen}, T.-W. and {Gillanders}, J.~H. and {Rhodes}, L. and {Smartt}, S.~J. and {Huber}, M.~E. and {Aryan}, A. and {Yang}, S. and {Beri}, A. and {Cooper}, A.~J. and {Nicholl}, M. and {Smith}, K.~W. and {Stevance}, H.~F. and {Carotenuto}, F. and {Chambers}, K.~C. and {Aamer}, A. and {Angus}, C.~R. and {Fulton}, M.~D. and {Moore}, T. and {Smith}, I.~A. and {Young}, D.~R. and {de Boer}, T. and {Gao}, H. and {Lin}, C.-C. and {Lowe}, T. and {Magnier}, E.~A. and {Minguez}, P. and {Pan}, Y.-C. and {Wainscoat}, R.~J.},
        title = "{Identification of the Optical Counterpart of the Fast X-Ray Transient EP240414a}",
      journal = {\apjl},
         year = 2025,
        month = jan,
       volume = {978},
       number = {2},
          eid = {L21},
        pages = {L21},
          doi = {10.3847/2041-8213/ad9c75},
archivePrefix = {arXiv},
       eprint = {2409.19070},
 primaryClass = {astro-ph.HE},
       adsurl = {https://ui.adsabs.harvard.edu/abs/2025ApJ...978L..21S}
}

@ARTICLE{Starling2011,
       author = {{Starling}, R.~L.~C. and {Wiersema}, K. and {Levan}, A.~J. and {Sakamoto}, T. and {Bersier}, D. and {Goldoni}, P. and {Oates}, S.~R. and {Rowlinson}, A. and {Campana}, S. and {Sollerman}, J. and {Tanvir}, N.~R. and {Malesani}, D. and {Fynbo}, J.~P.~U. and {Covino}, S. and {D'Avanzo}, P. and {O'Brien}, P.~T. and {Page}, K.~L. and {Osborne}, J.~P. and {Vergani}, S.~D. and {Barthelmy}, S. and {Burrows}, D.~N. and {Cano}, Z. and {Curran}, P.~A. and {de Pasquale}, M. and {D'Elia}, V. and {Evans}, P.~A. and {Flores}, H. and {Fruchter}, A.~S. and {Garnavich}, P. and {Gehrels}, N. and {Gorosabel}, J. and {Hjorth}, J. and {Holland}, S.~T. and {van der Horst}, A.~J. and {Hurkett}, C.~P. and {Jakobsson}, P. and {Kamble}, A.~P. and {Kouveliotou}, C. and {Kuin}, N.~P.~M. and {Kaper}, L. and {Mazzali}, P.~A. and {Nugent}, P.~E. and {Pian}, E. and {Stamatikos}, M. and {Th{\"o}ne}, C.~C. and {Woosley}, S.~E.},
        title = "{Discovery of the nearby long, soft GRB 100316D with an associated supernova}",
      journal = {\mnras},
         year = 2011,
        month = mar,
       volume = {411},
       number = {4},
        pages = {2792-2803},
          doi = {10.1111/j.1365-2966.2010.17879.x},
archivePrefix = {arXiv},
       eprint = {1004.2919},
 primaryClass = {astro-ph.CO},
       adsurl = {https://ui.adsabs.harvard.edu/abs/2011MNRAS.411.2792S}
}

@ARTICLE{Steeghs22,
       author = {{Steeghs}, D. and {Galloway}, D.~K. and {Ackley}, K. and {Dyer}, M.~J. and {Lyman}, J. and {Ulaczyk}, K. and {Cutter}, R. and {Mong}, Y.-L. and {Dhillon}, V. and {O'Brien}, P. and {Ramsay}, G. and {Poshyachinda}, S. and {Kotak}, R. and {Nuttall}, L.~K. and {Pall{\'e}}, E. and {Breton}, R.~P. and {Pollacco}, D. and {Thrane}, E. and {Aukkaravittayapun}, S. and {Awiphan}, S. and {Burhanudin}, U. and {Chote}, P. and {Chrimes}, A. and {Daw}, E. and {Duffy}, C. and {Eyles-Ferris}, R. and {Gompertz}, B. and {Heikkil{\"a}}, T. and {Irawati}, P. and {Kennedy}, M.~R. and {Killestein}, T. and {Kuncarayakti}, H. and {Levan}, A.~J. and {Littlefair}, S. and {Makrygianni}, L. and {Marsh}, T. and {Mata-Sanchez}, D. and {Mattila}, S. and {Maund}, J. and {McCormac}, J. and {Mkrtichian}, D. and {Mullaney}, J. and {Noysena}, K. and {Patel}, M. and {Rol}, E. and {Sawangwit}, U. and {Stanway}, E.~R. and {Starling}, R. and {Str{\o}m}, P. and {Tooke}, S. and {West}, R. and {White}, D.~J. and {Wiersema}, K.},
        title = "{The Gravitational-wave Optical Transient Observer (GOTO): prototype performance and prospects for transient science}",
      journal = {\mnras},
         year = 2022,
        month = apr,
       volume = {511},
       number = {2},
        pages = {2405-2422},
          doi = {10.1093/mnras/stac013},
archivePrefix = {arXiv},
       eprint = {2110.05539},
 primaryClass = {astro-ph.IM},
       adsurl = {https://ui.adsabs.harvard.edu/abs/2022MNRAS.511.2405S}
}

@ARTICLE{stritzinger26,
       author = {{Stritzinger}, M.~D. and {Moriya}, T.~J. and {Bose}, S. and {Mazzali}, P.~A. and {Lundqvist}, P. and {Karamehmetoglu}, E. and {Arndt}, L.~S. and {Ashall}, C. and {Galbany}, L. and {Hoogendam}, W.~B. and {Baron}, E. and {DerKacy}, J.~M. and {Elias-Rosa}, N. and {Hsiao}, E.~Y. and {H{\"o}flich}, P. and {Pian}, E. and {Jensen}, E.~A.~M. and {Moran}, S. and {Pastorello}, A. and {Shahbandeh}, M. and {Valerin}, G.},
        title = "{The broad-lined type Ic supernova 2020lao experienced an energetic explosion with no central-engine signatures}",
      journal = {\aap},
         year = 2026,
        month = apr,
       volume = {708},
          eid = {A305},
        pages = {A305},
          doi = {10.1051/0004-6361/202558378},
       adsurl = {https://ui.adsabs.harvard.edu/abs/2026A&A...708A.305S}
}

@ARTICLE{Sun25,
       author = {{Sun}, H. and {Li}, W.-X. and {Liu}, L.-D. and {Gao}, H. and {Wang}, X.-F. and {Yuan}, W. and {Zhang}, B. and {Filippenko}, A.~V. and {Xu}, D. and {An}, T. and {Ai}, S. and {Brink}, T.~G. and {Liu}, Y. and {Liu}, Y.-Q. and {Wang}, C.-Y. and {Wu}, Q.-Y. and {Wu}, X.-F. and {Yang}, Y. and {Zhang}, B.-B. and {Zheng}, W.-K. and {Ahumada}, T. and {Dai}, Z.-G. and {Delaunay}, J. and {Elias-Rosa}, N. and {Benetti}, S. and {Fu}, S.-Y. and {Howell}, D.~A. and {Huang}, Y.-F. and {Kasliwal}, M.~M. and {Karambelkar}, V. and {Stein}, R. and {Lei}, W.-H. and {Lian}, T.-Y. and {Peng}, Z.-K. and {Frederiks}, D.~D. and {Ridnaia}, A.~V. and {Svinkin}, D.~S. and {Wang}, X.-Y. and {Wang}, A.-L. and {Wei}, D.-M. and {An}, J. and {Andrews}, M. and {Bai}, J.-M. and {Dai}, C.-Y. and {Ehgamberdiev}, S.~A. and {Fan}, Z. and {Farah}, J. and {Feng}, H.-C. and {Fynbo}, J.~P.~U. and {Guo}, W.-J. and {Guo}, Z. and {Hu}, M.-K. and {Hu}, J.-W. and {Jiang}, S.-Q. and {Jin}, J.-J. and {Li}, A. and {Li}, J.-D. and {Li}, R.-Z. and {Liang}, Y.-F. and {Ling}, Z.-X. and {Liu}, X. and {Mao}, J.-R. and {McCully}, C. and {Mirzaqulov}, D. and {Newsome}, M. and {Padilla Gonzalez}, E. and {Pan}, X. and {Terreran}, G. and {Tinyanont}, S. and {Wang}, B.-T. and {Wang}, L.-Z. and {Wen}, X.-D. and {Xiang}, D.-F. and {Xue}, S.-J. and {Yang}, J. and {Zhu}, Z.-P. and {Cai}, Z.-M. and {Castro-Tirado}, A.~J. and {Chen}, F.-S. and {Chen}, H.-L. and {Chen}, T.-X. and {Chen}, W. and {Chen}, Y.-H. and {Chen}, Y.-F. and {Chen}, Y. and {Cheng}, H.-Q. and {Cordier}, B. and {Cui}, C.-Z. and {Cui}, W.-W. and {Dai}, Y.-F. and {Fan}, D.-W. and {Feng}, H. and {Guan}, J. and {Han}, D.-W. and {Hou}, D.-J. and {Hu}, H.-B. and {Huang}, M.-H. and {Huo}, J. and {Jia}, S.-M. and {Jia}, Z.-Q. and {Jiang}, B.-W. and {Jin}, C.-C. and {Jin}, G. and {Kuulkers}, E. and {Li}, C.-K. and {Li}, D.-Y. and {Li}, J.-F. and {Li}, L.-H. and {Li}, M.-S. and {Li}, W. and {Li}, Z.-D. and {Liu}, C.-Z. and {Liu}, H.-Y. and {Liu}, H.-Q. and {Liu}, M.-J. and {Lu}, F.-J. and {Luo}, L.-D. and {Ma}, J. and {Mao}, X. and {Nandra}, K. and {O'Brien}, P. and {Pan}, H.-W. and {Rau}, A. and {Rea}, N. and {Sanders}, J. and {Song}, L.-M. and {Sun}, S.-L. and {Sun}, X.-J. and {Tan}, Y.-Y. and {Tang}, Q.-J. and {Tao}, Y.-H. and {Wang}, H. and {Wang}, J. and {Wang}, L. and {Wang}, W.-X. and {Wang}, Y.-L. and {Wang}, Y.-S. and {Xiong}, D.-R. and {Xu}, H.-T. and {Xu}, J.-J. and {Xu}, X.-P. and {Xu}, Y.-F. and {Xu}, Z. and {Xue}, C.-B. and {Xue}, Y.-L. and {Yan}, A.-L. and {Yang}, H.-N. and {Yang}, X.-T. and {Yang}, Y.-J. and {Zhang}, C. and {Zhang}, J. and {Zhang}, M. and {Zhang}, S.-N. and {Zhang}, W.-D. and {Zhang}, W.-J. and {Zhang}, Y.-H. and {Zhang}, Z. and {Zhang}, Z. and {Zhang}, Z.-L. and {Zhao}, D.-H. and {Zhao}, H.-S. and {Zhao}, X.-F. and {Zhao}, Z.-J. and {Zhou}, Y.-L. and {Zhu}, Y.-X. and {Zhu}, Z.-C. and {Zou}, H.},
        title = "{A fast X-ray transient from a weak relativistic jet associated with a type Ic-BL supernova}",
      journal = {Nature Astronomy},
         year = 2025,
        month = jul,
       volume = {9},
        pages = {1073-1085},
          doi = {10.1038/s41550-025-02571-1},
archivePrefix = {arXiv},
       eprint = {2410.02315},
 primaryClass = {astro-ph.HE},
       adsurl = {https://ui.adsabs.harvard.edu/abs/2025NatAs...9.1073S}
}

@ARTICLE{Taddia19,
       author = {{Taddia}, F. and {Sollerman}, J. and {Fremling}, C. and {Barbarino}, C. and {Karamehmetoglu}, E. and {Arcavi}, I. and {Cenko}, S.~B. and {Filippenko}, A.~V. and {Gal-Yam}, A. and {Hiramatsu}, D. and {Hosseinzadeh}, G. and {Howell}, D.~A. and {Kulkarni}, S.~R. and {Laher}, R. and {Lunnan}, R. and {Masci}, F. and {Nugent}, P.~E. and {Nyholm}, A. and {Perley}, D.~A. and {Quimby}, R. and {Silverman}, J.~M.},
        title = "{Analysis of broad-lined Type Ic supernovae from the (intermediate) Palomar Transient Factory}",
      journal = {\aap},
         year = 2019,
        month = jan,
       volume = {621},
          eid = {A71},
        pages = {A71},
          doi = {10.1051/0004-6361/201834429},
archivePrefix = {arXiv},
       eprint = {1811.09544},
 primaryClass = {astro-ph.HE},
       adsurl = {https://ui.adsabs.harvard.edu/abs/2019A&A...621A..71T}
}

@ARTICLE{Terreran19,
       author = {{Terreran}, G. and {Margutti}, R. and {Bersier}, D. and {Brimacombe}, J. and {Caprioli}, D. and {Challis}, P. and {Chornock}, R. and {Coppejans}, D.~L. and {Dong}, Subo and {Guidorzi}, C. and {Hurley}, K. and {Kirshner}, R. and {Migliori}, G. and {Milisavljevic}, D. and {Palmer}, D.~M. and {Prieto}, J.~L. and {Tomasella}, L. and {Marchant}, P. and {Pastorello}, A. and {Shappee}, B.~J. and {Stanek}, K.~Z. and {Stritzinger}, M.~D. and {Benetti}, S. and {Chen}, Ping and {DeMarchi}, L. and {Elias-Rosa}, N. and {Gall}, C. and {Harmanen}, J. and {Mattila}, S.},
        title = "{SN 2016coi (ASASSN-16fp): An Energetic H-stripped Core-collapse Supernova from a Massive Stellar Progenitor with Large Mass Loss}",
      journal = {\apj},
         year = 2019,
        month = oct,
       volume = {883},
       number = {2},
          eid = {147},
        pages = {147},
          doi = {10.3847/1538-4357/ab3e37},
archivePrefix = {arXiv},
       eprint = {1905.02226},
 primaryClass = {astro-ph.HE},
       adsurl = {https://ui.adsabs.harvard.edu/abs/2019ApJ...883..147T}
}

@ARTICLE{Thoene14,
   author = {{Th{\"o}ne}, C.~C. and {Christensen}, L. and {Prochaska}, J.~X. and 
	{Bloom}, J.~S. and {Gorosabel}, J. and {Fynbo}, J.~P.~U. and 
	{Jakobsson}, P. and {Fruchter}, A.~S.},
    title = "{The host of the SN-less GRB 060505 in high resolution}",
  journal = {\mnras},
archivePrefix = "arXiv",
   eprint = {1404.0881},
 primaryClass = "astro-ph.HE",
     year = 2014,
    month = jul,
   volume = 441,
    pages = {2034-2048},
      doi = {10.1093/mnras/stu711},
   adsurl = {http://adsabs.harvard.edu/abs/2014MNRAS.441.2034T}
}

@ARTICLE{Thoene2024,
       author = {{Th{\"o}ne}, C.~C. and {de Ugarte Postigo}, A. and {Izzo}, L. and {Michalowski}, M.~J. and {Levan}, A.~J. and {Leung}, J.~K. and {Ag{\"u}{\'\i} Fern{\'a}ndez}, J.~F. and {G{\'e}ron}, T. and {Friesen}, R. and {Christensen}, L. and {Covino}, S. and {D'Elia}, V. and {Hartmann}, D.~H. and {Jakobsson}, P. and {De Pasquale}, M. and {Pugliese}, G. and {Rossi}, A. and {Schady}, P. and {Wiersema}, K. and {Zafar}, T.},
        title = "{The host of GRB 171205A in 3D: A resolved multiwavelength study of a rare grand-design spiral GRB host}",
      journal = {\aap},
         year = 2024,
        month = oct,
       volume = {690},
          eid = {A66},
        pages = {A66},
          doi = {10.1051/0004-6361/202348141},
archivePrefix = {arXiv},
       eprint = {2406.16725},
 primaryClass = {astro-ph.HE},
       adsurl = {https://ui.adsabs.harvard.edu/abs/2024A&A...690A..66T}
}

@article{Valeckas2025,
         doi = {10.21105/joss.09264},
         url = {https://doi.org/10.21105/joss.09264},
         year = {2025}, publisher = {The Open Journal},
         volume = {10}, number = {116}, pages = {9264},
         author = {Valeckas, Kostas and Fynbo,
         Johan Peter Uldall and Krogager,
         Jens-Kristian and Heintz, Kasper Elm},
         title = {PyLongslit: a simple manual Python pipeline for processing of astronomical long-slit spectra recorded with CCD detectors},
         journal = {Journal of Open Source Software}
}

@ARTICLE{vanDalen25,
       author = {{van Dalen}, Joyce N.~D. and {Levan}, Andrew J. and {Jonker}, Peter G. and {Malesani}, Daniele Bj{\o}rn and {Izzo}, Luca and {Sarin}, Nikhil and {Quirola-V{\'a}squez}, Jonathan and {Mata S{\'a}nchez}, Daniel and {de Ugarte Postigo}, Antonio and {van Hoof}, Agnes P.~C. and {Torres}, Manuel A.~P. and {Schulze}, Steve and {Littlefair}, Stuart P. and {Chrimes}, Ashley and {Ravasio}, Maria E. and {Bauer}, Franz E. and {Martin-Carrillo}, Antonio and {Fraser}, Morgan and {van der Horst}, Alexander J. and {Jakobsson}, Pall and {O'Brien}, Paul and {De Pasquale}, Massimiliano and {Pugliese}, Giovanna and {Sollerman}, Jesper and {Tanvir}, Nial R. and {Zafar}, Tayyaba and {Anderson}, Joseph P. and {Galbany}, Llu{\'\i}s and {Gal-Yam}, Avishay and {Gromadzki}, Mariusz and {M{\"u}ller-Bravo}, Tom{\'a}s E. and {Ragosta}, Fabio and {Terwel}, Jacco H.},
        title = "{The Einstein Probe Transient EP240414a: Linking Fast X-Ray Transients, Gamma-Ray Bursts, and Luminous Fast Blue Optical Transients}",
      journal = {\apjl},
         year = 2025,
        month = apr,
       volume = {982},
       number = {2},
          eid = {L47},
        pages = {L47},
          doi = {10.3847/2041-8213/adbc7e},
archivePrefix = {arXiv},
       eprint = {2409.19056},
 primaryClass = {astro-ph.HE},
       adsurl = {https://ui.adsabs.harvard.edu/abs/2025ApJ...982L..47V}
}

@ARTICLE{villar17,
       author = {{Villar}, V. Ashley and {Berger}, Edo and {Metzger}, Brian D. and {Guillochon}, James},
        title = "{Theoretical Models of Optical Transients. I. A Broad Exploration of the Duration-Luminosity Phase Space}",
      journal = {\apj},
         year = 2017,
        month = nov,
       volume = {849},
       number = {1},
          eid = {70},
        pages = {70},
          doi = {10.3847/1538-4357/aa8fcb},
archivePrefix = {arXiv},
       eprint = {1707.08132},
 primaryClass = {astro-ph.HE},
       adsurl = {https://ui.adsabs.harvard.edu/abs/2017ApJ...849...70V}
}

@article{Williams21,
   author = "Williams, Michael J. and Veitch, John and Messenger, Chris",
   title = "{Nested sampling with normalizing flows for gravitational-wave inference}",
   eprint = "2102.11056",
   archivePrefix = "arXiv",
   primaryClass = "gr-qc",
   doi = "10.1103/PhysRevD.103.103006",
   journal = "Phys. Rev. D",
   volume = "103",
   number = "10",
   pages = "103006",
   year = "2021"
}

@article{Williams23,
   author = "Williams, Michael J. and Veitch, John and Messenger, Chris",
   title = "{Importance nested sampling with normalising flows}",
   eprint = "2302.08526",
   archivePrefix = "arXiv",
   primaryClass = "astro-ph.IM",
   reportNumber = "LIGO-P2200283",
   doi = "10.1088/2632-2153/acd5aa",
   journal = "Mach. Learn. Sci. Tech.",
   volume = "4",
   number = "3",
   pages = "035011",
   year = "2023"
}

@ARTICLE{woosley2002,
       author = {{Woosley}, S.~E. and {Heger}, A. and {Weaver}, T.~A.},
        title = "{The evolution and explosion of massive stars}",
      journal = {Reviews of Modern Physics},
         year = 2002,
        month = nov,
       volume = {74},
       number = {4},
        pages = {1015-1071},
          doi = {10.1103/RevModPhys.74.1015},
       adsurl = {https://ui.adsabs.harvard.edu/abs/2002RvMP...74.1015W}
}

@ARTICLE{yamanaka17,
       author = {{Yamanaka}, Masayuki and {Nakaoka}, Tatsuya and {Tanaka}, Masaomi and {Maeda}, Keiichi and {Honda}, Satoshi and {Hanayama}, Hidekazu and {Morokuma}, Tomoki and {Imai}, Masataka and {Kinugasa}, Kenzo and {Murata}, Katsuhiro L. and {Nishimori}, Takefumi and {Hashimoto}, Osamu and {Gima}, Hirotaka and {Hosoya}, Kensuke and {Ito}, Ayano and {Karita}, Mayu and {Kawabata}, Miho and {Morihana}, Kumiko and {Morikawa}, Yuto and {Murakami}, Kotone and {Nagayama}, Takahiro and {Ono}, Tatsuharu and {Onozato}, Hiroki and {Sarugaku}, Yuki and {Sato}, Mitsuteru and {Suzuki}, Daisuke and {Takahashi}, Jun and {Takayama}, Masaki and {Yaguchi}, Hijiri and {Akitaya}, Hiroshi and {Asakura}, Yuichiro and {Kawabata}, Koji S. and {Kuroda}, Daisuke and {Nogami}, Daisaku and {Oasa}, Yumiko and {Omodaka}, Toshihiro and {Saito}, Yoshihiko and {Sekiguchi}, Kazuhiro and {Tominaga}, Nozomu and {Uemura}, Makoto and {Watanabe}, Makoto},
        title = "{Broad-lined Supernova 2016coi with a Helium Envelope}",
      journal = {\apj},
         year = 2017,
        month = mar,
       volume = {837},
       number = {1},
          eid = {1},
        pages = {1},
          doi = {10.3847/1538-4357/aa5f57},
archivePrefix = {arXiv},
       eprint = {1702.02035},
 primaryClass = {astro-ph.HE},
       adsurl = {https://ui.adsabs.harvard.edu/abs/2017ApJ...837....1Y}
}

@ARTICLE{zheng25,
       author = {{Zheng}, Jian-He and {Zhu}, Jin-Ping and {Lu}, Wenbin and {Zhang}, Bing},
        title = "{EP240414a: Off-axis View of a Jet-cocoon System from an Expanded Progenitor Star}",
      journal = {\apj},
         year = 2025,
        month = may,
       volume = {985},
       number = {1},
          eid = {21},
        pages = {21},
          doi = {10.3847/1538-4357/adc993},
archivePrefix = {arXiv},
       eprint = {2503.24266},
 primaryClass = {astro-ph.HE},
       adsurl = {https://ui.adsabs.harvard.edu/abs/2025ApJ...985...21Z}
}

@ARTICLE{zheng26,
       author = {{Zheng}, Jian-He and {Lu}, Wenbin},
        title = "{Fast X-Ray Transients Produced by Off-axis Jet Cocoons from Long Gamma-Ray Bursts}",
      journal = {\apjl},
         year = 2026,
        month = may,
       volume = {1003},
       number = {1},
          eid = {L19},
        pages = {L19},
          doi = {10.3847/2041-8213/ae67f2},
archivePrefix = {arXiv},
       eprint = {2603.09674},
 primaryClass = {astro-ph.HE},
       adsurl = {https://ui.adsabs.harvard.edu/abs/2026ApJ..1003L..19Z}
}

@ARTICLE{Thoene15,
       author = {{Thone}, C.~C. and {de Ugarte Postigo}, A. and {Garcia-Benito}, R. and {Leloudas}, G. and {Schulze}, S. and {Amorin}, R.},
        title = "{A young stellar environment for the superluminous supernova PTF12dam.}",
      journal = {\mnras},
         year = 2015,
        month = jul,
       volume = {451},
        pages = {L65-L69},
          doi = {10.1093/mnrasl/slv051},
archivePrefix = {arXiv},
       eprint = {1411.1104},
 primaryClass = {astro-ph.GA},
       adsurl = {https://ui.adsabs.harvard.edu/abs/2015MNRAS.451L..65T}
}

@ARTICLE{Chambers2016,
       author = {{Chambers}, K.~C. and {Magnier}, E.~A. and {Metcalfe}, N. and {Flewelling}, H.~A. and {Huber}, M.~E. and {Waters}, C.~Z. and {Denneau}, L. and {Draper}, P.~W. and {Farrow}, D. and {Finkbeiner}, D.~P. and {Holmberg}, C. and {Koppenhoefer}, J. and {Price}, P.~A. and {Rest}, A. and {Saglia}, R.~P. and {Schlafly}, E.~F. and {Smartt}, S.~J. and {Sweeney}, W. and {Wainscoat}, R.~J. and {Burgett}, W.~S. and {Chastel}, S. and {Grav}, T. and {Heasley}, J.~N. and {Hodapp}, K.~W. and {Jedicke}, R. and {Kaiser}, N. and {Kudritzki}, R.-P. and {Luppino}, G.~A. and {Lupton}, R.~H. and {Monet}, D.~G. and {Morgan}, J.~S. and {Onaka}, P.~M. and {Shiao}, B. and {Stubbs}, C.~W. and {Tonry}, J.~L. and {White}, R. and {Ba{\~n}ados}, E. and {Bell}, E.~F. and {Bender}, R. and {Bernard}, E.~J. and {Boegner}, M. and {Boffi}, F. and {Botticella}, M.~T. and {Calamida}, A. and {Casertano}, S. and {Chen}, W.-P. and {Chen}, X. and {Cole}, S. and {Deacon}, N. and {Frenk}, C. and {Fitzsimmons}, A. and {Gezari}, S. and {Gibbs}, V. and {Goessl}, C. and {Goggia}, T. and {Gourgue}, R. and {Goldman}, B. and {Grant}, P. and {Grebel}, E.~K. and {Hambly}, N.~C. and {Hasinger}, G. and {Heavens}, A.~F. and {Heckman}, T.~M. and {Henderson}, R. and {Henning}, T. and {Holman}, M. and {Hopp}, U. and {Ip}, W.-H. and {Isani}, S. and {Jackson}, M. and {Keyes}, C.~D. and {Koekemoer}, A.~M. and {Kotak}, R. and {Le}, D. and {Liska}, D. and {Long}, K.~S. and {Lucey}, J.~R. and {Liu}, M. and {Martin}, N.~F. and {Masci}, G. and {McLean}, B. and {Mindel}, E. and {Misra}, P. and {Morganson}, E. and {Murphy}, D.~N.~A. and {Obaika}, A. and {Narayan}, G. and {Nieto-Santisteban}, M.~A. and {Norberg}, P. and {Peacock}, J.~A. and {Pier}, E.~A. and {Postman}, M. and {Primak}, N. and {Rae}, C. and {Rai}, A. and {Riess}, A. and {Riffeser}, A. and {Rix}, H.~W. and {R{\"o}ser}, S. and {Russel}, R. and {Rutz}, L. and {Schilbach}, E. and {Schultz}, A.~S.~B. and {Scolnic}, D. and {Strolger}, L. and {Szalay}, A. and {Seitz}, S. and {Small}, E. and {Smith}, K.~W. and {Soderblom}, D.~R. and {Taylor}, P. and {Thomson}, R. and {Taylor}, A.~N. and {Thakar}, A.~R. and {Thiel}, J. and {Thilker}, D. and {Unger}, D. and {Urata}, Y. and {Valenti}, J. and {Wagner}, J. and {Walder}, T. and {Walter}, F. and {Watters}, S.~P. and {Werner}, S. and {Wood-Vasey}, W.~M. and {Wyse}, R.},
        title = "{The Pan-STARRS1 Surveys}",
      journal = {arXiv e-prints},
         year = 2016,
        month = dec,
          eid = {arXiv:1612.05560},
        pages = {arXiv:1612.05560},
          doi = {10.48550/arXiv.1612.05560},
archivePrefix = {arXiv},
       eprint = {1612.05560},
 primaryClass = {astro-ph.IM},
       adsurl = {https://ui.adsabs.harvard.edu/abs/2016arXiv161205560C}
}

@ARTICLE{stdweb,
       author = {{Karpov}, Sergey},
        title = "{STDweb: simple transient detection pipeline for the web}",
      journal = {Acta Polytechnica},
         year = 2025,
        month = mar,
       volume = {65},
       number = {1},
        pages = {50-64},
          doi = {10.14311/AP.2025.65.0050},
archivePrefix = {arXiv},
       eprint = {2411.16470},
 primaryClass = {astro-ph.IM},
       adsurl = {https://ui.adsabs.harvard.edu/abs/2025AcPol..65...50K}
}

@ARTICLE{BC03,
       author = {{Bruzual}, G. and {Charlot}, S.},
        title = "{Stellar population synthesis at the resolution of 2003}",
      journal = {\mnras},
         year = 2003,
        month = oct,
       volume = {344},
       number = {4},
        pages = {1000-1028},
          doi = {10.1046/j.1365-8711.2003.06897.x},
archivePrefix = {arXiv},
       eprint = {astro-ph/0309134},
 primaryClass = {astro-ph},
       adsurl = {https://ui.adsabs.harvard.edu/abs/2003MNRAS.344.1000B}
}

@ARTICLE{Dale14,
       author = {{Dale}, Daniel A. and {Helou}, George and {Magdis}, Georgios E. and {Armus}, Lee and {D{\'\i}az-Santos}, Tanio and {Shi}, Yong},
        title = "{A Two-parameter Model for the Infrared/Submillimeter/Radio Spectral Energy Distributions of Galaxies and Active Galactic Nuclei}",
      journal = {\apj},
         year = 2014,
        month = mar,
       volume = {784},
       number = {1},
          eid = {83},
        pages = {83},
          doi = {10.1088/0004-637X/784/1/83},
archivePrefix = {arXiv},
       eprint = {1402.1495},
 primaryClass = {astro-ph.GA},
       adsurl = {https://ui.adsabs.harvard.edu/abs/2014ApJ...784...83D}
}

@ARTICLE{Inserra2013,
       author = {{Inserra}, C. and {Pastorello}, A. and {Turatto}, M. and {Pumo}, M.~L. and {Benetti}, S. and {Cappellaro}, E. and {Botticella}, M.~T. and {Bufano}, F. and {Elias-Rosa}, N. and {Harutyunyan}, A. and {Taubenberger}, S. and {Valenti}, S. and {Zampieri}, L.},
        title = "{Moderately luminous Type II supernovae}",
      journal = {\aap},
         year = 2013,
        month = jul,
       volume = {555},
          eid = {A142},
        pages = {A142},
          doi = {10.1051/0004-6361/201220496},
archivePrefix = {arXiv},
       eprint = {1210.1411},
 primaryClass = {astro-ph.CO},
       adsurl = {https://ui.adsabs.harvard.edu/abs/2013A&A...555A.142I}
}

@ARTICLE{Nagy2018,
       author = {{Nagy}, Andrea P.},
        title = "{Average Opacity Calculation for Core-collapse Supernovae}",
      journal = {\apj},
         year = 2018,
        month = aug,
       volume = {862},
       number = {2},
          eid = {143},
        pages = {143},
          doi = {10.3847/1538-4357/aace56},
archivePrefix = {arXiv},
       eprint = {1806.07188},
 primaryClass = {astro-ph.HE},
       adsurl = {https://ui.adsabs.harvard.edu/abs/2018ApJ...862..143N}
}

@ARTICLE{Wheeler2015,
       author = {{Wheeler}, J. Craig and {Johnson}, V. and {Clocchiatti}, A.},
        title = "{Analysis of late-time light curves of Type IIb, Ib and Ic supernovae}",
      journal = {\mnras},
         year = 2015,
        month = jun,
       volume = {450},
       number = {2},
        pages = {1295-1307},
          doi = {10.1093/mnras/stv650},
archivePrefix = {arXiv},
       eprint = {1411.5975},
 primaryClass = {astro-ph.SR},
       adsurl = {https://ui.adsabs.harvard.edu/abs/2015MNRAS.450.1295W}
}

@ARTICLE{Mazzali08,
       author = {{Mazzali}, Paolo A. and {Valenti}, Stefano and {Della Valle}, Massimo and {Chincarini}, Guido and {Sauer}, Daniel N. and {Benetti}, Stefano and {Pian}, Elena and {Piran}, Tsvi and {D'Elia}, Valerio and {Elias-Rosa}, Nancy and {Margutti}, Raffaella and {Pasotti}, Francesco and {Antonelli}, L. Angelo and {Bufano}, Filomena and {Campana}, Sergio and {Cappellaro}, Enrico and {Covino}, Stefano and {D'Avanzo}, Paolo and {Fiore}, Fabrizio and {Fugazza}, Dino and {Gilmozzi}, Roberto and {Hunter}, Deborah and {Maguire}, Kate and {Maiorano}, Elisabetta and {Marziani}, Paola and {Masetti}, Nicola and {Mirabel}, Felix and {Navasardyan}, Hripsime and {Nomoto}, Ken'ichi and {Palazzi}, Eliana and {Pastorello}, Andrea and {Panagia}, Nino and {Pellizza}, L.~J. and {Sari}, Re'em and {Smartt}, Stephen and {Tagliaferri}, Gianpiero and {Tanaka}, Masaomi and {Taubenberger}, Stefan and {Tominaga}, Nozomu and {Trundle}, Carrie and {Turatto}, Massimo},
        title = "{The Metamorphosis of Supernova SN 2008D/XRF 080109: A Link Between Supernovae and GRBs/Hypernovae}",
      journal = {Science},
         year = 2008,
        month = aug,
       volume = {321},
       number = {5893},
        pages = {1185},
          doi = {10.1126/science.1158088},
archivePrefix = {arXiv},
       eprint = {0807.1695},
 primaryClass = {astro-ph},
       adsurl = {https://ui.adsabs.harvard.edu/abs/2008Sci...321.1185M}
}

@ARTICLE{Soderberg08,
       author = {{Soderberg}, A.~M. and {Berger}, E. and {Page}, K.~L. and {Schady}, P. and {Parrent}, J. and {Pooley}, D. and {Wang}, X.-Y. and {Ofek}, E.~O. and {Cucchiara}, A. and {Rau}, A. and {Waxman}, E. and {Simon}, J.~D. and {Bock}, D.~C.-J. and {Milne}, P.~A. and {Page}, M.~J. and {Barentine}, J.~C. and {Barthelmy}, S.~D. and {Beardmore}, A.~P. and {Bietenholz}, M.~F. and {Brown}, P. and {Burrows}, A. and {Burrows}, D.~N. and {Byrngelson}, G. and {Cenko}, S.~B. and {Chandra}, P. and {Cummings}, J.~R. and {Fox}, D.~B. and {Gal-Yam}, A. and {Gehrels}, N. and {Immler}, S. and {Kasliwal}, M. and {Kong}, A.~K.~H. and {Krimm}, H.~A. and {Kulkarni}, S.~R. and {Maccarone}, T.~J. and {M{\'e}sz{\'a}ros}, P. and {Nakar}, E. and {O'Brien}, P.~T. and {Overzier}, R.~A. and {de Pasquale}, M. and {Racusin}, J. and {Rea}, N. and {York}, D.~G.},
        title = "{An extremely luminous X-ray outburst at the birth of a supernova}",
      journal = {\nat},
         year = 2008,
        month = may,
       volume = {453},
       number = {7194},
        pages = {469-474},
          doi = {10.1038/nature06997},
archivePrefix = {arXiv},
       eprint = {0802.1712},
 primaryClass = {astro-ph},
       adsurl = {https://ui.adsabs.harvard.edu/abs/2008Natur.453..469S}
}

@ARTICLE{Corcoran26,
       author = {{Corcoran}, Gregory and {others}},
        title = "{EP250827b}",
      journal = {\apj},
         year = 2026,
        month = dec,
}

@ARTICLE{Yuan26,
       author = {{Yuan}, Weimin and {et al.}},
        title = "{Thermal X-rays breaking out from pre-explosion ejecta of a dying massive star}",
      journal = {submitted},
         year = 2026,
        month = jun,
}

@ARTICLE{Sarin2026_csm,
       author = {{Sarin}, Nikhil and {Hirai}, Ryosuke},
        title = "{From mass-loss histories to lightcurves: a generalised framework for interaction-powered transients}",
      journal = {arXiv e-prints},
         year = 2026,
        month = may,
          eid = {arXiv:2605.19571},
        pages = {arXiv:2605.19571},
          doi = {10.48550/arXiv.2605.19571},
archivePrefix = {arXiv},
       eprint = {2605.19571},
 primaryClass = {astro-ph.HE},
       adsurl = {https://ui.adsabs.harvard.edu/abs/2026arXiv260519571S}
}

@ARTICLE{ashton19,
       author = {{Ashton}, Gregory and {H{\"u}bner}, Moritz and {Lasky}, Paul D. and {Talbot}, Colm and {Ackley}, Kendall and {Biscoveanu}, Sylvia and {Chu}, Qi and {Divakarla}, Atul and {Easter}, Paul J. and {Goncharov}, Boris and {Hernandez Vivanco}, Francisco and {Harms}, Jan and {Lower}, Marcus E. and {Meadors}, Grant D. and {Melchor}, Denyz and {Payne}, Ethan and {Pitkin}, Matthew D. and {Powell}, Jade and {Sarin}, Nikhil and {Smith}, Rory J.~E. and {Thrane}, Eric},
        title = "{BILBY: A User-friendly Bayesian Inference Library for Gravitational-wave Astronomy}",
      journal = {\apjs},
         year = 2019,
        month = apr,
       volume = {241},
       number = {2},
          eid = {27},
        pages = {27},
          doi = {10.3847/1538-4365/ab06fc},
archivePrefix = {arXiv},
       eprint = {1811.02042},
 primaryClass = {astro-ph.IM},
       adsurl = {https://ui.adsabs.harvard.edu/abs/2019ApJS..241...27A}
}

@ARTICLE{Brown13,
       author = {{Brown}, T.~M. and {Baliber}, N. and {Bianco}, F.~B. and {Bowman}, M. and {Burleson}, B. and {Conway}, P. and {Crellin}, M. and {Depagne}, {\'E}. and {De Vera}, J. and {Dilday}, B. and {Dragomir}, D. and {Dubberley}, M. and {Eastman}, J.~D. and {Elphick}, M. and {Falarski}, M. and {Foale}, S. and {Ford}, M. and {Fulton}, B.~J. and {Garza}, J. and {Gomez}, E.~L. and {Graham}, M. and {Greene}, R. and {Haldeman}, B. and {Hawkins}, E. and {Haworth}, B. and {Haynes}, R. and {Hidas}, M. and {Hjelstrom}, A.~E. and {Howell}, D.~A. and {Hygelund}, J. and {Lister}, T.~A. and {Lobdill}, R. and {Martinez}, J. and {Mullins}, D.~S. and {Norbury}, M. and {Parrent}, J. and {Paulson}, R. and {Petry}, D.~L. and {Pickles}, A. and {Posner}, V. and {Rosing}, W.~E. and {Ross}, R. and {Sand}, D.~J. and {Saunders}, E.~S. and {Shobbrook}, J. and {Shporer}, A. and {Street}, R.~A. and {Thomas}, D. and {Tsapras}, Y. and {Tufts}, J.~R. and {Valenti}, S. and {Vander Horst}, K. and {Walker}, Z. and {White}, G. and {Willis}, M.},
        title = "{Las Cumbres Observatory Global Telescope Network}",
      journal = {\pasp},
         year = 2013,
        month = sep,
       volume = {125},
       number = {931},
        pages = {1031},
          doi = {10.1086/673168},
archivePrefix = {arXiv},
       eprint = {1305.2437},
 primaryClass = {astro-ph.IM},
       adsurl = {https://ui.adsabs.harvard.edu/abs/2013PASP..125.1031B}
}

@INPROCEEDINGS{steele2004,
       author = {{Steele}, Iain A. and {Smith}, Robert J. and {Rees}, Paul C. and {Baker}, Ian P. and {Bates}, S.~D. and {Bode}, Michael F. and {Bowman}, Mark K. and {Carter}, Dave and {Etherton}, Jason and {Ford}, Martyn J. and {Fraser}, Stephen N. and {Gomboc}, A. and {Lett}, Robert D.~J. and {Mansfield}, Anthony G. and {Marchant}, Jonathon M. and {Medrano-Cerda}, Gustavo A. and {Mottram}, Christopher J. and {Raback}, D. and {Scott}, A.~B. and {Tomlinson}, M.~D. and {Zamanov}, R.},
        title = "{The Liverpool Telescope: performance and first results}",
    booktitle = {Ground-based Telescopes},
         year = 2004,
       editor = {{Oschmann}, Jr., Jacobus M.},
       series = {Society of Photo-Optical Instrumentation Engineers (SPIE) Conference Series},
       volume = {5489},
        month = oct,
        pages = {679-692},
          doi = {10.1117/12.551456},
       adsurl = {https://ui.adsabs.harvard.edu/abs/2004SPIE.5489..679S}
}

@ARTICLE{zerbi01,
       author = {{Zerbi}, R.~M. and {Chincarini}, G. and {Ghisellini}, G. and {Rondon{\'o}}, M. and {Tosti}, G. and {Antonelli}, L.~A. and {Conconi}, P. and {Covino}, S. and {Cutispoto}, G. and {Molinari}, E. and {Nicastro}, L. and {Palazzi}, E. and {Akerlof}, C. and {Burderi}, L. and {Campana}, S. and {Crimi}, G. and {Danzinger}, J. and {di Paola}, A. and {Fernandez-Soto}, A. and {Fiore}, F. and {Frontera}, F. and {Fugazza}, D. and {Gentile}, G. and {Goldoni}, P. and {Israel}, G. and {Jordan}, B. and {Lorenzetti}, D. and {McBreen}, B. and {Martinetti}, E. and {Mazzoleni}, R. and {Masetti}, N. and {Messina}, S. and {Meurs}, E. and {Monfardini}, A. and {Nucciarelli}, G. and {Orlandini}, M. and {Paul}, J. and {Pian}, E. and {Saracco}, P. and {Sardone}, S. and {Stella}, L. and {Tagliaferri}, L. and {Tavani}, M. and {Testa}, V. and {Vitali}, F.},
        title = "{The REM telescope: detecting the near infra-red counterparts of Gamma-Ray Bursts and the prompt behavior of their optical continuum}",
      journal = {Astronomische Nachrichten},
         year = 2001,
        month = dec,
       volume = {322},
        pages = {275-285},
          doi = {10.1002/1521-3994(200112)322:5/6<275::AID-ASNA275>3.0.CO;2-N},
       adsurl = {https://ui.adsabs.harvard.edu/abs/2001AN....322..275Z}
}

@ARTICLE{Piro21,
       author = {{Piro}, Anthony L. and {Haynie}, Annastasia and {Yao}, Yuhan},
        title = "{Shock Cooling Emission from Extended Material Revisited}",
      journal = {\apj},
         year = 2021,
        month = mar,
       volume = {909},
       number = {2},
          eid = {209},
        pages = {209},
          doi = {10.3847/1538-4357/abe2b1},
archivePrefix = {arXiv},
       eprint = {2007.08543},
 primaryClass = {astro-ph.HE},
       adsurl = {https://ui.adsabs.harvard.edu/abs/2021ApJ...909..209P}
}

@ARTICLE{granot2002,
       author = {{Granot}, Jonathan and {Sari}, Re'em},
        title = "{The Shape of Spectral Breaks in Gamma-Ray Burst Afterglows}",
      journal = {\apj},
         year = 2002,
        month = apr,
       volume = {568},
       number = {2},
        pages = {820-829},
          doi = {10.1086/338966},
archivePrefix = {arXiv},
       eprint = {astro-ph/0108027},
 primaryClass = {astro-ph},
       adsurl = {https://ui.adsabs.harvard.edu/abs/2002ApJ...568..820G}
}

@ARTICLE{zhang24,
       author = {{Zhang}, Lu-Lu and {Zhong}, Shu-Qing and {Xin}, Li-Ping and {Liang}, En-Wei},
        title = "{A Comprehensive Analysis of Textbook-version Afterglow Light Curves of Gamma-Ray Bursts and Implication for Universal Radiation Physics of Baryonic Jets}",
      journal = {\apj},
         year = 2024,
        month = sep,
       volume = {972},
       number = {2},
          eid = {170},
        pages = {170},
          doi = {10.3847/1538-4357/ad5f92},
archivePrefix = {arXiv},
       eprint = {2408.06694},
 primaryClass = {astro-ph.HE},
       adsurl = {https://ui.adsabs.harvard.edu/abs/2024ApJ...972..170Z}
}

@ARTICLE{gavin18,
       author = {{Lamb}, Gavin P. and {Mandel}, Ilya and {Resmi}, Lekshmi},
        title = "{Late-time evolution of afterglows from off-axis neutron star mergers}",
      journal = {\mnras},
         year = 2018,
        month = dec,
       volume = {481},
       number = {2},
        pages = {2581-2589},
          doi = {10.1093/mnras/sty2196},
archivePrefix = {arXiv},
       eprint = {1806.03843},
 primaryClass = {astro-ph.HE},
       adsurl = {https://ui.adsabs.harvard.edu/abs/2018MNRAS.481.2581L}
}

\end{document}